\documentclass[aps,showpacs,preprintnumbers,amsmath, amssymb]{revtex4}

\usepackage{float}
\usepackage{graphics,epsfig}
\usepackage{graphicx}
\usepackage{dcolumn}
\usepackage{bm}

\usepackage{amsmath}
\usepackage{graphicx,subfigure}
\usepackage[usenames]{color}
\usepackage[normalem]{ulem}
\usepackage{textcomp}
\usepackage{gensymb}
\usepackage{xspace}
\usepackage{verbatim}

\begin{document}

\baselineskip=0.8 cm
\title{{\bf Thermodynamic phase transition of Euler-Heisenberg-AdS black hole on free energy landscape}}

\author{Heng Dai$^{1}$, Zixu Zhao$^{1}$\footnote{Corresponding author. zhao$_{-}$zixu@yeah.net}, Shuhang Zhang$^{1}$}
\affiliation{$^{1}$School of Science, Xi'an University of Posts and Telecommunications, Xi'an 710121, China}

\vspace*{0.2cm}
\begin{abstract}
\baselineskip=0.6 cm
\begin{center}
{\bf Abstract}
\end{center}

We study the first order phase transition of Euler-Heisenberg-AdS black hole based on free energy landscape. By solving the Fokker-Planck equation, we research the probability distribution of the system states. The small (large) black hole can have the chance to switch to the large (small) black hole due to the change of the temperature $T$ or Euler-Heisenberg parameter $a$. A higher (lower) $T$ corresponds to a larger probability for a large (small) black hole. The coexistent small and large black hole states can be acquired for some conditions. For $0< a\leq \frac{32}{7} Q^2 $, the small-large black hole phase transition can be acquired with a small $a$. The probability of small (large) black holes will decrease to zero for a large $a$. For a small $a$, a higher peak of the first passage time can be acquired for higher (lower) $T$ or smaller (larger) $a$ with the initial small (large) black hole state. For $a<0$, a smaller (larger) $a$ corresponds to a larger probability for a large (small) black hole. A higher peak of the first passage time can also be obtained for higher (lower) $T$ or smaller (larger) $a$ with initial small (large) black hole state.

\end{abstract}

\maketitle
\newpage
\vspace*{0.2cm}

\section{Introduction}

From the view of information theory, Bekenstein showed that the black hole entropy is proportional to the area \cite{Bekenstein1973}. Bardeen, Carter, and Hawking \cite{Bardeen1973} revealed the formulation of four laws of black hole mechanics. Hawking showed that quantum mechanical effects cause black holes to create and emit particles \cite{Hawking1975}. The laws of black hole mechanics were considered as the laws of thermodynamics. Hawking and Page discovered a phase transition between the Schwarzschild AdS black hole and the thermal AdS space \cite{Hawking1983}. There has been a great interest to include the variation of the cosmological constant in the first law of black hole thermodynamics \cite{Caldarelli,Kastor,Dolan}. The cosmological constant as a pressure and its thermodynamically conjugate variable as a volume, then the black hole mass is identified with enthalpy rather than internal energy \cite{Kastor}. Dolan showed that the Euclidean action is associated with a bridge equation for the Gibbs free energy and not the Helmholtz free energy \cite{Dolan}.

Studying the phase transition of black hole thermodynamics seems to be quite natural since black holes behave as ordinary thermodynamic systems in many respects. Besides Hawking-Page phase transition \cite{Hawking1983}, Davies found that Kerr-Newman black holes undergo a phase transition, where the heat capacity has an infinite discontinuity \cite{Davies}. Hut considered a charged black hole in a box in equilibrium with neutral thermal radiation \cite{Hut}. Treating the cosmological constant as a thermodynamic pressure and its conjugate quantity as a thermodynamic volume, Kubiz\v{n}\'{a}k and Mann studied the $P-V$ criticality of charged AdS black holes and calculated the critical exponents, and showed they are consistent with the van der Waals system \cite{Kubiznak}. The black hole thermodynamic behaviors have been studied widely \cite{Banerjee,Banerjee2011,Lala,zhao2014,Nguyen2015,Hendi2015,Belhaj2015,Hendi2017,Hendi769,Miao2018,Wei044014,Wei044013,Wei104011,Dehyadegari,Mansoori,Zhou,Wei024,Wei084,Xu}.

Recently, Li and Wang introduced a new way to study the dynamic process of phase transition on the free energy landscape, and showed that a black hole or AdS space has a chance to escape from one phase to another due to the thermal fluctuations \cite{Li}. By numerically calculating the Fokker-Planck equation, Li, Zhang, and Wang studied the van der Waals type phase transition in RN-AdS black holes from the viewpoint of the free energy landscape, and obtained the probability distribution of states and time distribution of the first passage kinetic process of black hole state switching \cite{Li2020}. These results revealed some interesting dynamic properties of the phase transition, therefore the dynamic process of the phase transition on the free energy landscape has attracted significant attention recently \cite{Wei2020,Lan,weiNB,Yang}.

On the other hand, according to Dirac electron-positrons theory, Heisenberg and Euler gave the effective Lagrangian of nonlinear electromagnetic fields \cite{Heisenberg}. A renormalization of field strength and charge has been applied to the modified Lagrange function for constant fields by Schwinger \cite{Schwinger}. Yajima and Tamaki constructed static and spherically symmetric black hole solutions in the Einstein-Euler-Heisenberg system \cite{Yajima}. Ruffini, Wu, and Xue \cite{Ruffini} formulated the Einstein-Euler-Heisenberg theory and studied the solutions of nonrotating black holes with electric and magnetic charges in spherical geometry by taking into account the Euler-Heisenberg effective Lagrangian of one-loop nonperturbative quantum electrodynamics contributions. Magos and Breton generalized the solution to the Euler-Heisenberg theory coupled to gravity that represents a nonlinearly charged static black hole by introducing the cosmological constant, and showed that the consistency between the Smarr formula and the first law of black hole thermodynamics \cite{Magos}. More recently, a complementary paper \cite{Ye} appeared. In this work, on the free energy landscape, we study the phase transition in the Euler-Heisenberg-AdS black hole.

The organization of the work is as follows. In Sec. II, we present the basic formula of thermodynamics for the Euler-Heisenberg-AdS black hole. In Sec. III, we study the small-large black hole phase transition on the free energy landscape. We will conclude in the last section of our main results.

\section{The Euler-Heisenberg-AdS black hole}

We consider the background of the Euler-Heisenberg-AdS black hole \cite{Magos}
\begin{eqnarray}\label{metric}
ds^2=-f\left ( r \right )dt^2
+f\left ( r \right )^{-1}dr^2
+r^2\left ( \mathrm{d}\theta ^{2} +\sin ^{2} \theta \mathrm{d}\phi^{2} \right ),
\end{eqnarray}
with
\begin{eqnarray}
f\left ( r \right )=1-\frac{2M}{r}+\frac{Q^2}{r^2}-\frac{\Lambda r^{2}}{3} -\frac{a Q^{4} }{20r^{6} },
\end{eqnarray}
where $M$ is the mass of the black hole and $Q$ is the electric charge, and $\Lambda$ is the cosmological constant; $a=\frac{8{\alpha} ^{2}}{45m^{4}}$ is the Euler-Heisenberg parameter, where $\alpha$ is the fine structure constant and $m$ is the electron mass (we take $c=\hbar=1$). Although the parameter $a$ is positive as introduced by Heisenberg and Euler, as an independent nonlinear model in gravity theory, it is also meaningful to study the case that the parameter $a$ is negative. The case $a=0$ corresponds to the RN-AdS black hole solution.

The mass of the black hole can be written as
\begin{eqnarray}\label{Mass}
M=\frac{r_+}{2}-\frac{\Lambda r_+^{3}}{6}+\frac{Q^{2}}{2r_+}-\frac{a Q^{4}}{40r_+^{5}}.
\end{eqnarray}
The Hawking temperature is given by
\begin{eqnarray}\label{Hawkingtemperature}
T_H=\frac{1}{4\pi}f'(r)|_{r=r_+}=\frac{a Q^{4} }{16\pi r_+^{7}  } -\frac{Q^{2} }{4\pi r_+^{3} } +\frac{1}{4\pi r_+ } -\frac{\Lambda r_+}{4\pi}.
\end{eqnarray}
The thermodynamic pressure takes \cite{Kastor}
\begin{eqnarray}\label{Pressure}
P=-\frac{\Lambda}{8\pi}.
\end{eqnarray}
Therefore, the Hawking temperature can be rewritten as
\begin{eqnarray}\label{stateequation}
T_{H}=\frac{a Q^{4} }{16\pi r_+^{7}  } -\frac{Q^{2} }{4\pi r_+^{3} } +\frac{1}{4\pi r_+ } +2Pr_+.
\end{eqnarray}
The first law takes
\begin{equation}\label{fistlaw}
dM=T_{H} d\,S+V d\,P+\Phi d\,Q +\mathcal{A} d\,a,
\end{equation}
where $\mathcal{A}=\frac{\partial M}{\partial a}$ is the conjugate quantity to the Euler-Heisenberg parameter $a$,
\begin{eqnarray}
S = \pi r_{+}^{2},~~V =\frac{4}{3}\pi r_+^3,~~\Phi=\frac{Q}{r_+}-\frac{a Q^3}{10 r_+^5},~~\mathcal{A}&= -\frac{Q^4}{40 r_{+}^{5}}.
\end{eqnarray}
The Smarr formula takes the form
\begin{equation}
 M=2(T_{H}S-VP+\mathcal{A}a)-\Phi Q.
\end{equation}
We have
\begin{eqnarray}\label{CriticalPressure}
P=\frac{T_{H}}{2r_{+}} -\frac{1}{8\pi r_{+}^{2} } +\frac{Q^{2} }{8\pi r_{+}^{4} } -\frac{aQ^{4} }{32\pi r_{+}^{8} }.
\end{eqnarray}
Therefore, we can obtain the critical point from $\left({\partial P}/{\partial r_+} \right)_{Q,T_{H},a}=0$ and $\left( {\partial^{2}  P}/{\partial r_{_+}^{2} }\right)_{Q,T_{H},a}=0$.
These conditions lead to a third degree equation for $x\equiv 4r_{+c}^{2} $,
\begin{equation}
x^3-24 Q^2 x^2 +448 a Q^4=0,\label{ex}
\end{equation}
which admits three real roots if $ 0< a \leq \frac{32}{7} Q^2 $ and the roots are given by
  \begin{equation}
     x_k= 8Q^2 \left( 2\cos{\left[\frac{1}{3}\arccos{\left( 1-\frac{7
a}{16Q^2}\right)-\frac{2\pi k}{3}}\right]}+1\right), \hspace{.7cm} k=0,1,2.
\label{xk}
\end{equation}
It should be noted that only $x_{0}$ ($k=0$) and $x_{1}$ ($k=1$) are meaningful in the system. There exists only one real root for $a<0$. When $a > \frac{32 Q^2}{7}$, the three solutions of the equation (\ref{ex}) are nonphysical. The critical temperature and pressure can be written as,
 \begin{eqnarray}
     T_{ck} &=& \frac{1}{\pi \sqrt{x_k}} \left( 1-\frac{8Q^2}{x_k}+\frac{64 a
Q^4}{x_k^3}\right),\nonumber\\
     P_{ck} &=& \frac{1}{\pi x_k} \left( \frac{1}{2}-\frac{6Q^2}{x_k}+\frac{56 a
Q^4}{x_k^3}\right), \hspace{.8cm} k=0,1.\label{TcPc}
 \end{eqnarray}
\begin{figure}[htbp]
  \centering
  \includegraphics[width=7cm]{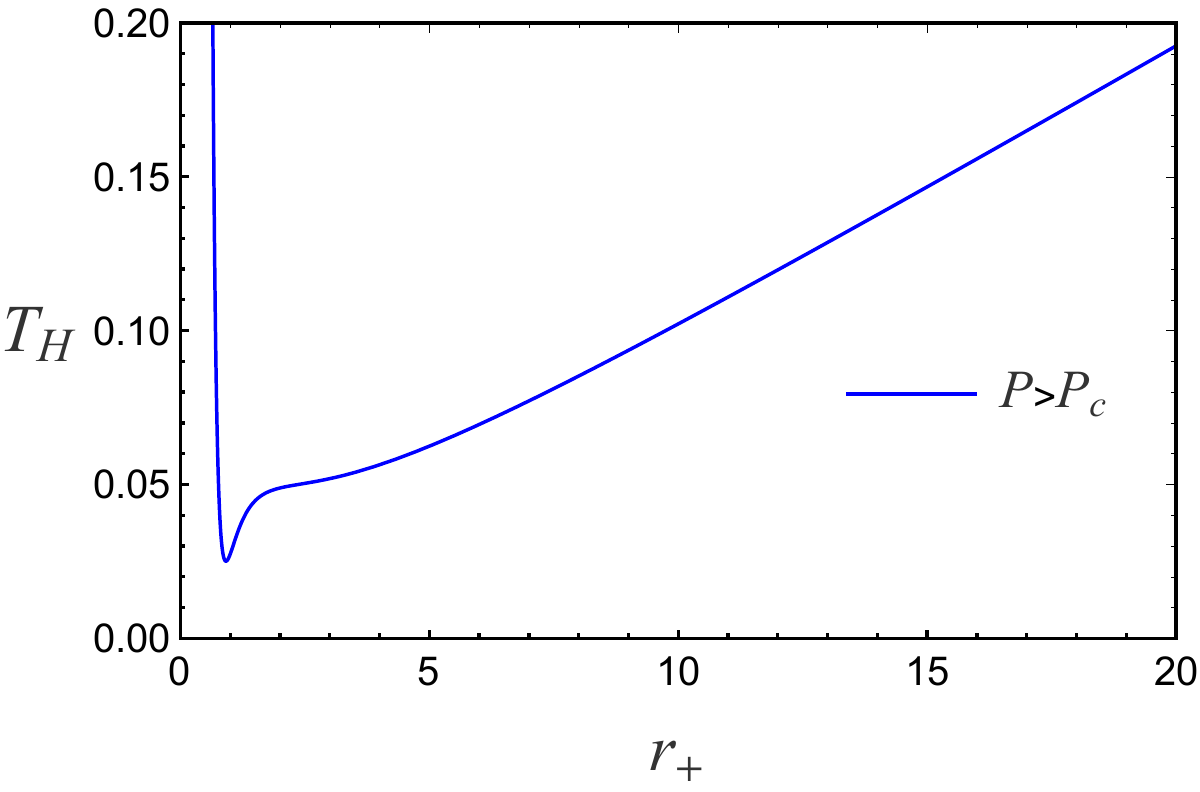}
  \includegraphics[width=7cm]{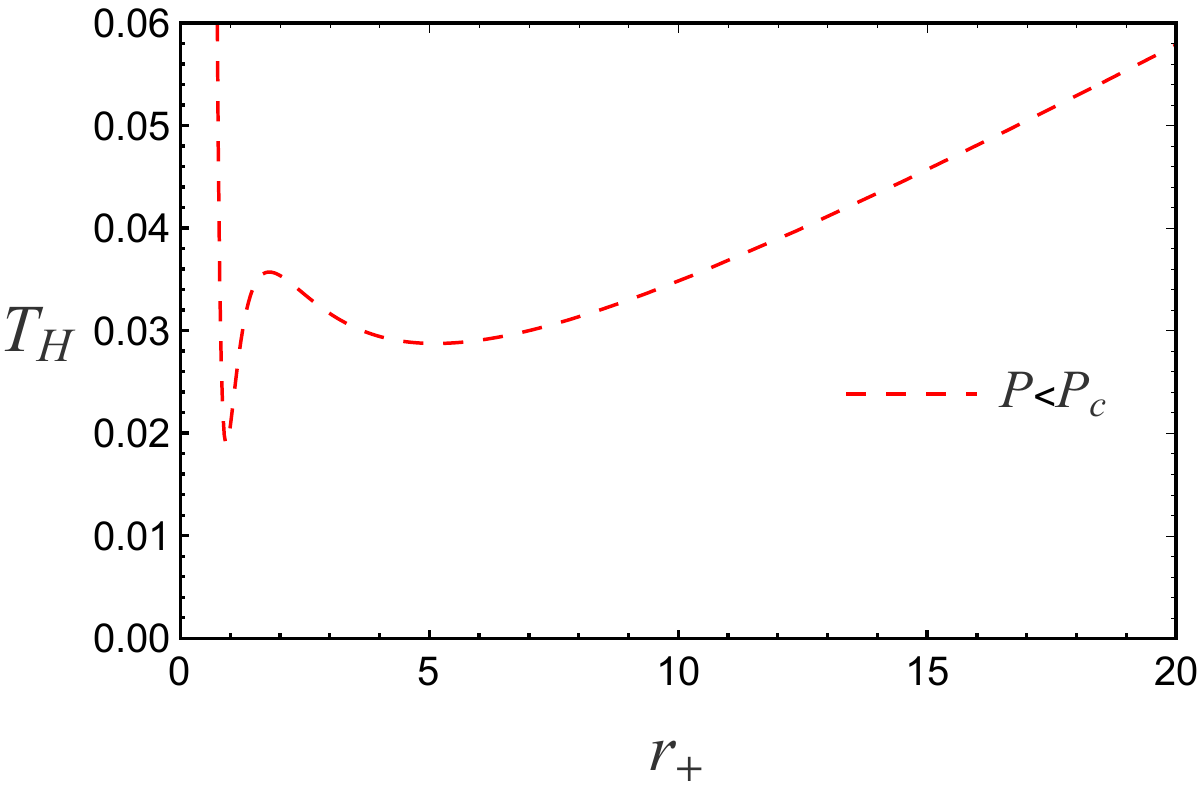}
  \caption{The temperature $T_{H}$ as a function of event horizon radius $r_+$ for $P>P_c$ (the left panel) and $P<P_c$ (the right panel) with $a=0.90$.}
  \label{TraZheng}
\end{figure}
\begin{figure}[htbp]
  \centering
  \includegraphics[width=7cm]{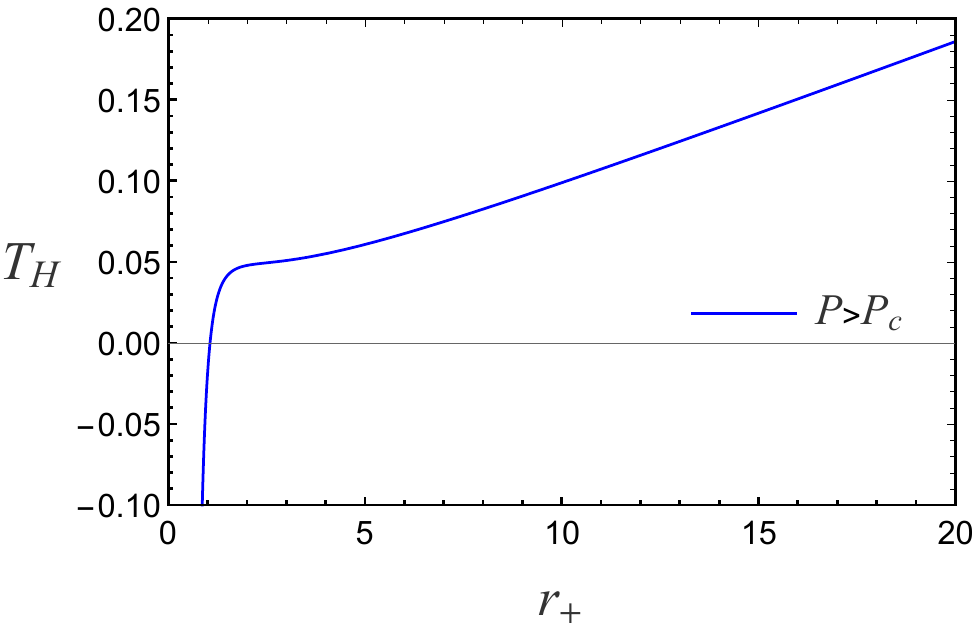}
  \includegraphics[width=7cm]{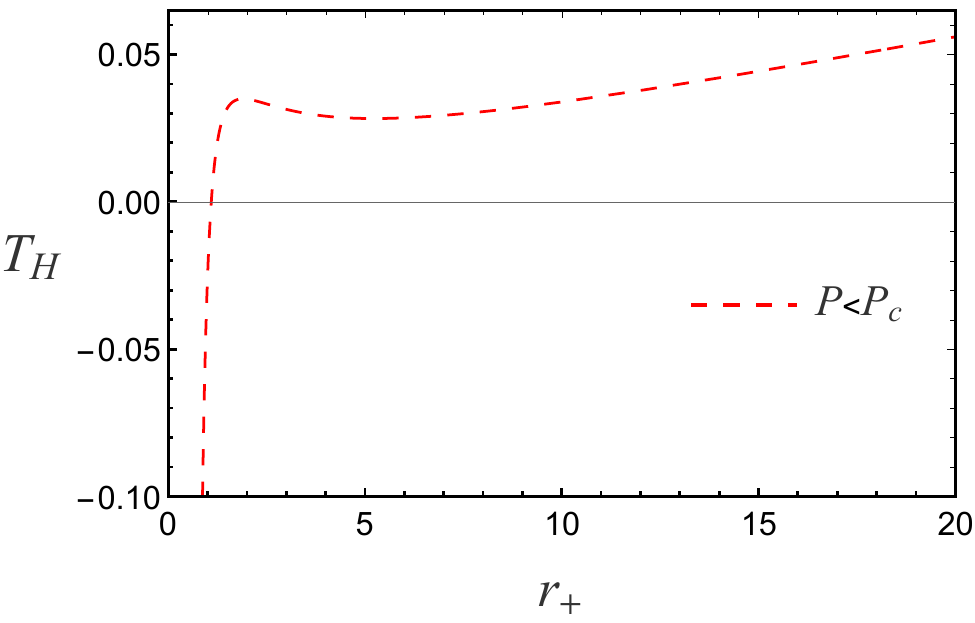}
  \caption{The temperature $T_{H}$ as a function of event horizon radius $r_+$ for $P>P_c$ (the left panel) and $P<P_c$ (the right panel) with $a=-1.35$.}
  \label{TraFu}
\end{figure}
We will specify the branch $k=0$ because the phase transition occurs in reference to the branch $k=0$ \cite{Magos}.  We plot the temperature $T_{H}$ as a function of event horizon radius $r_+$ with $a=0.90$ in Fig. \ref{TraZheng}, and with $a=-1.35$ in Fig. \ref{TraFu}. When $P<P_c$, for a given phase transition temperature $T$,  the local minimum and local maximum values of black hole temperature are determined by
\begin{eqnarray}
\frac{\partial T_H}{\partial r_+}=0\;,
\end{eqnarray}
which gives us the solutions
\begin{eqnarray}\label{rmaxmin}
r_{+_{\min/\max }}=\sqrt{\frac{1}{32J}+\frac{1}{2}\sqrt{\frac{1}{256J^{2}}-\frac{Q^{2}}{4J} + \frac{K}{^{2^{{2}/{3}}J N } }
+\frac{N}{L } }  \pm \frac{1}{2} W },
\end{eqnarray}
where $W=\sqrt{\frac{1}{128J^{2}} -\frac{Q^{2} }{2J } -\frac{K}{2^{{2}/{3} }J N }-\frac{N}{L} +\frac{X}{Y}}$, $X={\frac{1}{512J^{3}} -\frac{3Q^{2}}{16J^{2} } }$, $Y={4\sqrt{\frac{1}{256J^{2}}-\frac{Q^{2} }{4J } +\frac{K }{2^{{2}/{3}}J N} +\frac{N}{L}}}$, $K=3Q^{4}-56aP\pi Q^{4}$, $N=\left(U+\sqrt{-4\left(144Q^{4}-2688aP\pi Q^{4} \right )^{3} +U^{2}}  \right )^{{1}/{3} }$, $U=-3024aQ^{4} +3456Q^{6} +193536aP\pi Q^{6}$, $L=96\times2^{{1}/{3}}P\pi$ and $J=P\pi$.
Then, we can get the local minimum and local maximum values of the black hole temperature
\begin{eqnarray}\label{rmaxminT}
T_{\min / \max  } =\frac{aQ^{4} }{16\pi B^{{7}/{2} } } -\frac{Q^{2} }{4\pi B^{{3}/{2}}} +\frac{1}{4\pi \sqrt{B} }+2P\sqrt{B},
\end{eqnarray}
where $B=\frac{1}{32J} \pm \frac{1}{2}W+\frac{1}{2} \sqrt{\frac{1}{256J^{2} }-\frac{Q^{2} }{4J}+\frac{K}{2^{{2}/{3} }JN } +\frac{N}{L} }$. When $T_{min}<T_H<T_{max}$, the black hole state can have the chance of escape from one stable state to another.

\begin{figure}[H]
  \centering
  \includegraphics[width=8cm]{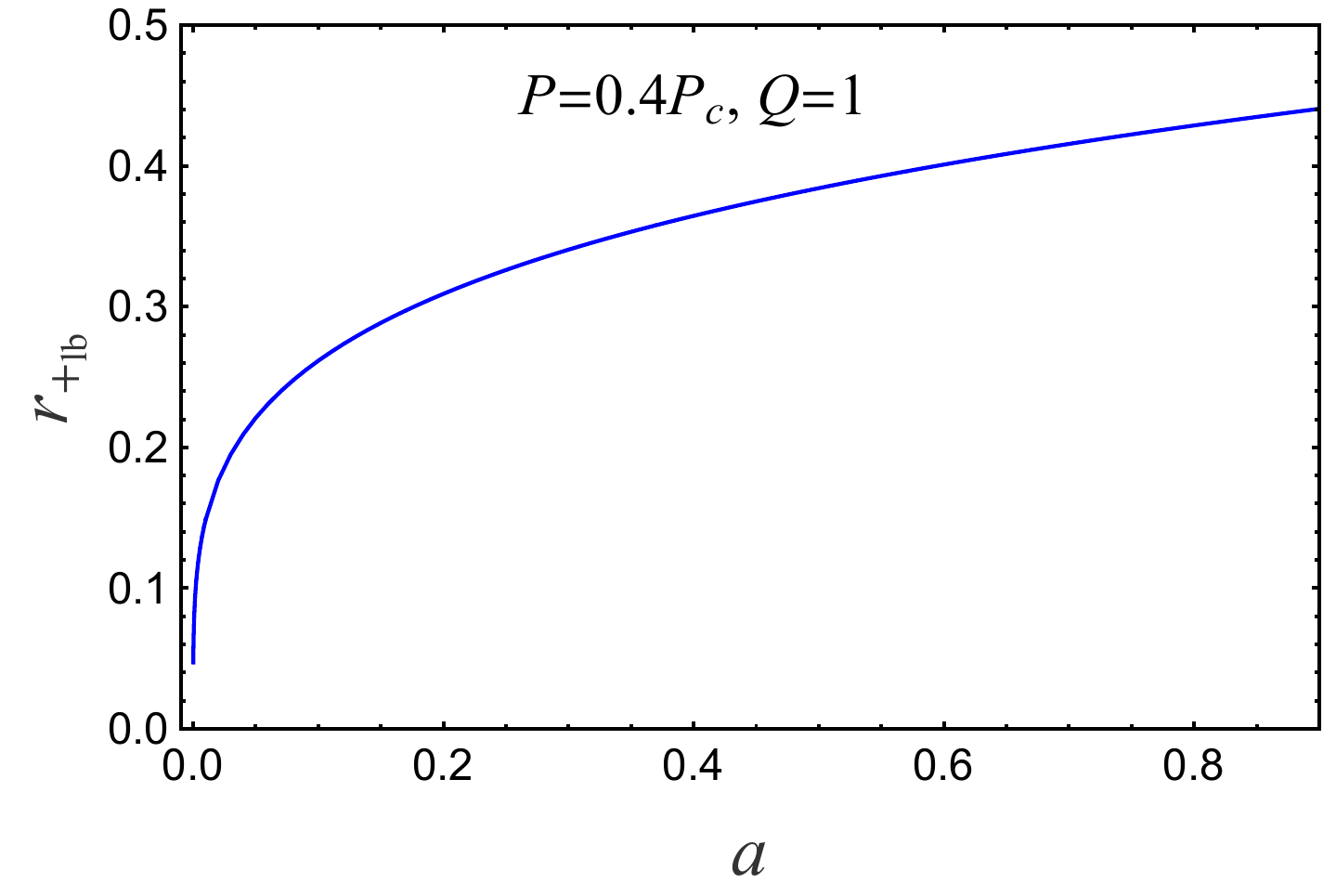}
  \caption{The $r_{+_{lb}}$ as a function of $a$ for $P=0.4P_c$ and $Q=1$.}
  \label{rlb}
\end{figure}

The positivity of the black hole mass can be used to restrict the range of the order parameter $r_+$. When $a=0$, we can take $r_+> 0$. Introducing the Euler-Heisenberg parameter $a$, we can obtain different results. When $a< 0$, we can also take $r_+> 0$. When $0< a\leq \frac{32}{7} Q^2 $, we can obtain the lower bound values of $r_+$ for given parameters. We take $Q=1$ in the following discussion. For example, the lower bound value of $r_+$ is $r_{+_{lb}}=0.288456$ for $a=0.15$, $P=0.4P_c$ and $Q=1$ as shown in Fig. \ref{rlb}.

\section{Phase transition of Euler-Heisenberg-AdS black hole on free energy landscape}

\subsection{The phase transition for the case of $0< a\leq \frac{32}{7} Q^2 $ }

The Gibbs free energy is an important thermodynamic quantity. The Gibbs free energy for Euler-Heisenberg AdS black hole can be given by the thermodynamic relationship $G = M-T_HS$. Replacing the Hawking temperature $T_H$ by the ensemble temperature $T_E$, the Gibbs free energy can be written as
\begin{eqnarray}\label{GibbsEq}
G_L=M-T_ES=\frac{-3a Q^{4}+60Q^{2} r_+^{4} +60r_+^{6} +160P\pi  r_+^{8} }{120r_+^{5} } -\pi T_Er_+^{2},
\end{eqnarray}
where the Gibbs free energy $G_L$ only describes a real black hole when $T_E = T_H$ \cite{Li,Li2020}. From now on, we use $T$ and $r$ to denote $T_E$ and $r_+$ for simplicity.
\begin{figure}[H]
  \centering
  \includegraphics[width=8cm]{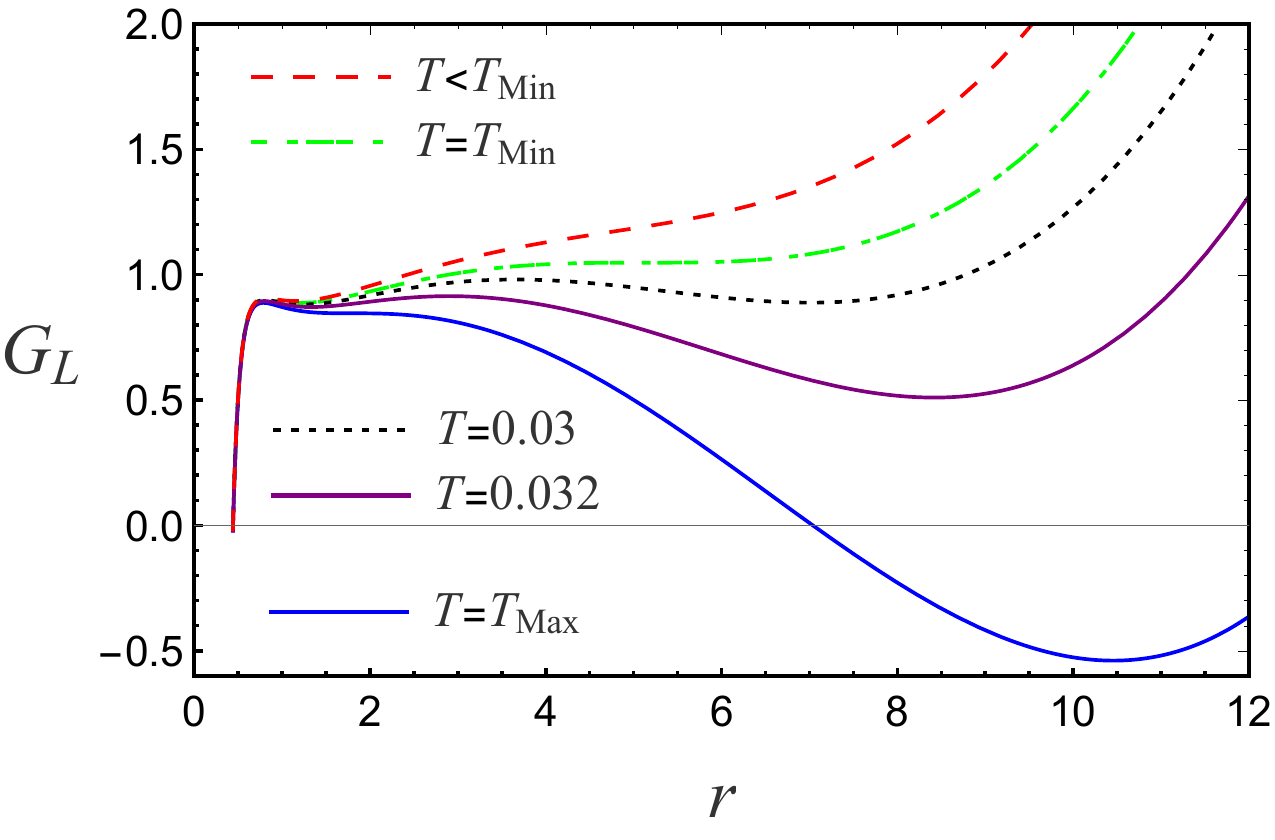}
  \caption{The Gibbs free energy as a function of $r$ for $P=0.4P_c$ and $a=0.90$ with different temperatures.}
  \label{GraZheng}
\end{figure}
In Fig. \ref{GraZheng}, we plot the Gibbs free energy as a function of $r$ for $P=0.4P_c$ and $a=0.90$ with different $T$ so as to quantify the free energy landscape. The two local minimum values for the Gibbs free energy correspond to the small and the large black holes, respectively.

\begin{figure}[H]
  \centering
  \subfigure[]{\includegraphics[width=14cm]{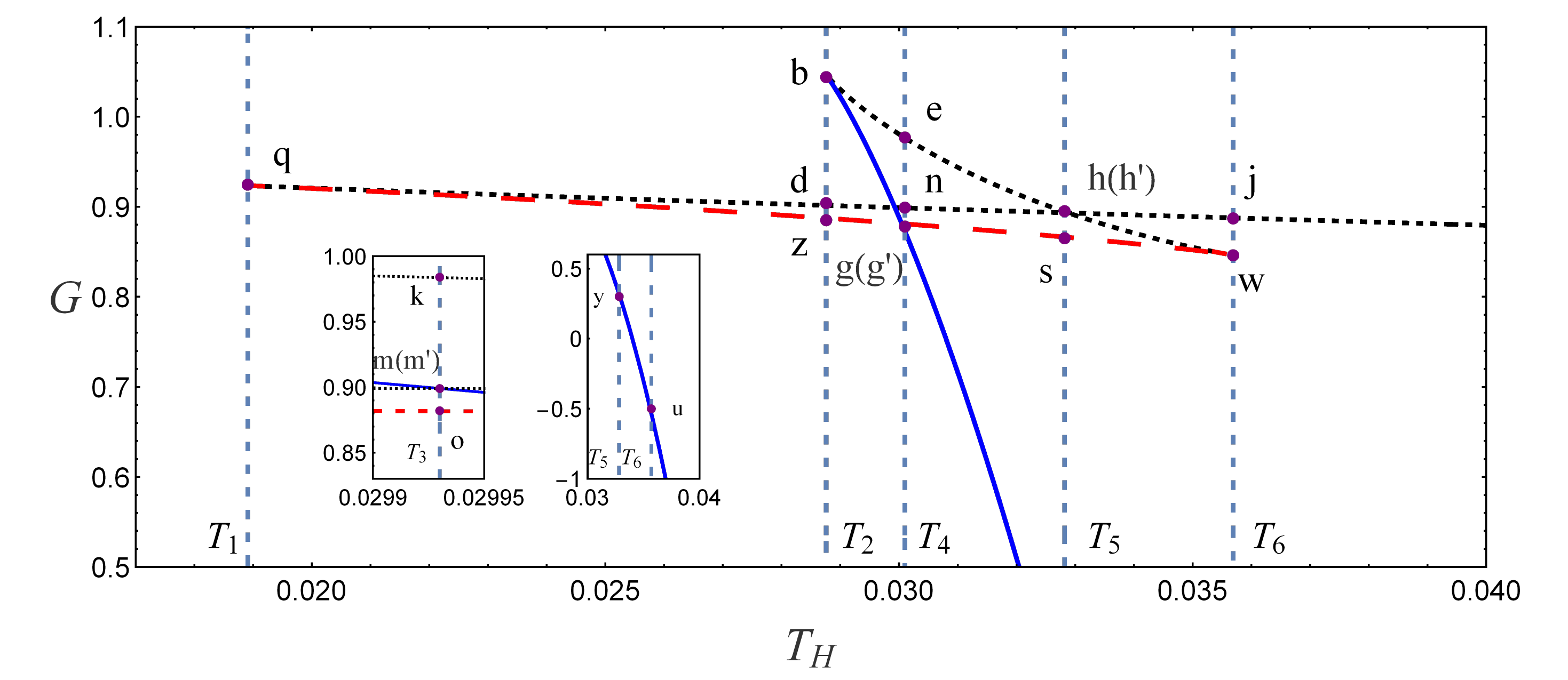}\label{Fig:off-shell0}} \\ \vspace{0.05cm}
 \subfigure[]{\includegraphics[width=6.5cm]{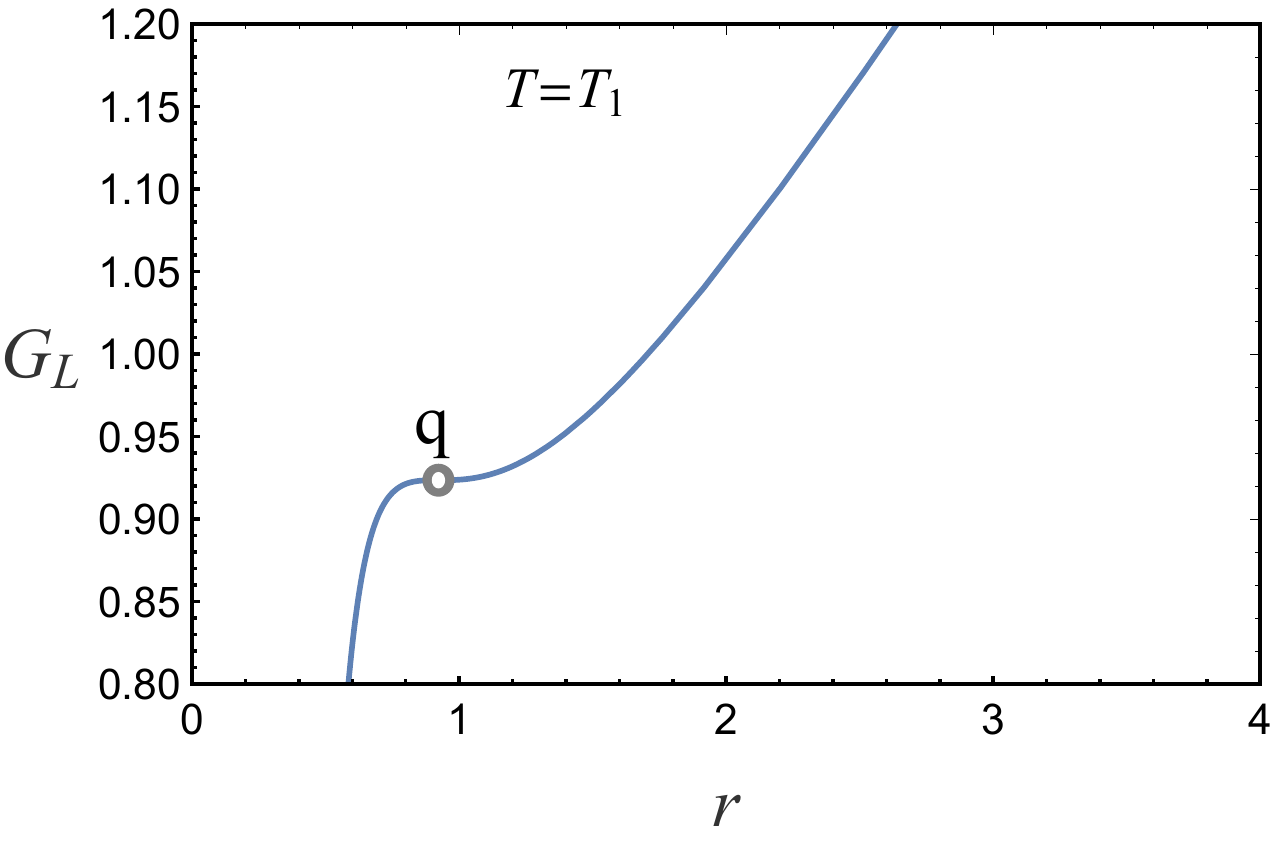}\label{Fig:off-shell1}}
  \subfigure[]{\includegraphics[width=6.5cm]{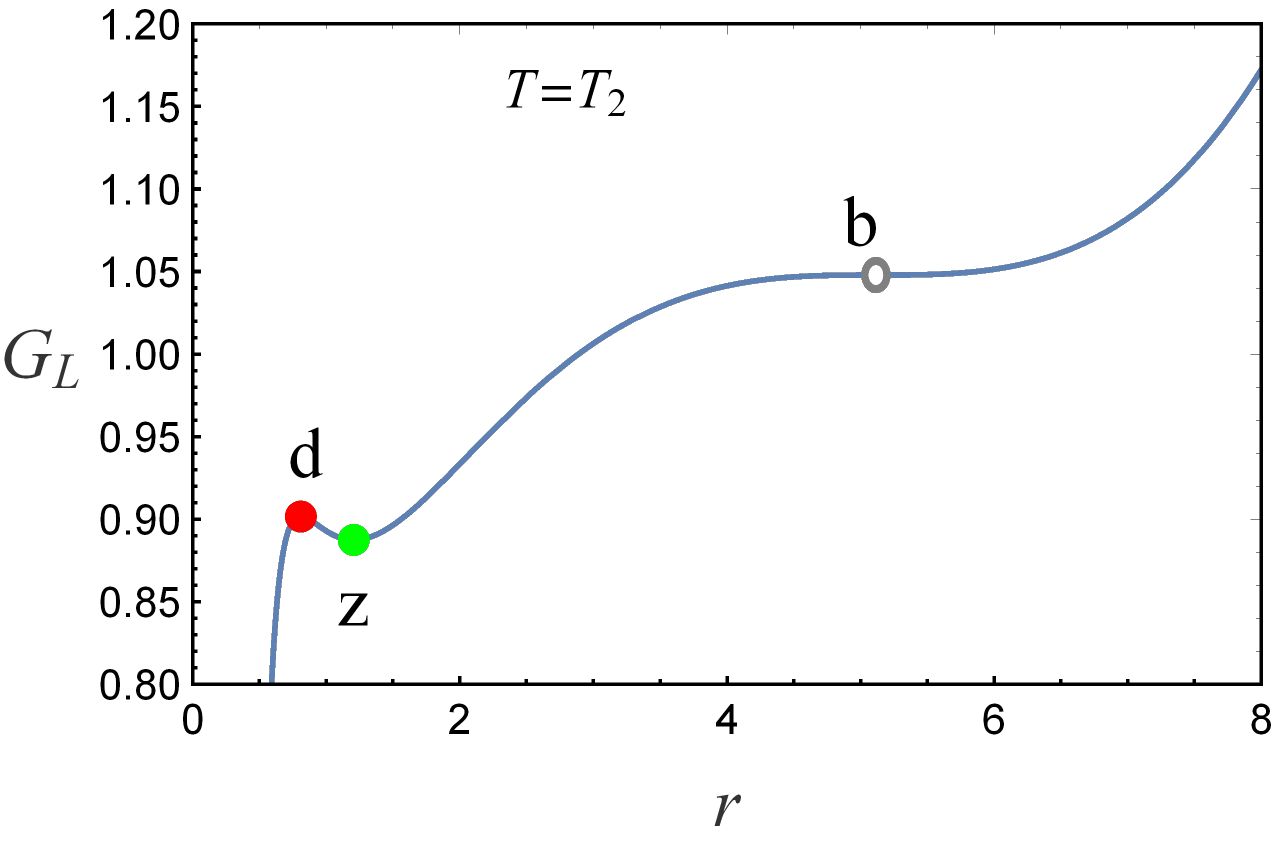}\label{Fig:off-shell2}} \\ \vspace{0.05cm}
  \subfigure[]{\includegraphics[width=6.5cm]{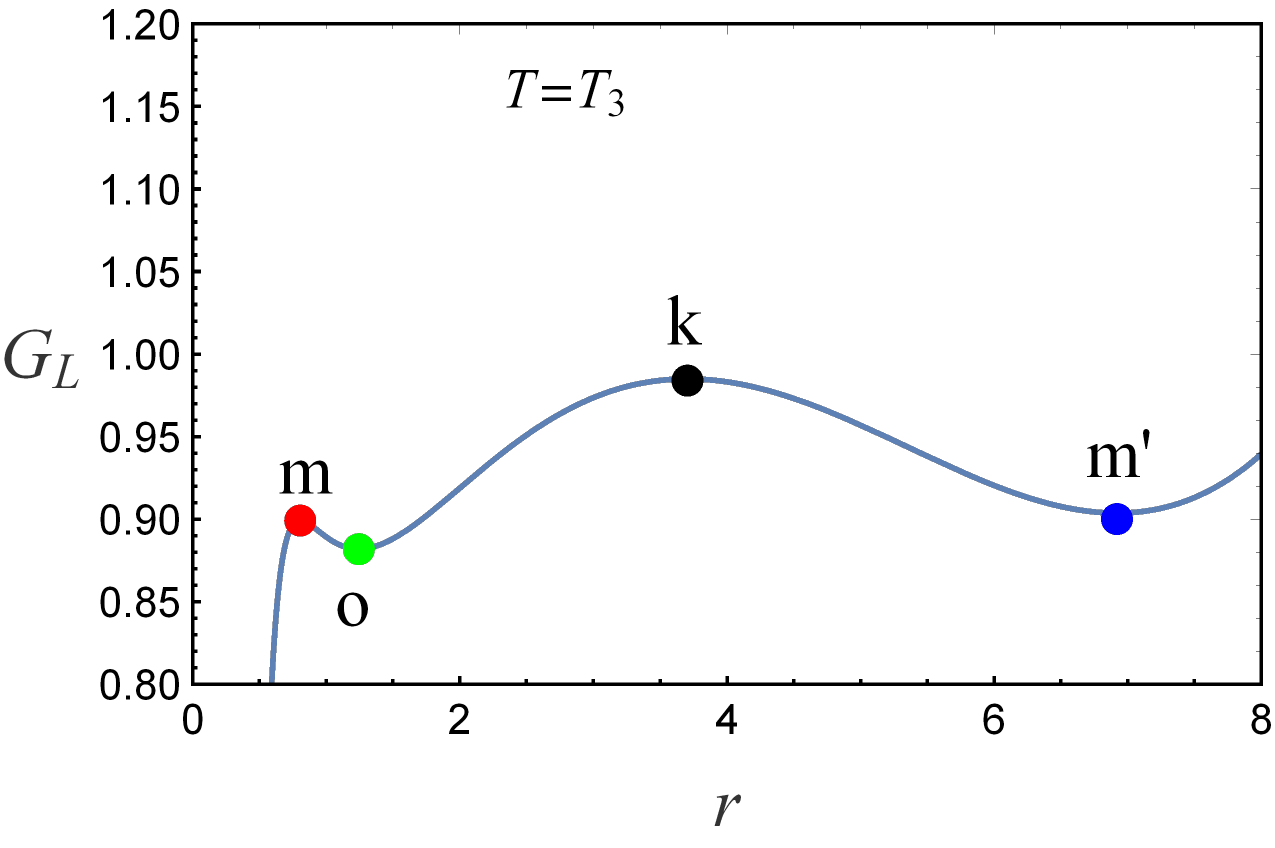}\label{Fig:off-shell3}}
  \subfigure[]{\includegraphics[width=6.5cm]{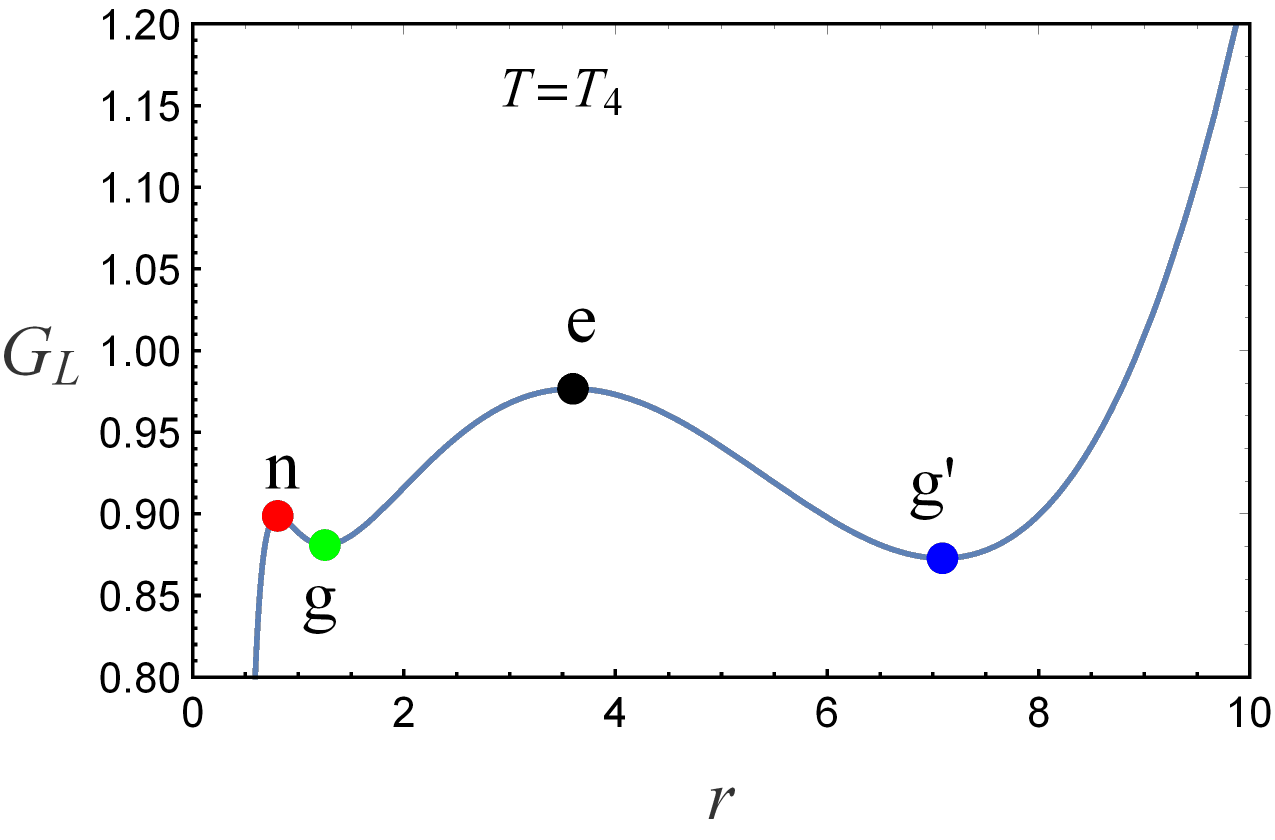}\label{Fig:off-shell4}} \\ \vspace{0.05cm}
  \subfigure[]{\includegraphics[width=6.5cm]{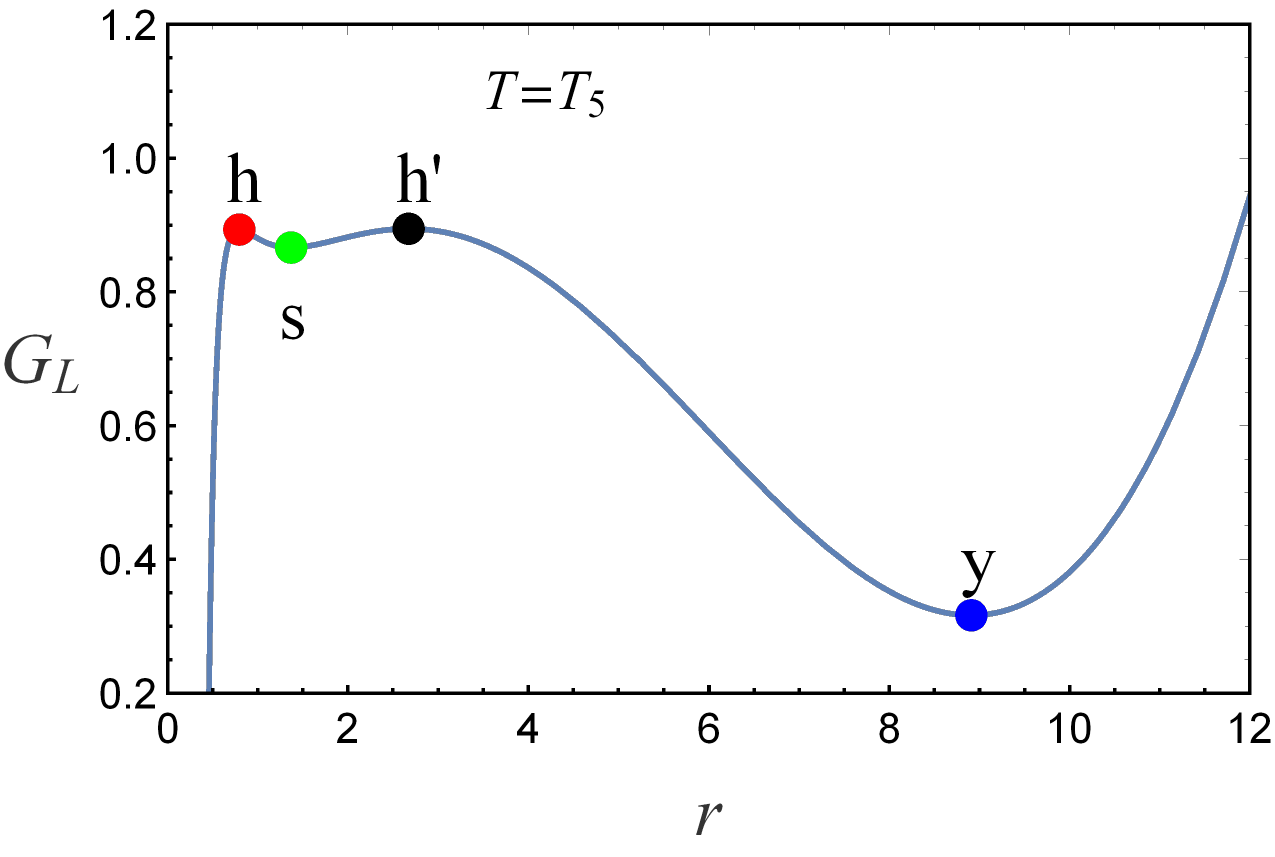}\label{Fig:off-shell5}}
  \subfigure[]{\includegraphics[width=6.5cm]{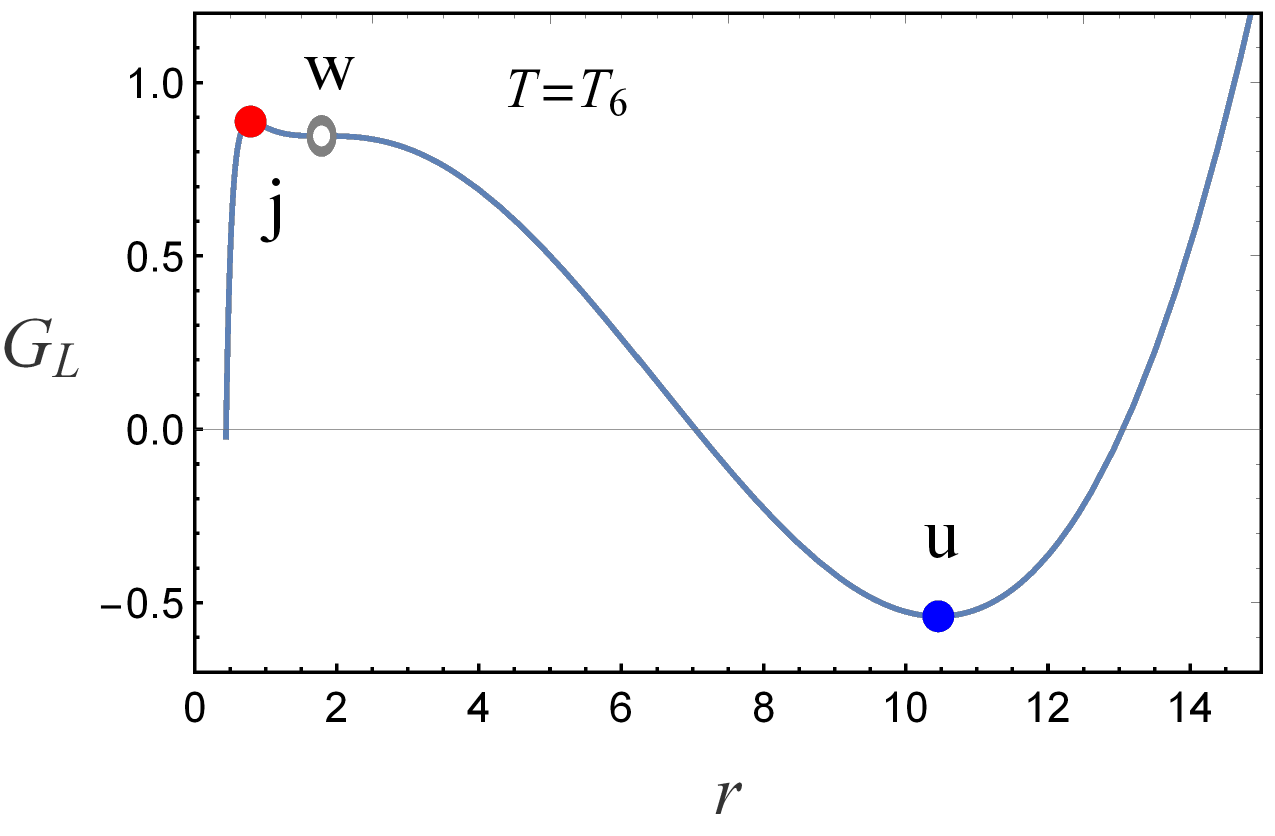}\label{Fig:off-shell6}}
\caption{The behaviors of the Gibbs free energy with $P=0.4P_c$ and $a=0.90$. (a) The Gibbs free energy $G$ vs the black hole temperature $T_{H}$. The solid (blue) and dashed (red) lines correspond to the large black hole and small black hole branches, respectively. (b) The Gibbs free energy $G_{L}$ vs the black hole event horizon $r$ with the temperature of the ensemble $T=T_1$. (c) $G_{L}$ vs $r$ with $T=T_2$. (d) $G_{L}$ vs $r$ with $T=T_3$. (e) $G_{L}$ vs $r$ with $T=T_4$. (f) $G_{L}$ vs $r$ with $T=T_5$. (g) $G_{L}$ vs $r$ with $T=T_6$.}
  \label{atyw}
\end{figure}

In Fig. \ref{Fig:off-shell0}, we show the Gibbs free energy $G$ as a function of the black hole temperature $T_{H}$. The large and small black hole branches are depicted by the solid (blue) and dashed (red) lines. With the increase of the temperature, the black hole system can have theoretical opportunity to go through states $z-o-g(g')-y-u$.

For the six different temperatures $T_1\sim T_6$, we display $G_L$ as a function of $r$ in Fig. \ref{atyw} $(b)\sim (g)$, respectively. When $T>T_1$, for the $T=T_2$, one local minimum point corresponds to the locally stable small black state, as shown in the states $z$ in Fig. \ref{Fig:off-shell2}. When $T=T_3$, four extremal points emerge, as shown in Fig. \ref{Fig:off-shell3}. Two local minimum points correspond to the locally stable small and large black hole states, as shown in the states $o$ and $m'$. Different from the case $T=T_3$, increasing the temperature such that $T=T_4$, although four extremal points are still given, the two local minimum points have the same Gibbs free energy. The states $g$ and $g'$ are the small and large black hole phases, respectively. It should be noted that the small and large black holes cannot exist stably in this case. For the case of $T=T_5$, the system will prefer the state $y$ which has the lower Gibbs free energy compared to state $s$. When $T=T_6$, only one local minimum point exists in the Gibbs free energy, which corresponds to a large black hole state.

\begin{figure}[H]
  \centering
  \includegraphics[width=8cm]{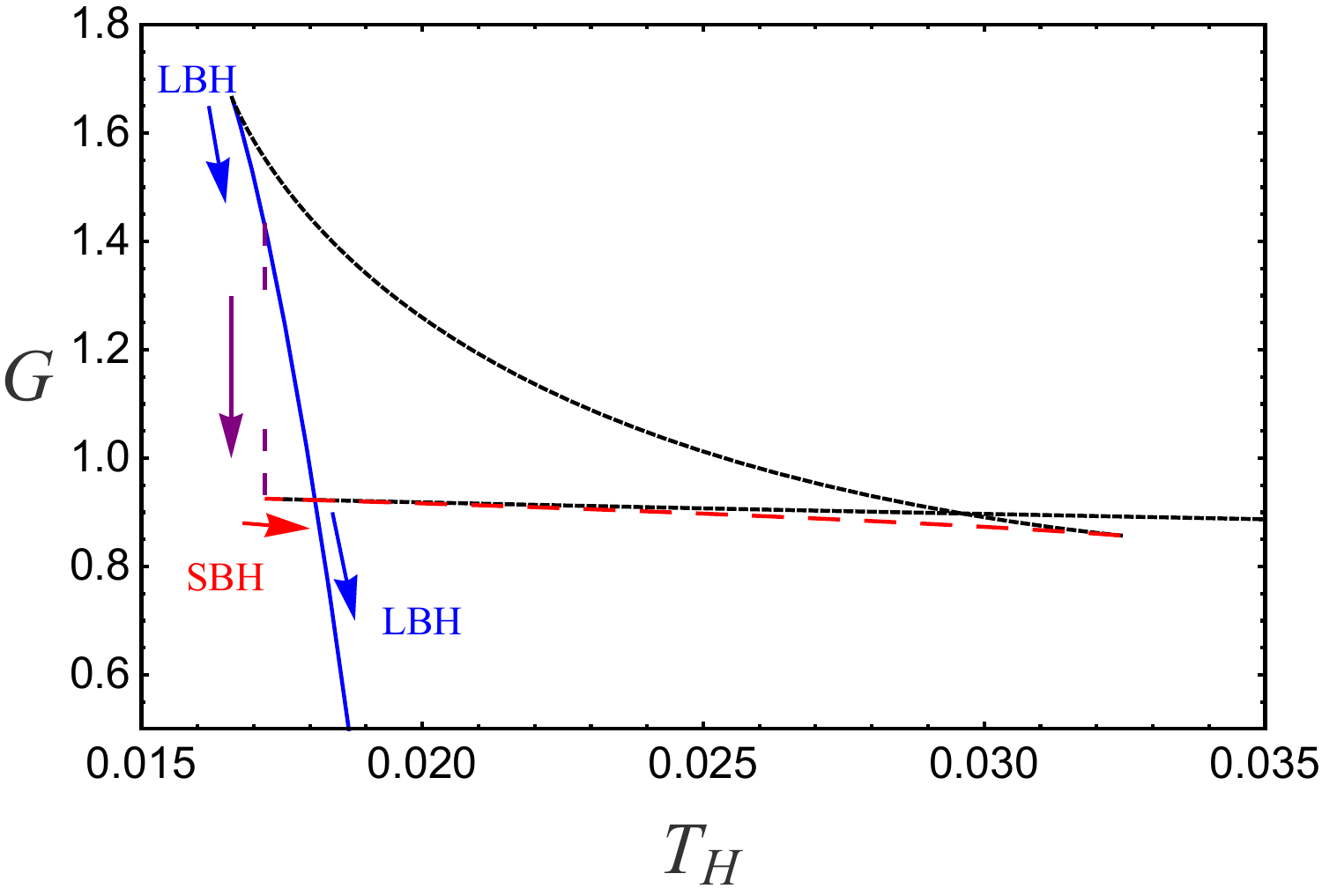}
  \includegraphics[width=8cm]{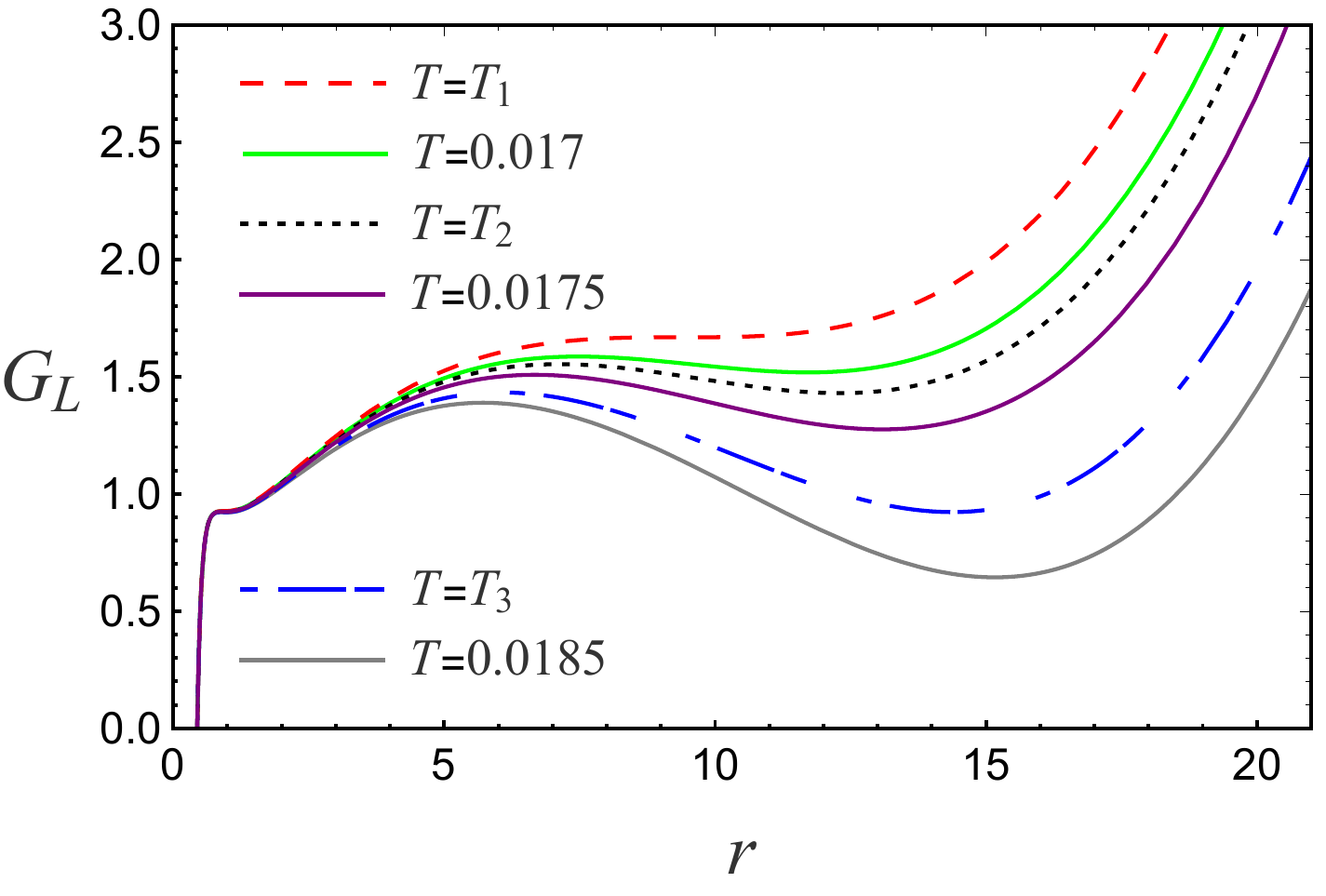}
  \caption{The behaviors of the Gibbs free energy as a function of $T_{H}$ with $P=0.13P_c$ and $a=0.90$ (the left panel) and the Gibbs free energy as a function of $r$ for $P=0.13P_c$ and $a=0.90$ with different temperatures (the right panel).}
  \label{re}
\end{figure}

In Fig. \ref{re}, we depict the Gibbs free energy as a function of $T_{H}$ for $P=0.13P_c$ and $a=0.90$ in the left panel, which shows that reentrant phase transition (large/small/large black hole phase transition) may exist. However, it is hard for reentrant phase transition to occur from the right panel. When $T>T_1$ ($T_1\approx0.016599$), for $T=0.017$, one local minimum point corresponds to the locally stable large black hole state, as shown in the right panel of Fig. \ref{re}. When $T_2<T<T_3$ ($T_2\approx0.017195$ and $T_3\approx0.018094$), two local minimum points correspond to the locally stable small and large black hole states, and the Gibbs free energy of small black hole state is lower than that of large black hole. When $T=T_3$, the two local minimum points have the same Gibbs free energy. After the system goes to a small black hole state from a large black hole state, it is hard for a small black hole state to go back to a large black hole state.

\begin{figure}[H]
  \centering
  \includegraphics[width=8cm]{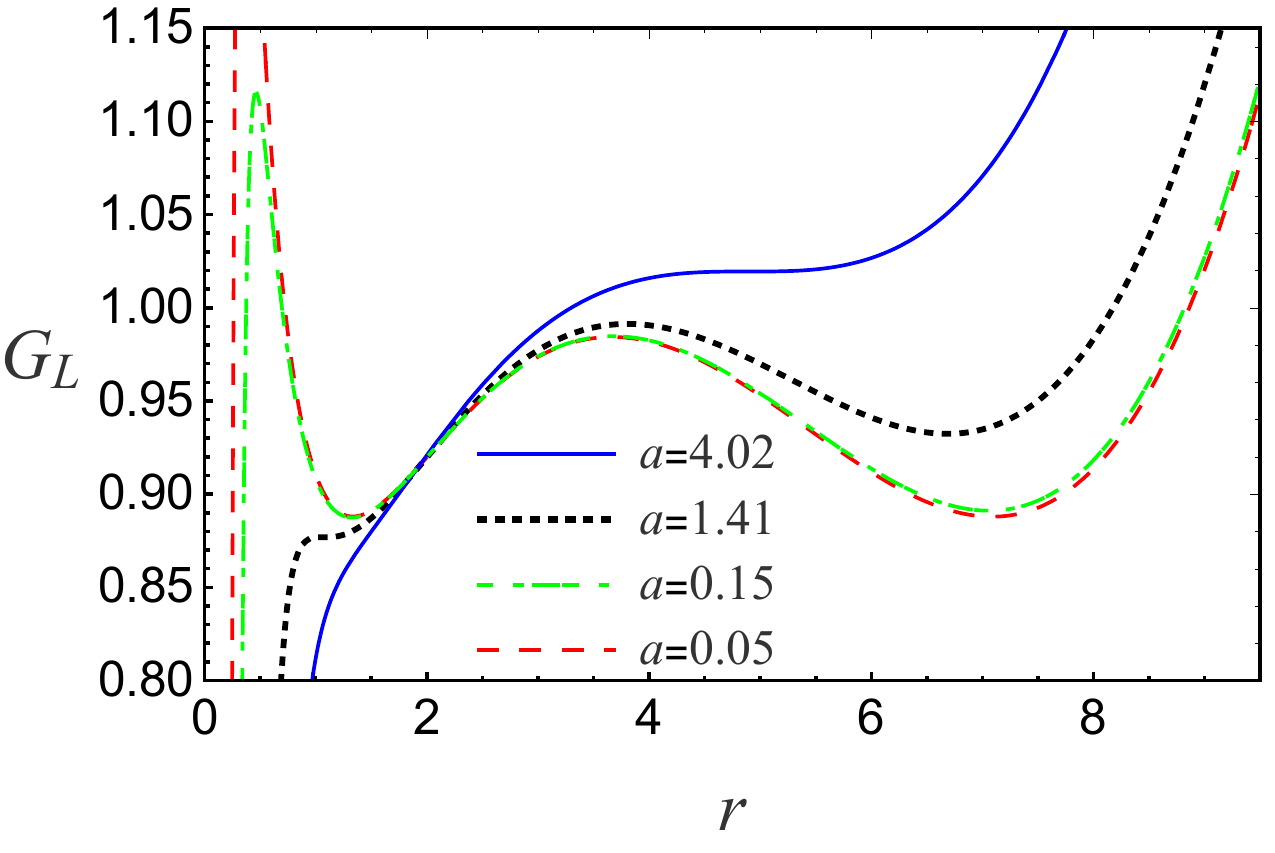}
  \caption{The Gibbs free energy as a function of $r$ for $P=0.4P_c$ and $T=0.02982$ with different Euler-Heisenberg parameters.}
  \label{GrTaZheng}
\end{figure}

For different Euler-Heisenberg parameters, we display $G_L$ as a function of $r$ with $P=0.4P_c$ and $T=0.02982$ in Fig. \ref{GrTaZheng}. When $a_{c_{1}}\leq a\leq 32/7 $ $(a_{c_{1}}\approx4.02)$, there is not extreme points. When $a_{c_{2}}\leq a<a_{c_{1}}$ $(a_{c_{2}}\approx1.41)$, there are two extreme points, and local minimum point corresponds to the locally stable large black hole state. When $0<a<a_{c_{2}}$, four extreme points are presented. Two local minimum points correspond to the locally stable small and large black hole states. When $a=0.15$, the system will prefer the small black hole state which has the lower Gibbs free energy than the large black hole. For $a=0.05$, the two local minimum points have the same Gibbs free energy. This case describes the small-large black hole phase transition, which is a coexistence phase of small and large black holes. From Fig. \ref{GrTaZheng}, we see different behaviors of the Gibbs free energy for different $a$, which can help us further understand small-large black hole phase transition from a new perspective. In order to obtain a stable small (large) black hole state, we need to choose a small $a$.

Now we will study the evolution of the system due to changes of $T$ and $a$. On the free energy landscape, the Gibbs free energy $G_L$ is a function of $r$, and the two local minimal for the Gibbs free energy correspond to the small and the large black holes, respectively. We denote the small and large black holes as $r_{s}$ and $r_{l}$. The probability distribution of the black hole states should be a function of $t$ and $r$. Therefore, the probability distribution of the spacetime state in the ensemble is denoted by $\rho(r, t)$.

The Fokker-Planck equation for the probabilistic evolution on the free energy landscape can be written as  \cite{Bryngelson,Zwanzig,Lee,Lee1,Wang}
\begin{eqnarray}\label{FPequation}
\frac{\partial \rho(r,t)}{\partial t}=D \frac{\partial}{\partial r}\left\{
e^{-\beta G_L(r)}\frac{\partial}{\partial r}\left[e^{\beta G_L(r)}\rho(r,t)\right]
\right\}\;.
\end{eqnarray}
In the above equation, the parameter $\beta=1/(kT)$ and the diffusion coefficient is  $D=kT/\zeta$ with $k$ being the Boltzman constant and $\zeta$ being dissipation coefficient. We set $k=\zeta=1$ in the following.

Solving the above Fokker-Planck equation, we should impose boundary conditions. For example, we show the boundary conditions at $r=r_0$.
Reflecting boundary condition is
\begin{eqnarray}\label{eb1}
\left.
e^{-\beta G_L(r)}\frac{\partial}{\partial r}\left[e^{\beta G(r)}\rho(r,t)\right]\right|_{r=r_0}=0\;.
\end{eqnarray}
It is equivalent to
\begin{eqnarray}\label{b1}
\left.\beta G_L'(r)\rho(r, t)+\rho'(r, t)\right|_{r=r_0}=0\;.
\end{eqnarray}
Absorbing boundary condition is
\begin{eqnarray}\label{b2}
\rho(r_0,t)=0\;.
\end{eqnarray}

We consider the time evolution of the probability of state distribution in the canonical ensemble, which is composed of a series of black hole spacetime for the radius ranging from $0$ to infinity. The reflecting boundary condition will preserve the probability conservation. The Gibbs free energy is divergent at $r=0$ and infinity. The system is in practice confined in a finite regime of the parameters from the certain finite small value to the certain finite large value. The positivity of the black hole mass can also be used to restrict the range of the order parameter $r$. In practice numerical computation, we set the reflecting boundary condition at $r=0.3$ and $r=12$ in order to avoid the numerical instability. We choose the initial condition
\begin{eqnarray}\label{initial}
\rho(r,0)=\frac{1}{\sqrt{\pi}\tilde{\alpha}} e^{-(r-r_i)^2/\tilde{\alpha}^2}\;,
\end{eqnarray}
with the parameter $\tilde{\alpha}=0.1$. This Gaussian wave packet is a good approximation of $\delta$-distribution for the numerical computation process. We take $r_i=r_s$ or $r_i=r_l$ as the initial condition of system, which means that the system initially is in the small or large black hole state.

\begin{figure}[H]
  \centering
  \includegraphics[width=6.5cm]{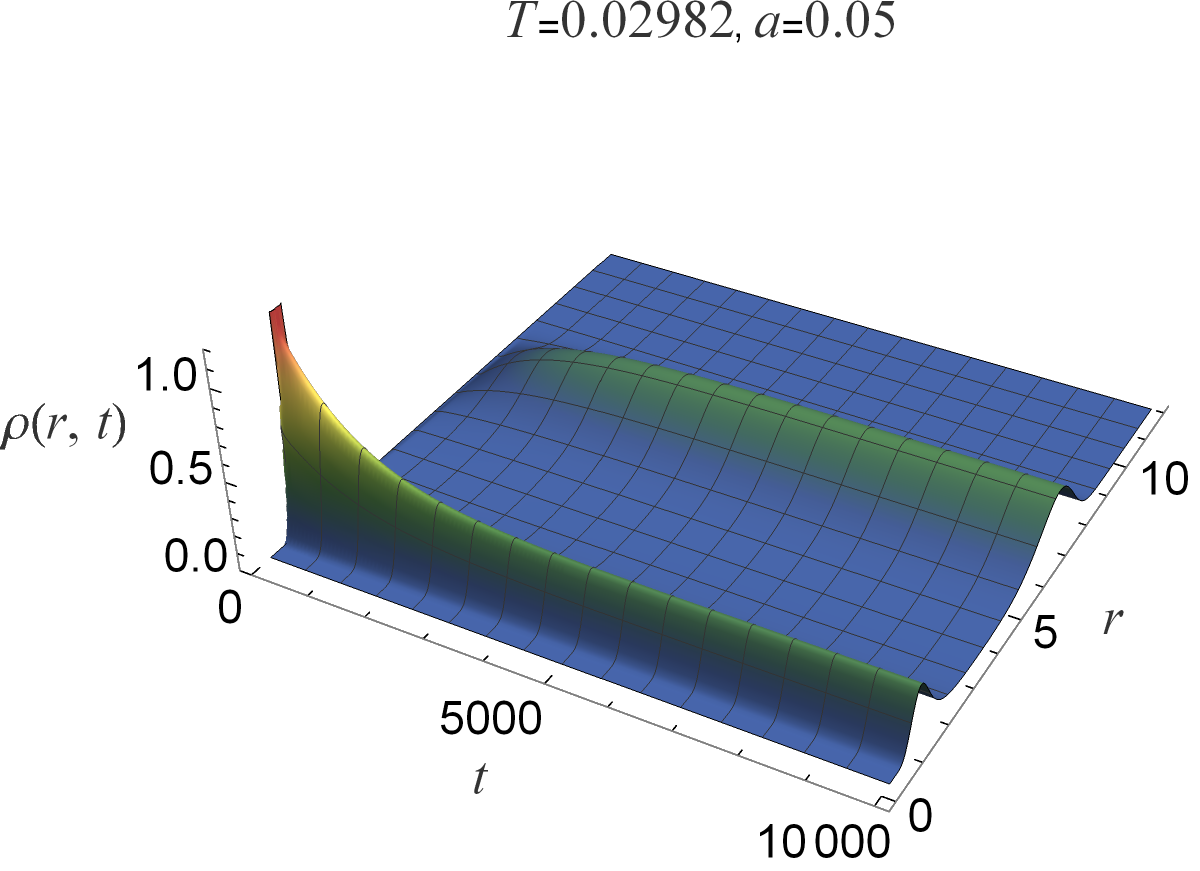}
  \includegraphics[width=6.5cm]{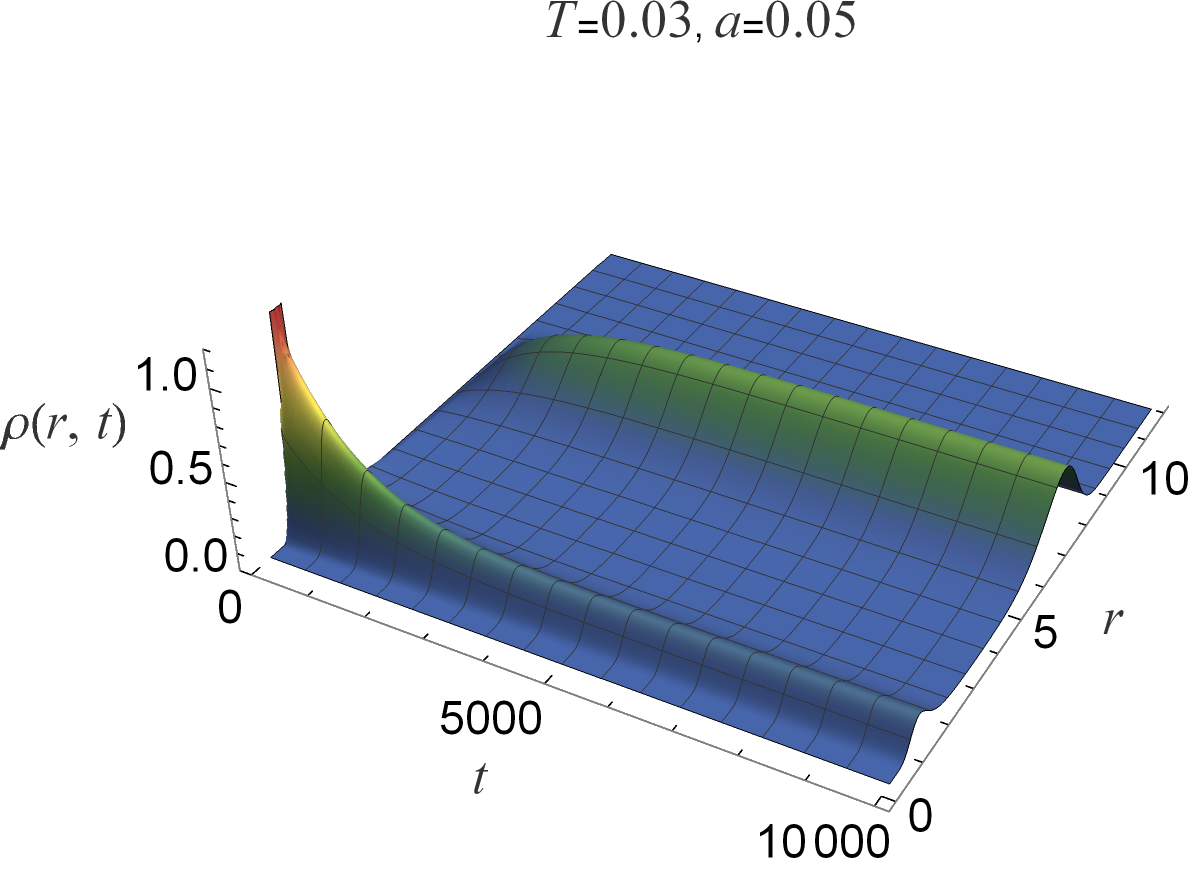}\\ \vspace{0.3cm}
  \includegraphics[width=6.5cm]{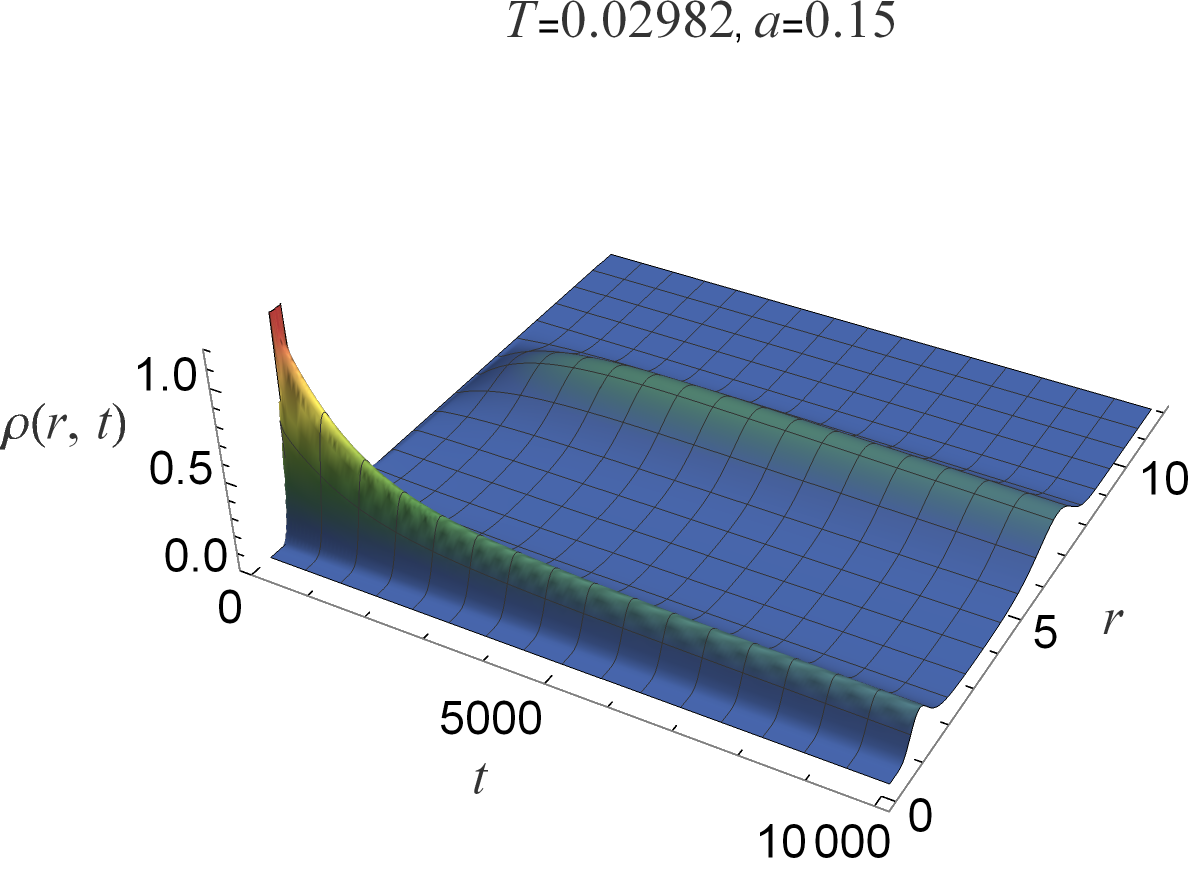}
  \includegraphics[width=6.5cm]{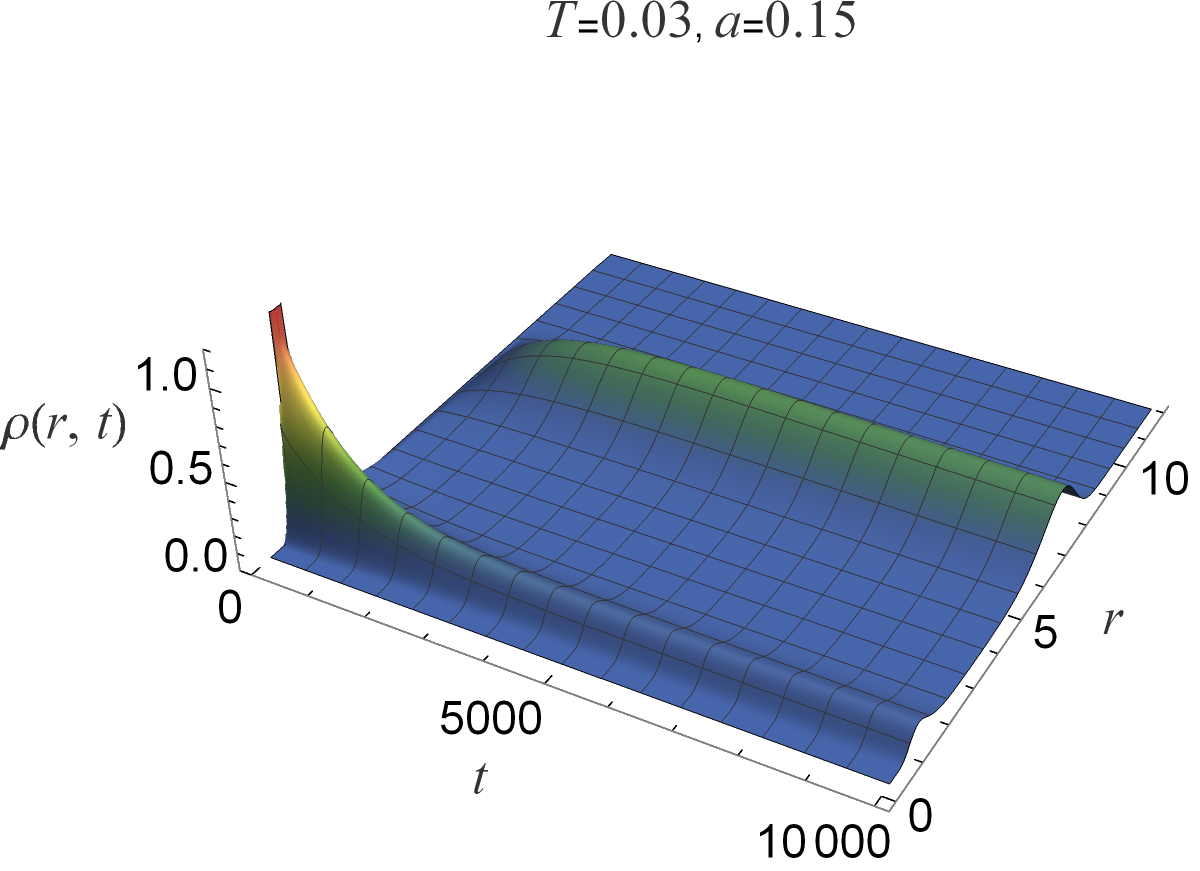}
  \caption{The distributions of probability $\rho(r, t)$ as a function of $r$ and $t$ for different $T$ and $a$. In the left and right columns, $T=0.02982$ and $0.03$. In the top and bottom rows, $a=0.05$ and $0.15$. The initial wave packet is located at the small black hole.}
  \label{aZheng3DS}
\end{figure}

\begin{figure}[H]
  \centering
  \includegraphics[width=6.5cm]{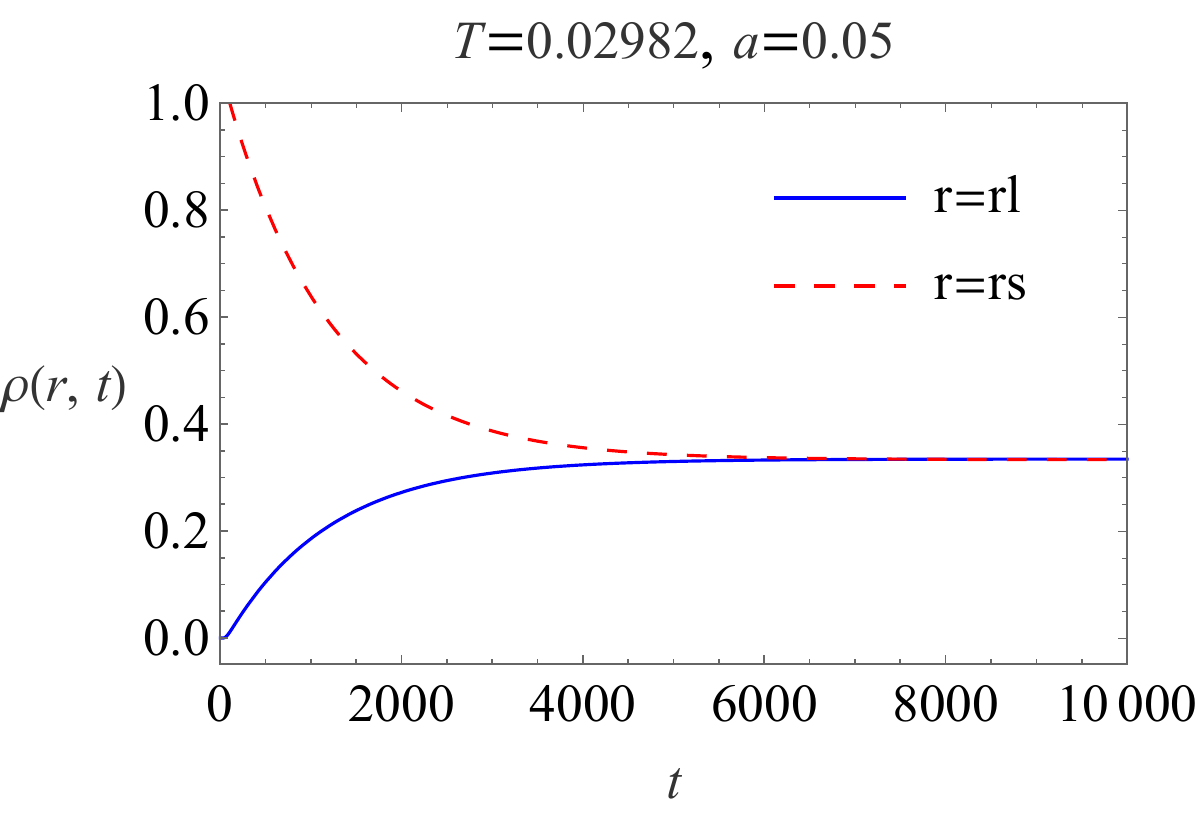}
  \includegraphics[width=6.5cm]{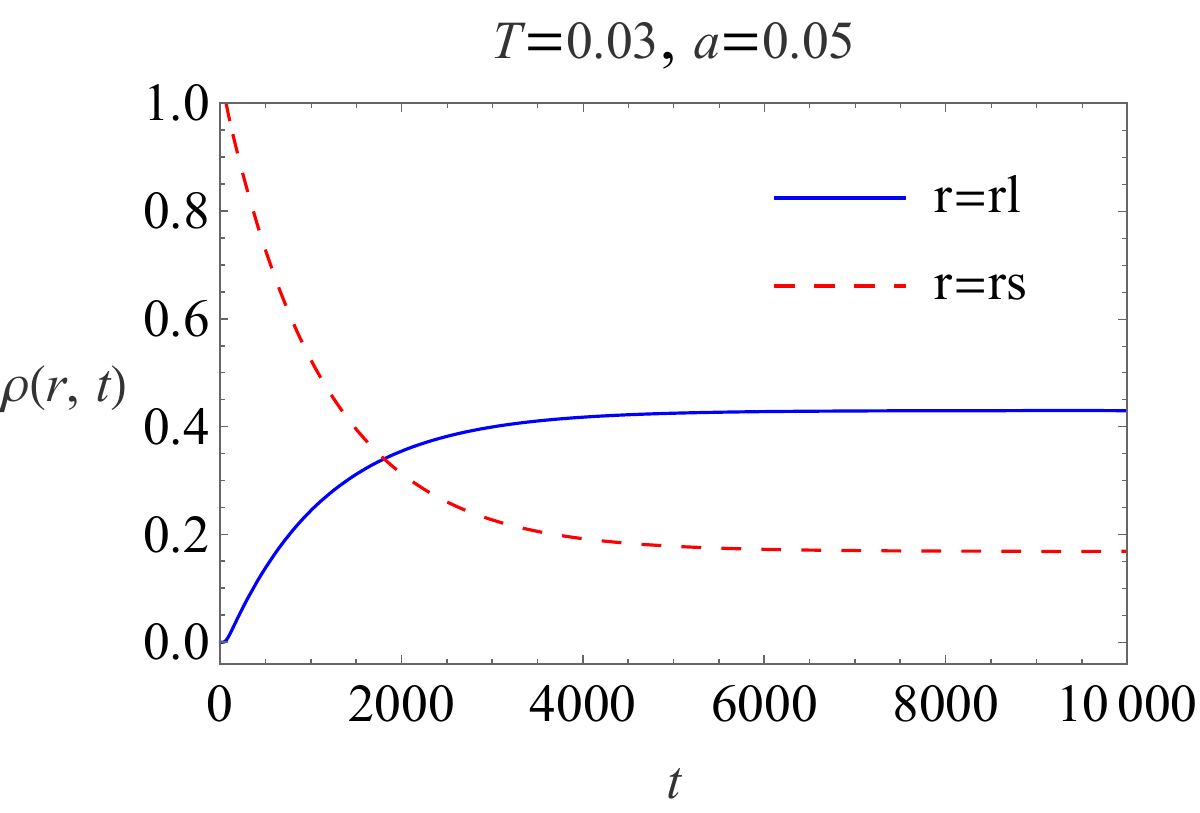}\\ \vspace{0.3cm}
    \includegraphics[width=6.5cm]{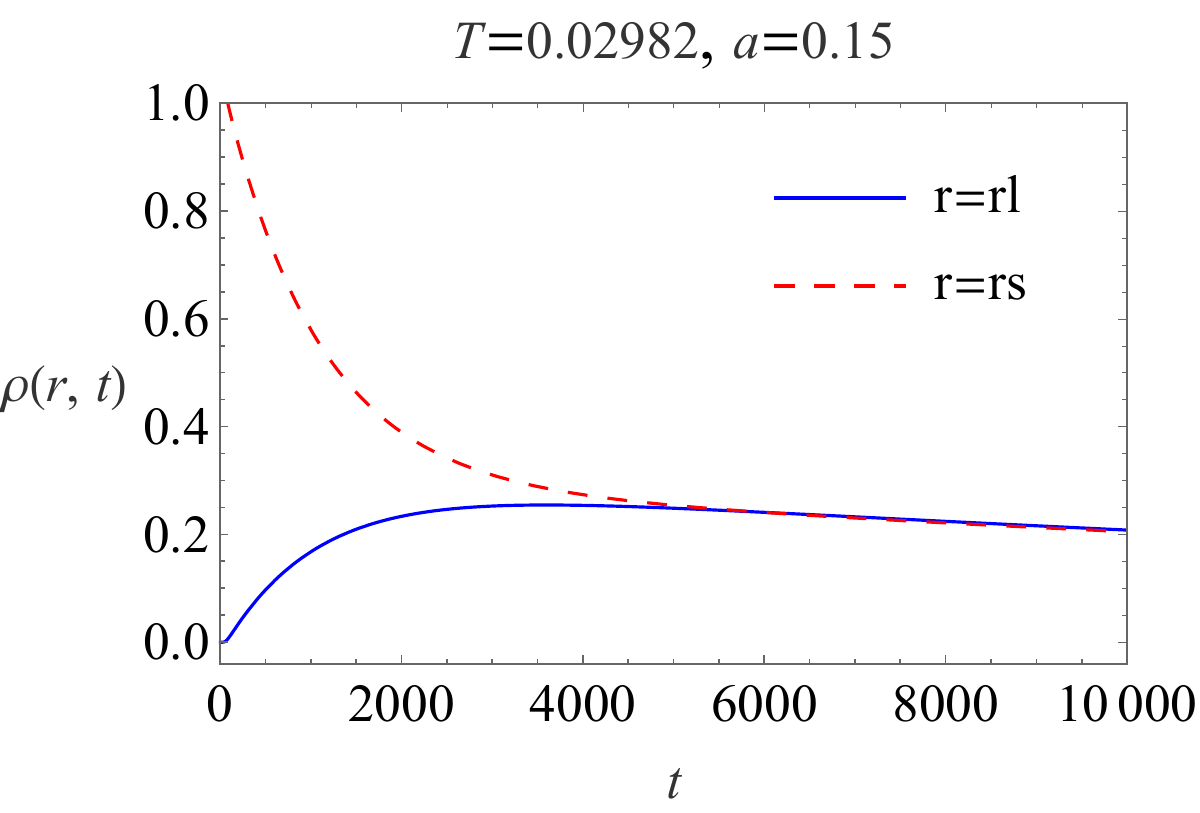}
  \includegraphics[width=6.5cm]{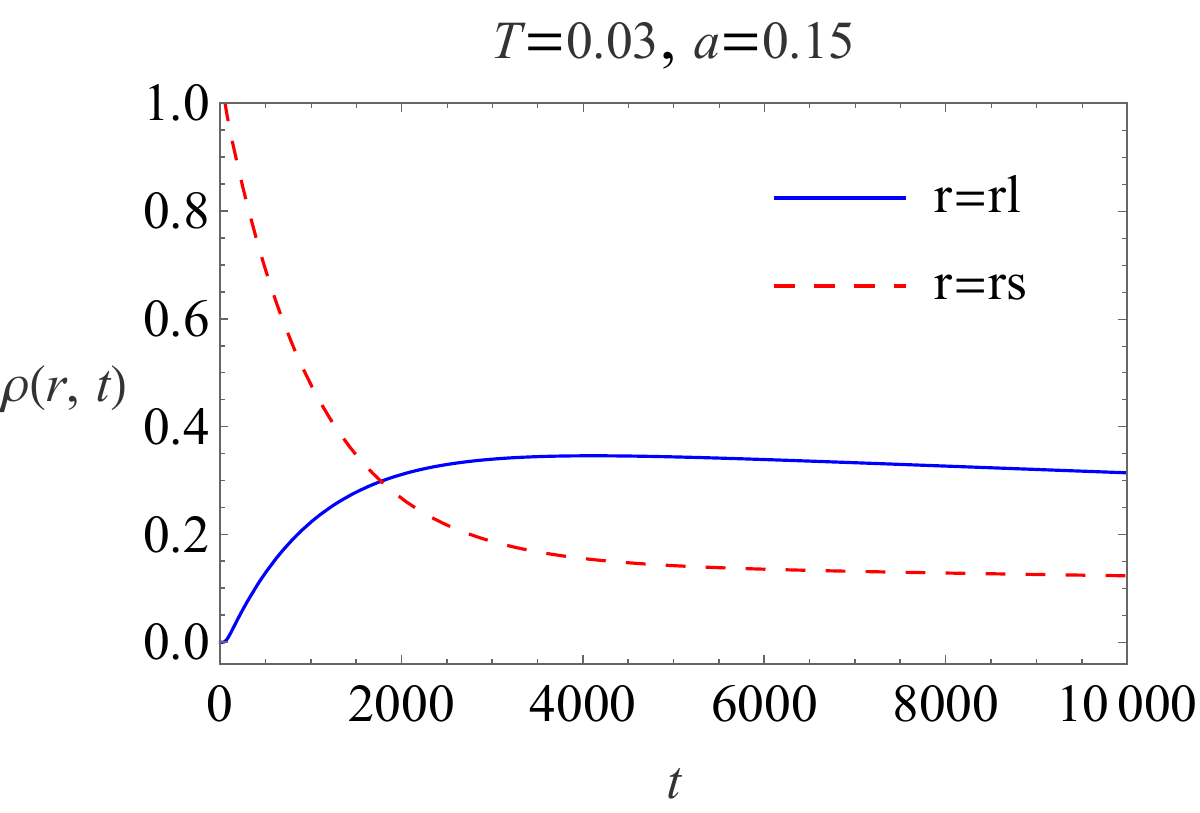}
  \caption{Behaviors of the probability $\rho(r, t)$ as a function of $t$ for different $T$ and $a$. In the left and right columns, $T=0.02982$ and $0.03$. In the top and bottom rows, $a=0.05$ and $0.15$. The initial wave packet is located at the small black hole.}
  \label{aZheng2DS}
\end{figure}

We discuss the case of the time evolution of the probability distribution $\rho(r, t)$. The initial wave packet is located at the small black hole. In Fig. \ref{aZheng3DS} and Fig. \ref{aZheng2DS}, we exhibit the behaviors of the probability distribution of the black holes at different $T$ and $a$. We observe that the probability of small black hole decreases with the evolution of time, and will reach a stable non-zero value for a smaller $a$. However, it will go to zero for a larger $a$. Furthermore, we see that it decreases faster for higher $T$ or larger $a$ in a very early time. The probability of large black hole first increases and then decreases for a larger $a$. However, it will increase from zero to the stable value for a smaller $a$. In addition, a larger stable value of the probability of large black holes can be acquired for higher $T$ or smaller $a$. The small black hole state can have the chance to switch to large black hole state. The coexistent small and large black hole states can be acquired for $T=0.02982$ and $a=0.05$.

\begin{figure}[H]
  \centering
 \includegraphics[width=6.5cm]{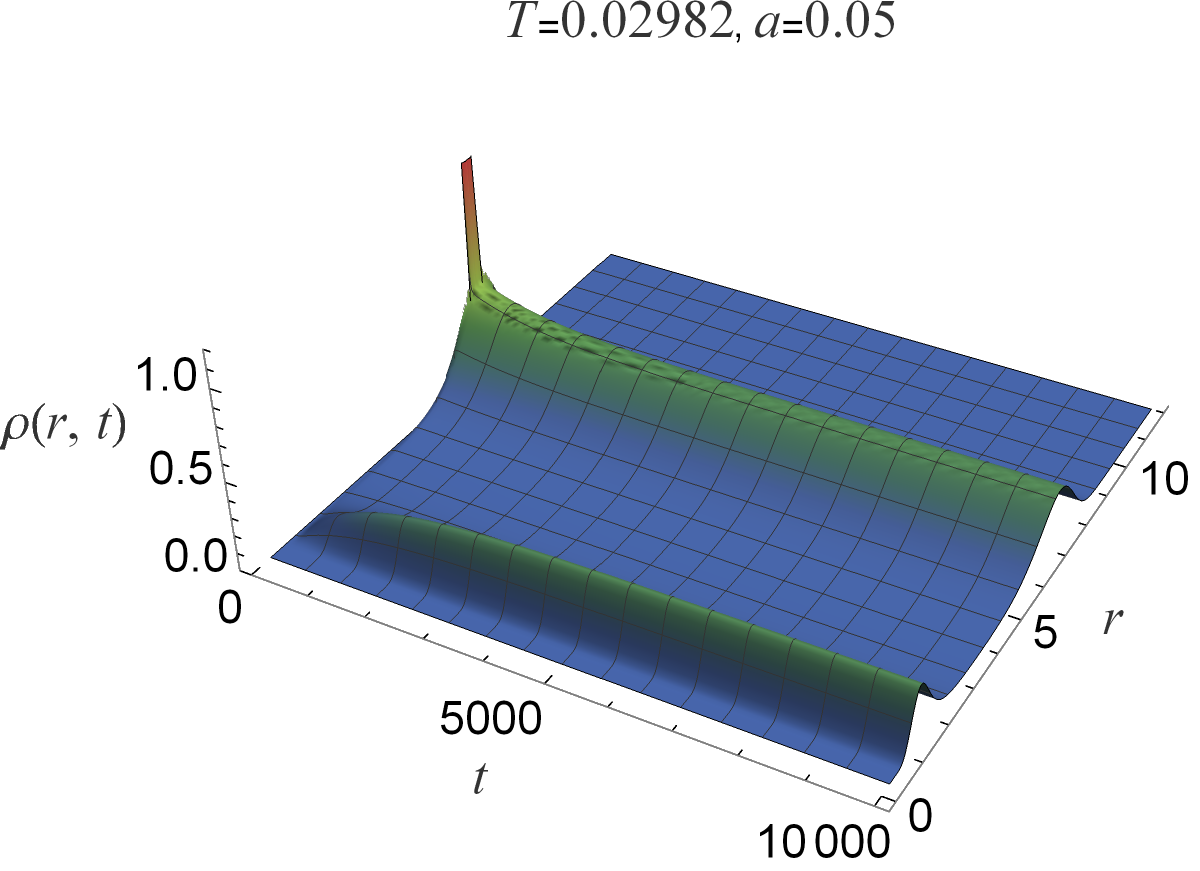}
  \includegraphics[width=6.5cm]{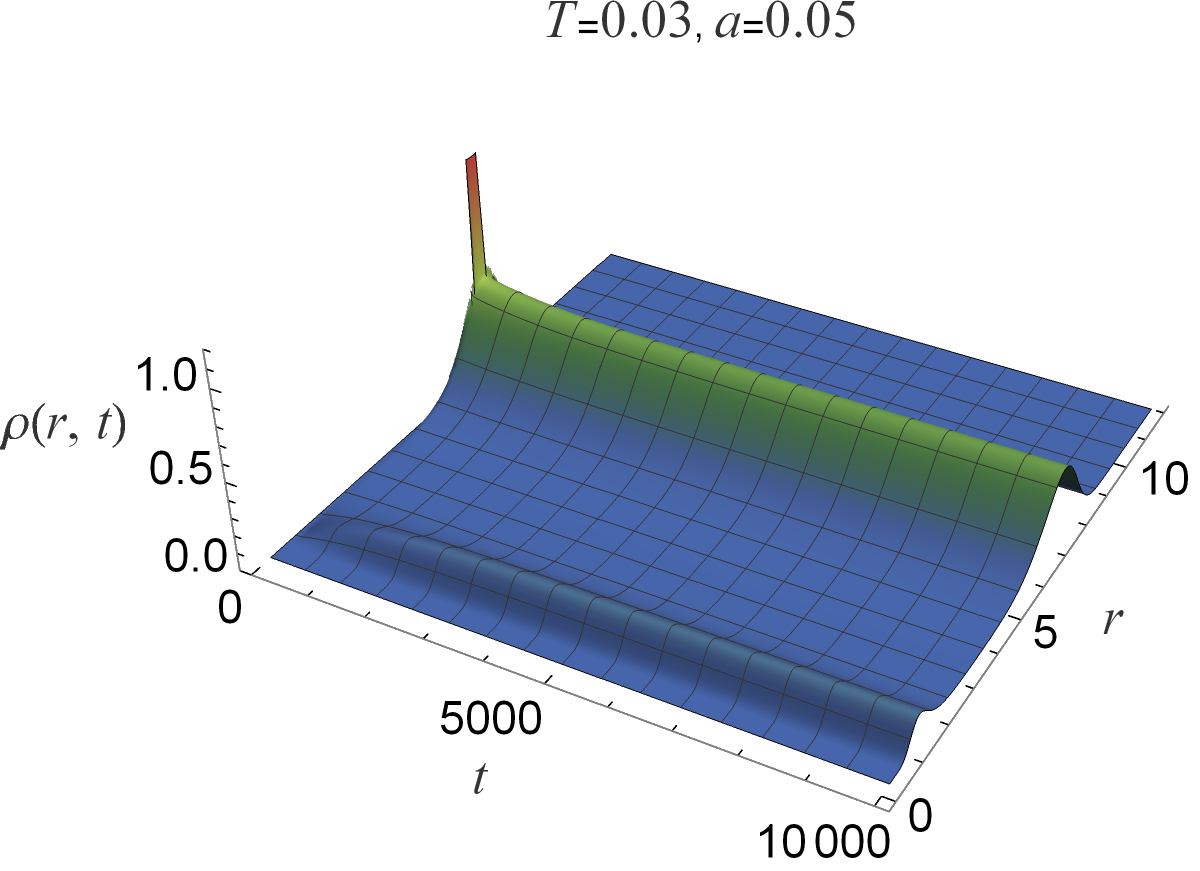}\\ \vspace{0.3cm}
    \includegraphics[width=6.5cm]{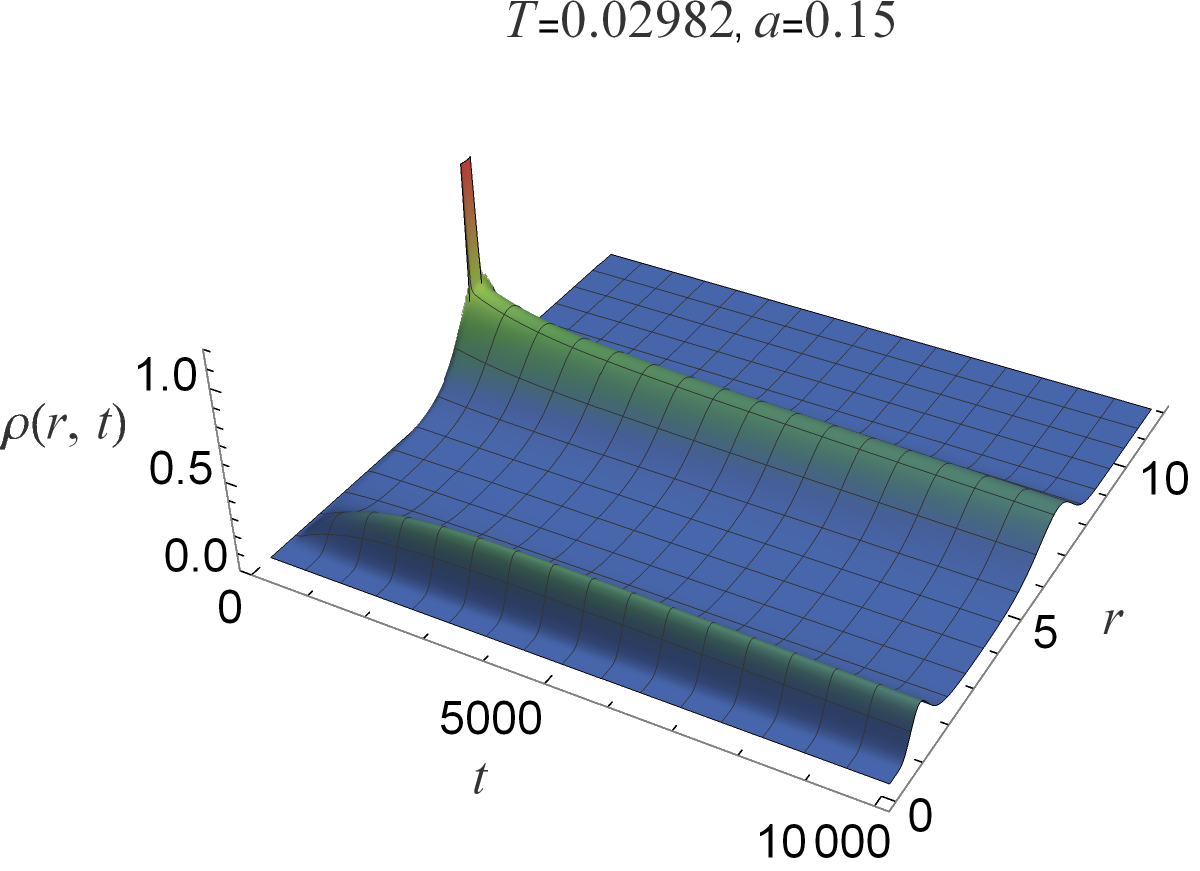}
  \includegraphics[width=6.5cm]{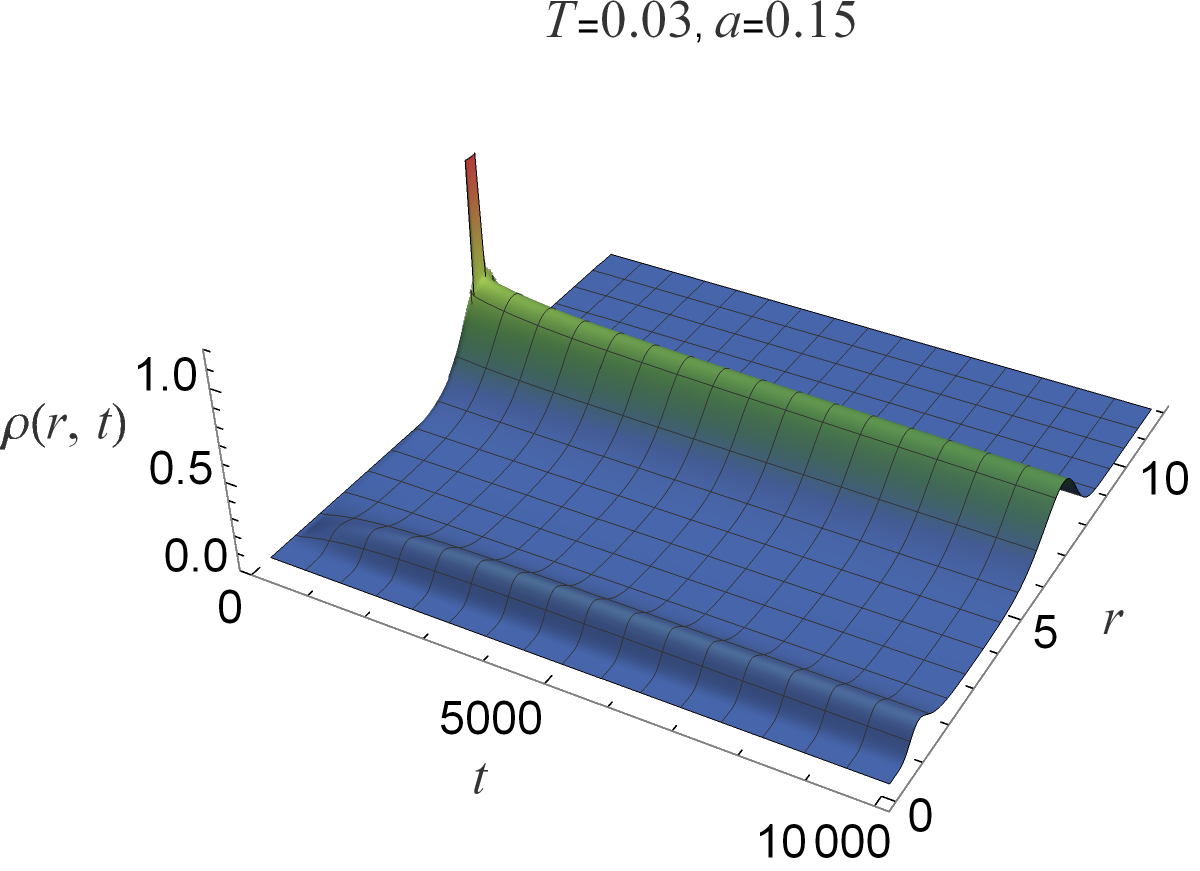}
  \caption{The distributions of probability $\rho(r, t)$ as a function of $r$ and $t$ for different $T$ and $a$. In the left and right columns, $T=0.02982$ and $0.03$. In the top and bottom rows, $a=0.05$ and $0.15$. The initial wave packet is located at the large black hole.}
  \label{aZheng3DL}
\end{figure}
\begin{figure}[H]
  \centering
 \includegraphics[width=6.5cm]{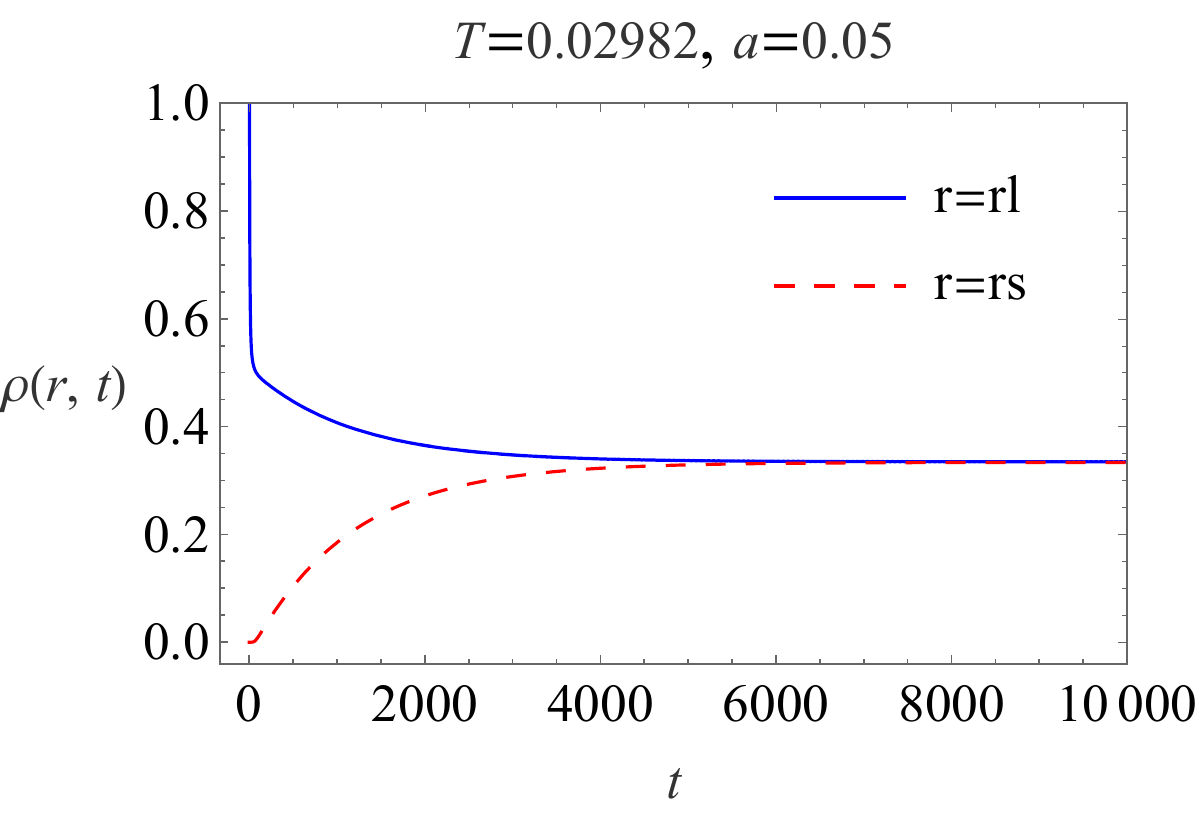}
  \includegraphics[width=6.5cm]{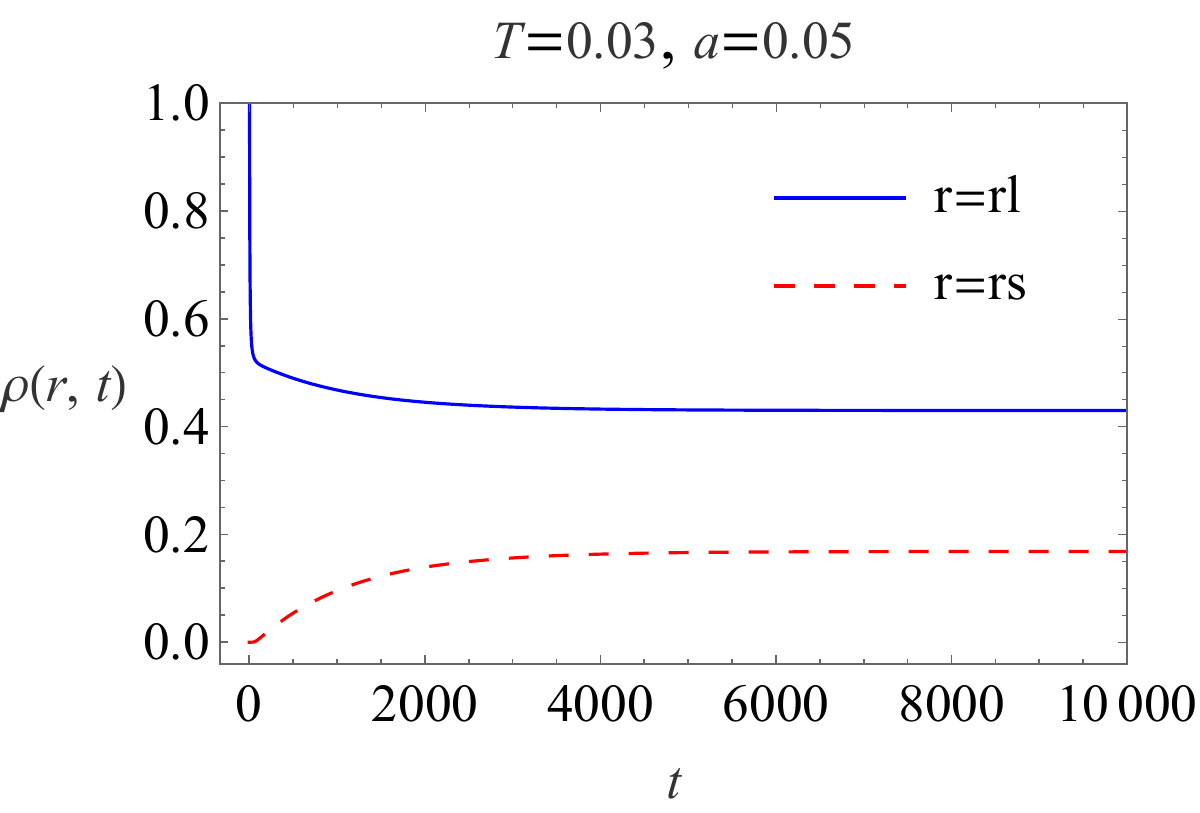}\\ \vspace{0.3cm}
    \includegraphics[width=6.5cm]{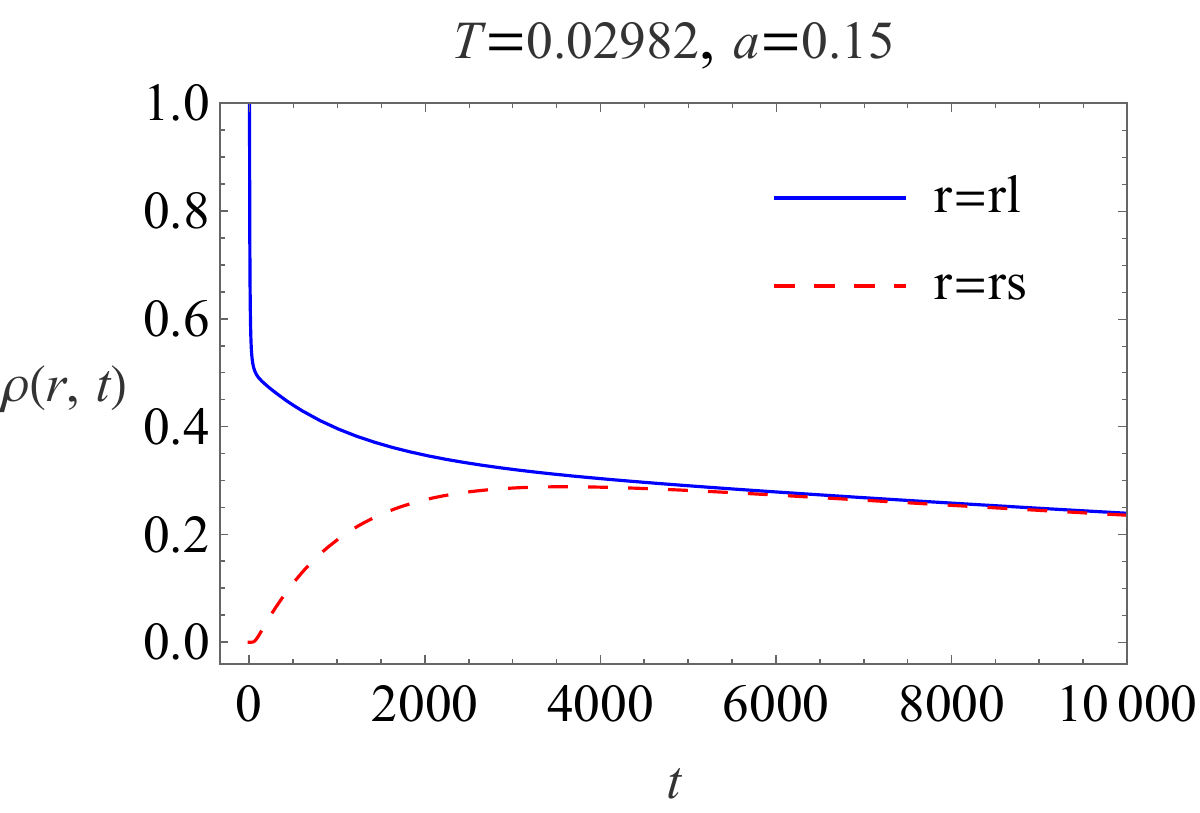}
  \includegraphics[width=6.5cm]{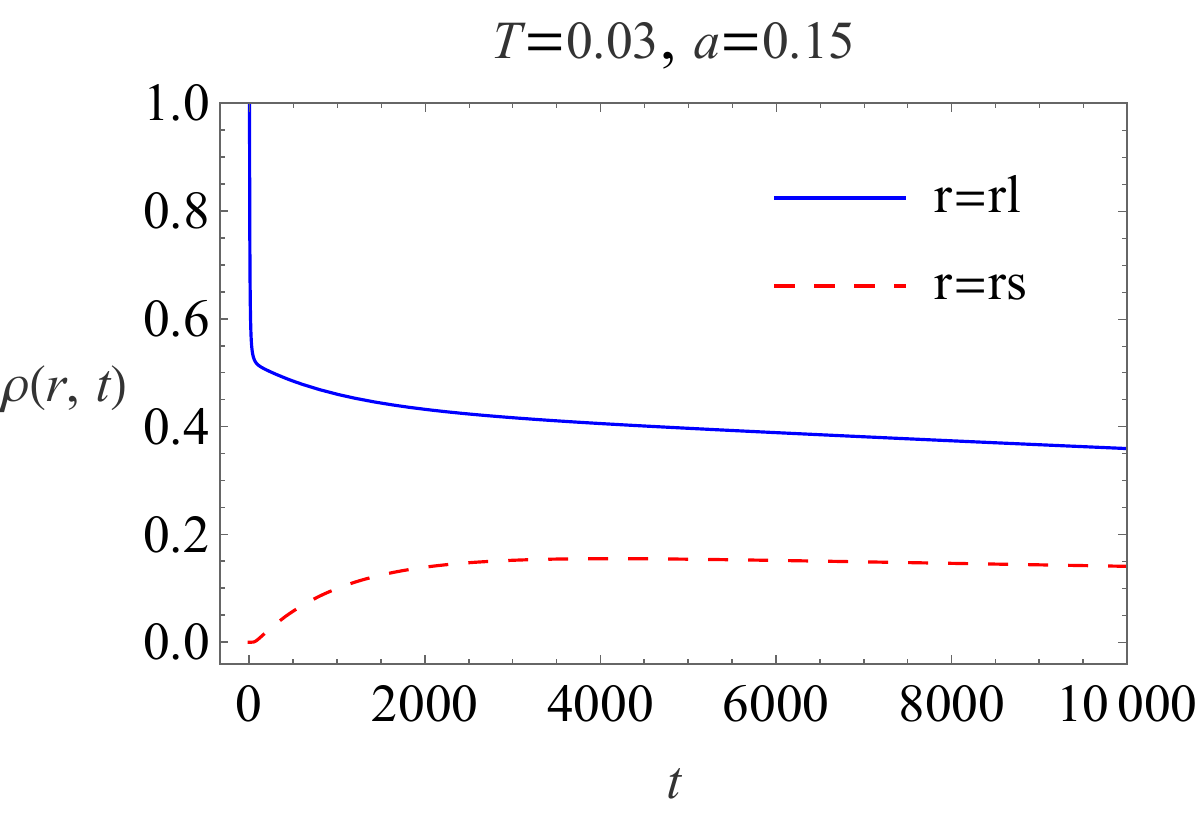}
  \caption{Behaviors of the probability $\rho(r, t)$ as a function of $t$ for different $T$ and $a$. In the left and right columns, $T=0.02982$ and $0.03$. In the top and bottom rows, $a=0.05$ and $0.15$. The initial wave packet is located at the large black hole.}
  \label{aZheng2DL}
\end{figure}

Now, the initial wave packet is located at the large black hole. In Fig. \ref{aZheng3DL} and Fig. \ref{aZheng2DL}, we describe the behaviors of the probability distribution of the black holes at different $T$ and $a$. We observe that the probability distribution of large black hole decreases with the evolution of time, and it will decrease to a stable non-zero value for a smaller $a$. The probability $\rho(r, t)$ of small black hole first increases and then decreases for a larger $a$. The probability of small black hole increases with the evolution of time and reaches a stable non-zero value for a smaller $a$. The large black hole state can have the chance to switch to the small black hole state in a suitable condition. The coexistent small and large black hole states also can be acquired for $T=0.02982$ and $a=0.05$.

Although some system state could theoretically exist, the probability is very small, which can be observed from the probability distribution of system states. For a larger $a$, the system goes to a non-black hole state (thermal radiation state), which is a complement to Ref. \cite{Magos}.

We will study the first passage time for the switching of the small and large black hole states. The first passage time is defined as the time that the present state of the black hole (the small or large black hole state) located at the well of $G_L$ reaches the intermediate transition state located at the barrier top of $G_L$ for the first time.

We denote $F_p(t)$ as the distribution of the first passage time, and define $\Sigma(t)$ to be the probability that the present state of the black hole has not made a first passage by time $t$. Therefore, $F_{\rm p}(t)$ and $\Sigma(t)$ are related by
\begin{eqnarray}\label{fptequation}
F_p(t)=-\frac{d\Sigma(t)}{dt}\;.
\end{eqnarray}
Obviously, $F_{\rm p}(t)dt$ is the probability that the present black hole state passes through the intermediate transition state located at the barrier top of $G_L$ for the first time in the time interval $(t, t+dt)$.

Considering the initial small black hole state, the probability that the small black hole has not made the first passage by time $t$ is given by
\begin{eqnarray}\label{Sigmat}
\Sigma(t)=\int_{0}^{r_m} \rho(r, t) dr\;,
\end{eqnarray}
where $r_m$ is the horizon radius for the intermediate black hole. By substituting Eq. (\ref{Sigmat}) into Eq. (\ref{fptequation}), and using the Fokker-Planck equation (\ref{FPequation}), we obtain \cite{Li2020}
\begin{eqnarray}\label{fpt}
F_p(t)&=&-\frac{d}{dt}\int_{0}^{r_m} \rho(r, t) dr\nonumber\\
&=&-\int_{0}^{r_m}\frac{\partial}{\partial t} \rho(r, t) dr\nonumber\\
&=&-D\int_{0}^{r_m}\frac{\partial}{\partial r}\left\{
e^{-\beta G_L(r)}\frac{\partial}{\partial r}\left[e^{\beta G_L(r)}\rho(r,t)\right]\right\}  dr\nonumber\\
&=&\left.-D e^{-\beta G_L(r)}\frac{\partial}{\partial r}\left[e^{\beta G_L(r)}\rho(r,t)\right]\right|_{0}^{r_m}\nonumber\\
&=&\left.-D\frac{\partial}{\partial r}\rho(r,t)\right|_{r=r_m}\;.
\end{eqnarray}
The reflecting boundary condition is imposed at $r=0$, and the absorbing boundary condition is imposed at $r=r_m$ (transition state at free energy barrier top). We invoke the numerical method, and the Gibbs free energy $G_L$ is divergent at $r=0$. We also need ensure the positivity of the black hole mass. Therefore, the reflecting boundary condition is imposed at $r=0.3$ here.

\begin{figure}[H]
  \centering
  \includegraphics[width=6.5cm]{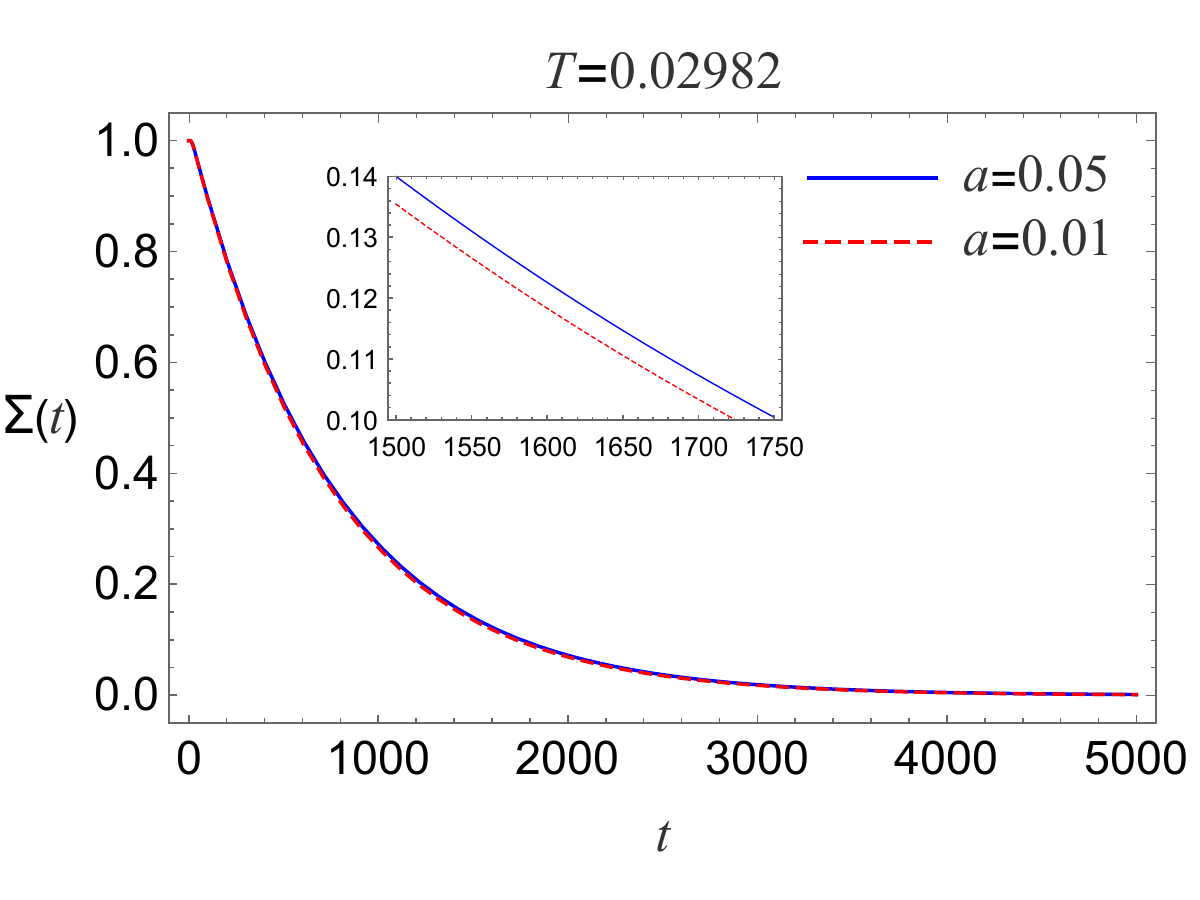}
  \includegraphics[width=6.5cm]{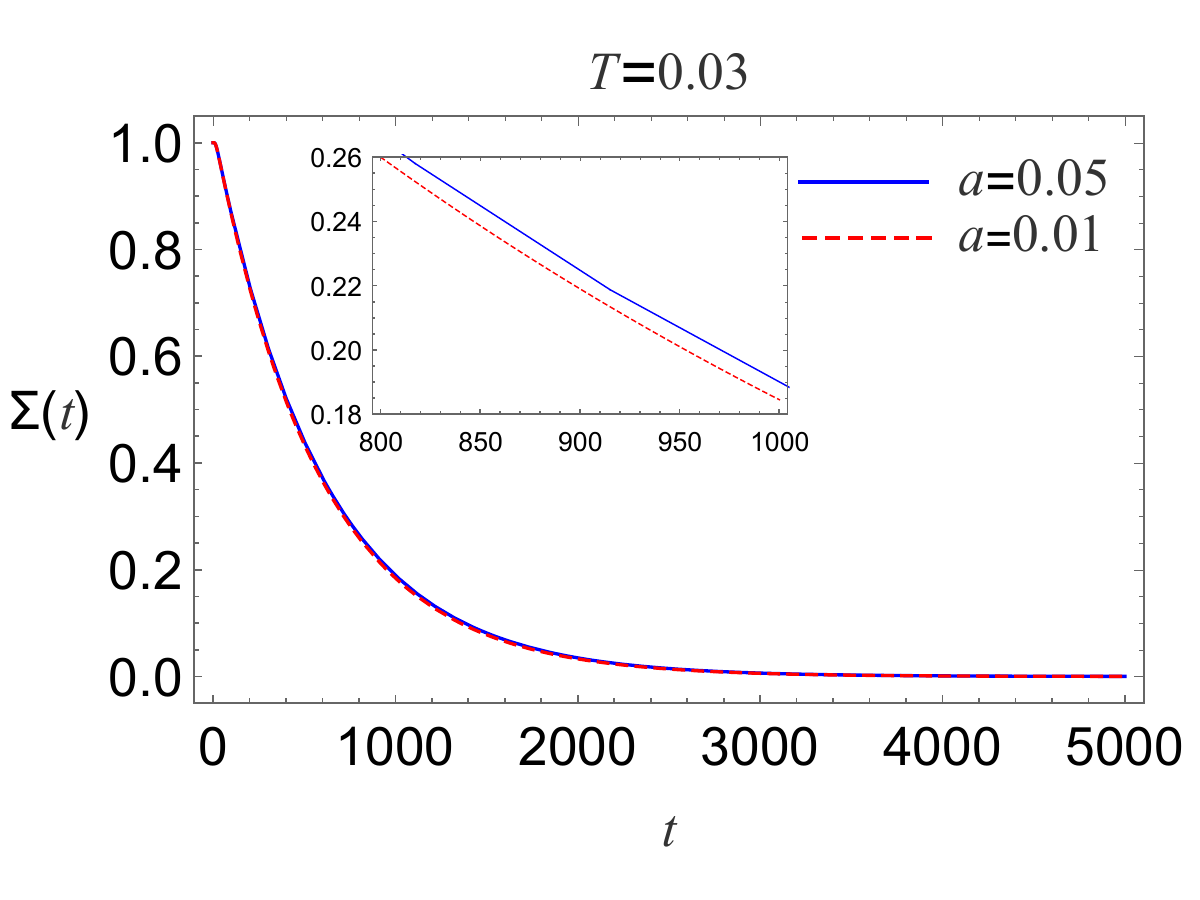}
  \caption{The probability $\Sigma(t)$ for different $T$ and $a$. The solid (blue) and dashed (red) lines represent $a=0.05$ and $0.01$. We take $T=0.02982$ (the left panel) and $T=0.03$ (the right panel). The initial wave packet is located at the small black hole.}
  \label{azhengx2d}
\end{figure}

We will discuss the first passage process. In order to obtain a stable small and large black hole, we choose small $a$ in this situation, where the left free energy barrier is higher. The initial wave packet is located at the small black hole. We see that the probability $\Sigma(t)$ decreases faster for higher temperature $T$ in Fig.~\ref{azhengx2d}. The smaller the Euler-Heisenberg parameter $a$ is, the faster the probability decreases.

\begin{figure}[H]
  \centering
    \includegraphics[width=6.5cm]{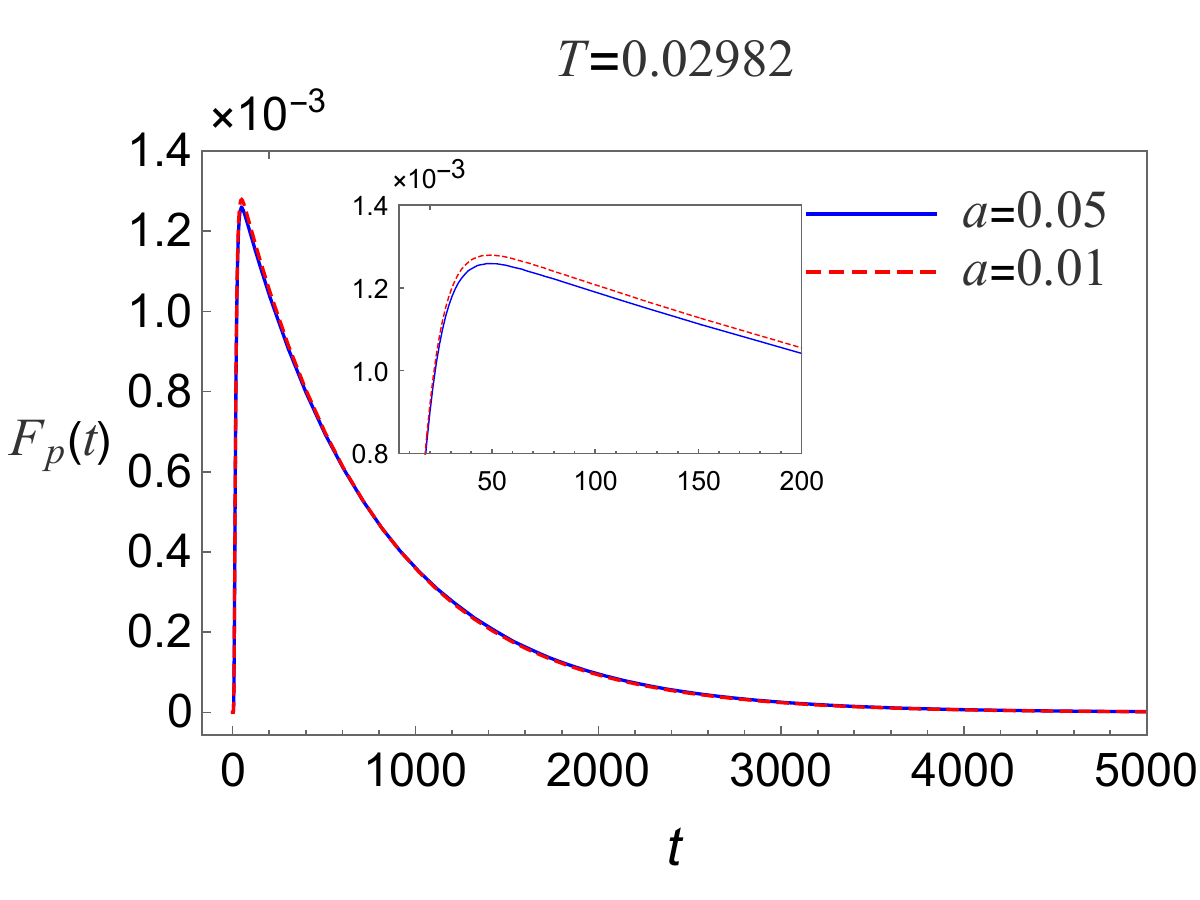}
  \includegraphics[width=6.5cm]{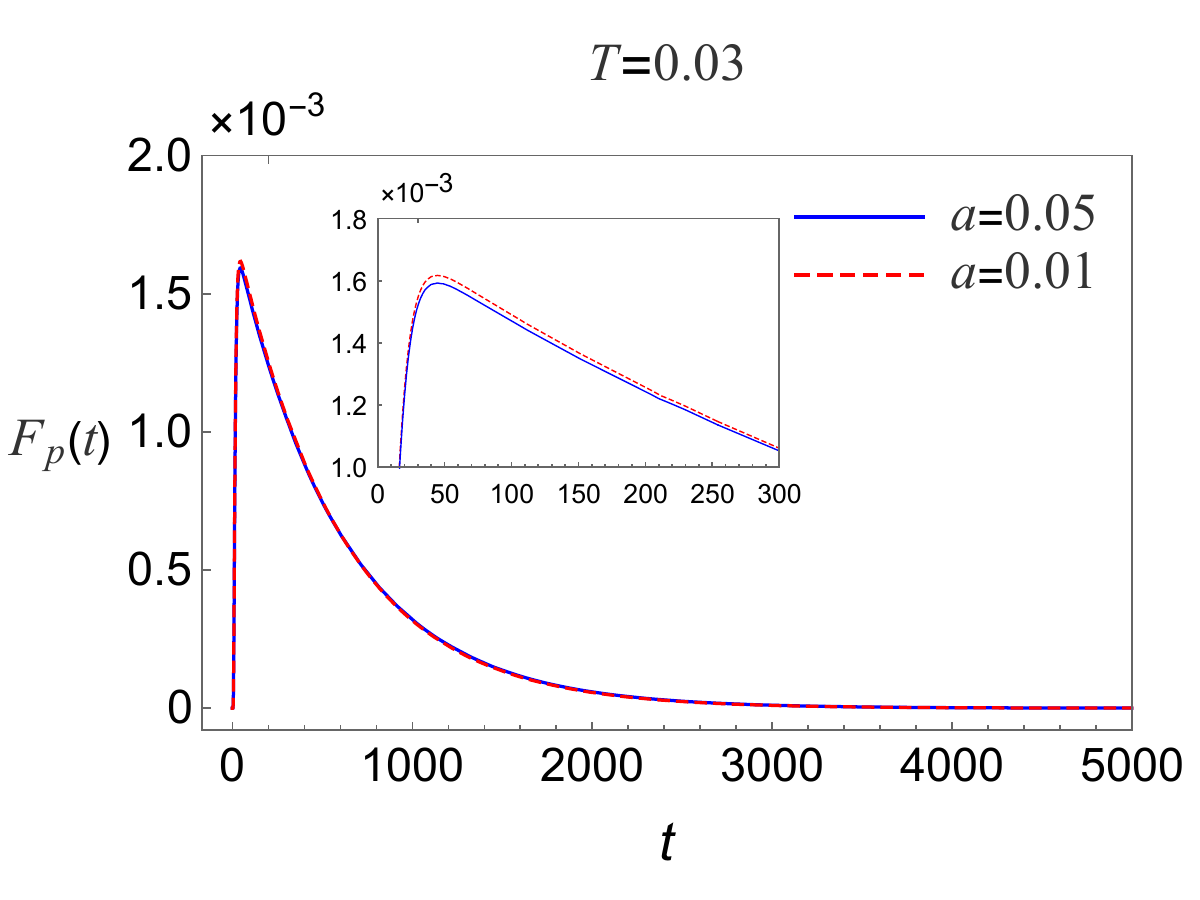}
  \caption{The distributions of the first passage time $F_{p}(t)$ for different $T$ and $a$. The solid (blue) and dashed (red) lines represent the Euler-Heisenberg parameters $a=0.05$ and $0.01$, and we take $T=0.02982$ (the left panel) and $T=0.03$ (the right panel). The initial wave packet is located at the small black hole.}
  \label{azhengx2df}
\end{figure}
In Fig. \ref{azhengx2df}, we present the distribution of the first passage time for different $T$ and $a$. There exists a single peak for each curve. When $a$ decreases or $T$ increases, the peak increases.

We will consider the first passage kinetic process for the initial large black hole state. We impose the absorbing boundary condition at $r_m$ (transition state at the free energy barrier top) and the reflecting boundary condition at infinity. Therefore, we have
\begin{gather}
 \Sigma(t) = \int^{+\infty}_{r_m}\rho(t,r)dr, \\
 F_p(t)= D\left.\frac{\partial \rho(t,r)}{\partial r}\right|_{r=r_m}.
\end{gather}

\begin{figure}[H]
  \centering
   \includegraphics[width=6.5cm]{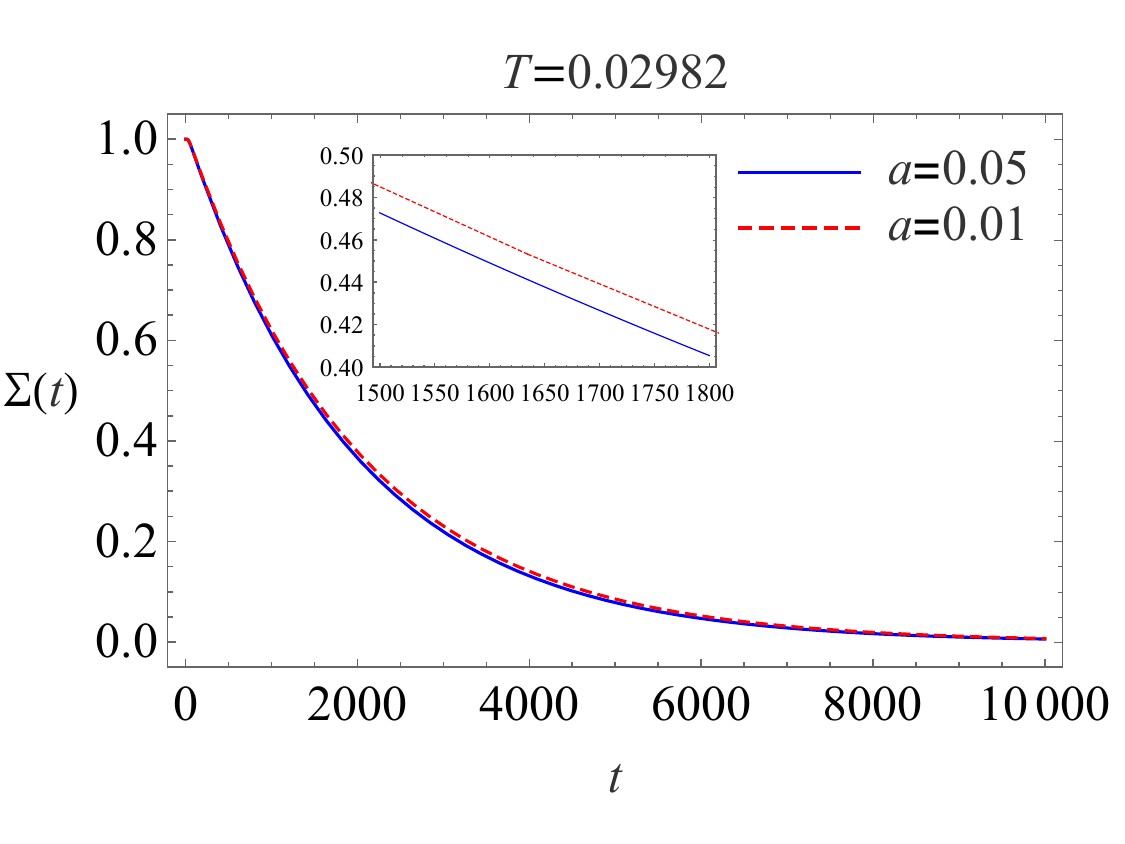}
  \includegraphics[width=6.5cm]{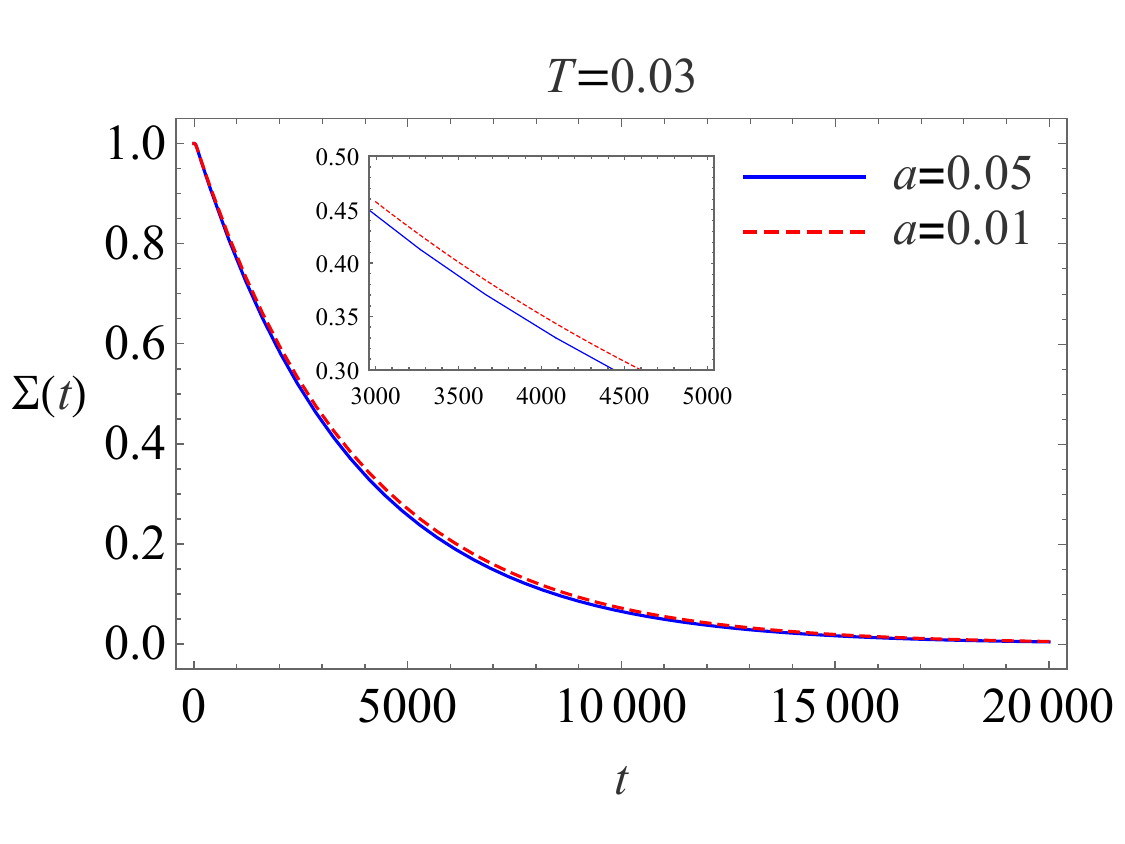}
  \caption{The probability $\Sigma(t)$ for different $T$ and $a$. The solid (blue) and dashed (red) lines represent $a=0.05$ and $0.01$. We take $T=0.02982$ (the left panel) and $T=0.03$ (the right panel). The initial wave packet is located at the large black hole.}
  \label{azhengd2d}
\end{figure}
The initial wave packet is located at the large black hole. From Fig.~\ref{azhengd2d}, we observe that the probability $\Sigma(t)$ decreases faster for lower temperature $T$. In addition, the larger the Euler-Heisenberg parameter $a$ is, the faster the probability decreases. This is quite different from the initial small black hole state.

\begin{figure}[H]
  \centering
  \includegraphics[width=6.5cm]{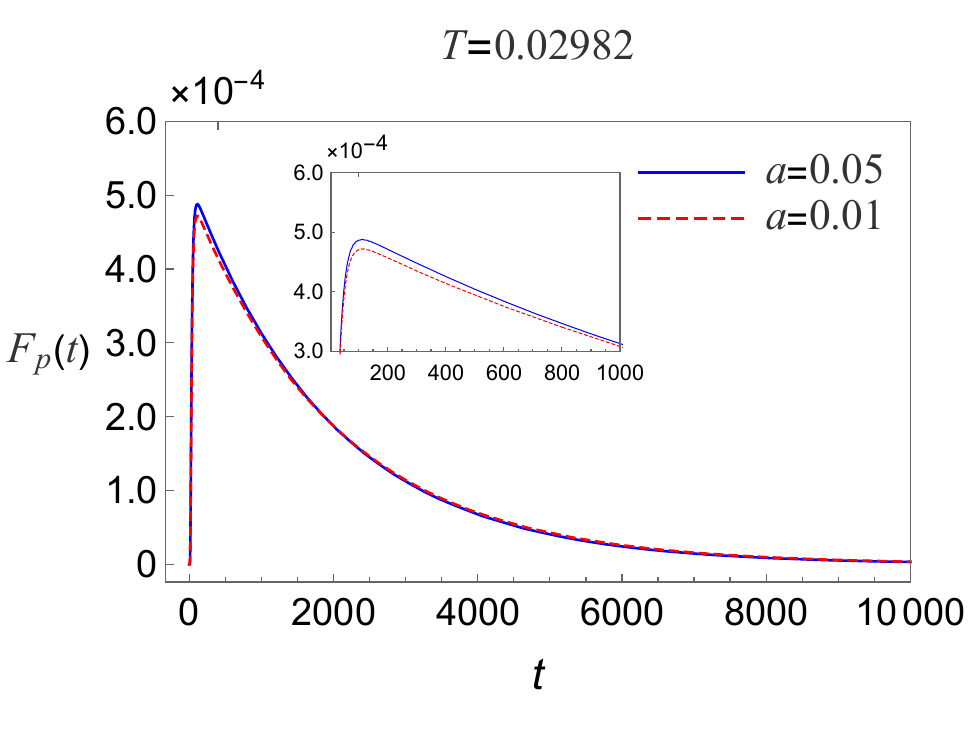}
  \includegraphics[width=6.5cm]{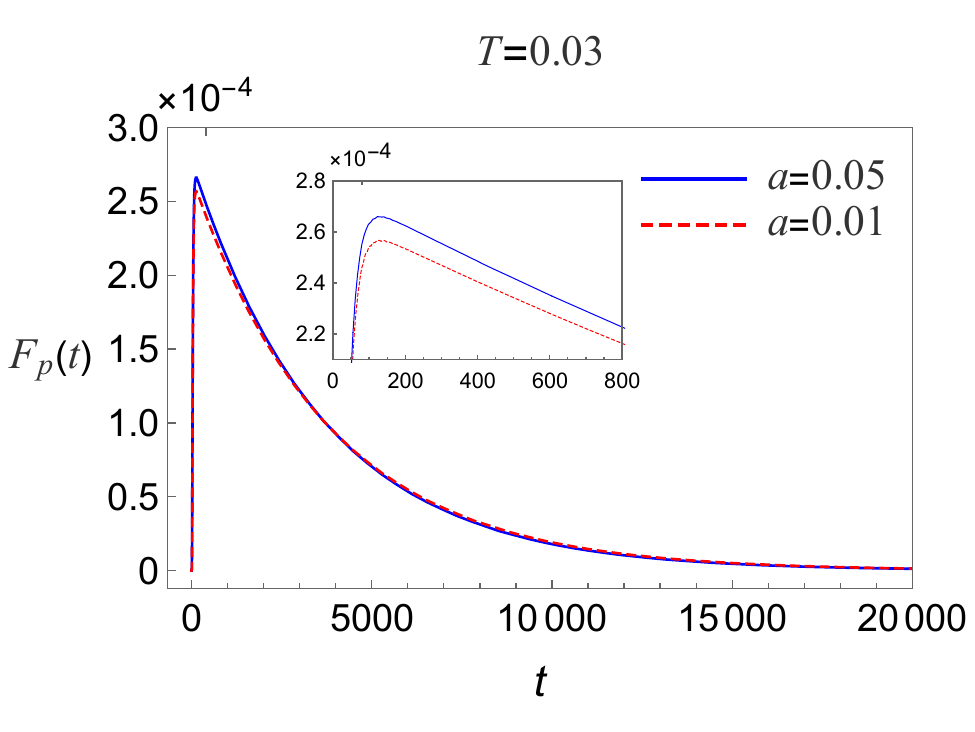}
  \caption{The distributions of the first passage time $F_{p}(t)$ for different $T$ and $a$. The solid (blue) and dashed (red) lines represent the Euler-Heisenberg parameters $a=0.05$ and $0.01$, and we take $T=0.02982$ (the left panel) and $T=0.03$ (the right panel). The initial wave packet is located at the large black hole.}
  \label{azhengd2df}
\end{figure}
In Fig. \ref{azhengd2df}, we show the distribution of the first passage time for different $T$ and $a$. There also exists a single peak for each curve at short time. When the temperature decreases or the Euler-Heisenberg parameter increases, the peak increases. The first passage process describes how fast a system undergoes a stochastic process for the first time, which is also a complement to Ref. \cite{Magos}.

Besides the temperature $T$, the parameter $a$ also play a crucial influence on the probability distribution of black hole states and the distribution of first passage times.

\subsection{The phase transition for the case of $a<0$ }

\begin{figure}[H]
  \centering
  \includegraphics[width=8cm]{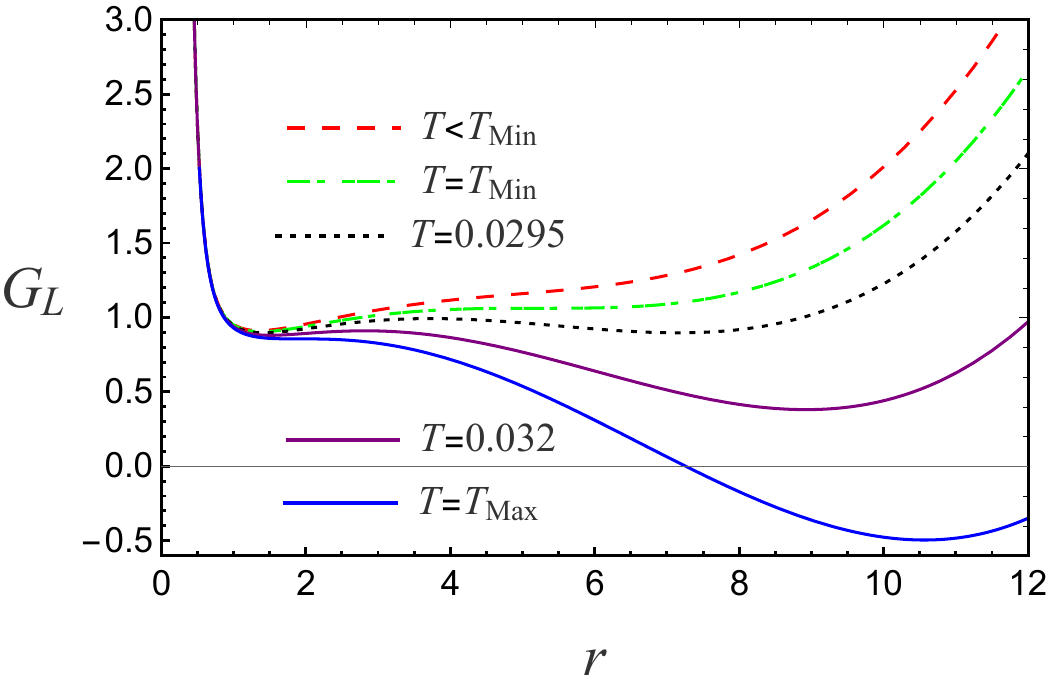}
  \caption{The Gibbs free energy as a function of $r$ for $P=0.4P_c$ and $a=-1.35$ with different temperatures.}
  \label{GraFu}
\end{figure}

We will study the phase transition for the case of $a<0$. In Fig. \ref{GraFu}, we plot the Gibbs free energy as a function of $r$ for $P=0.4P_c$ and $a=-1.35$ with different $T$ so as to quantify the free energy landscape. The two local minimum for the Gibbs free energy correspond to the small and the large black holes, respectively.

\begin{figure}[H]
  \centering
  \subfigure[]{\includegraphics[width=7.0cm]{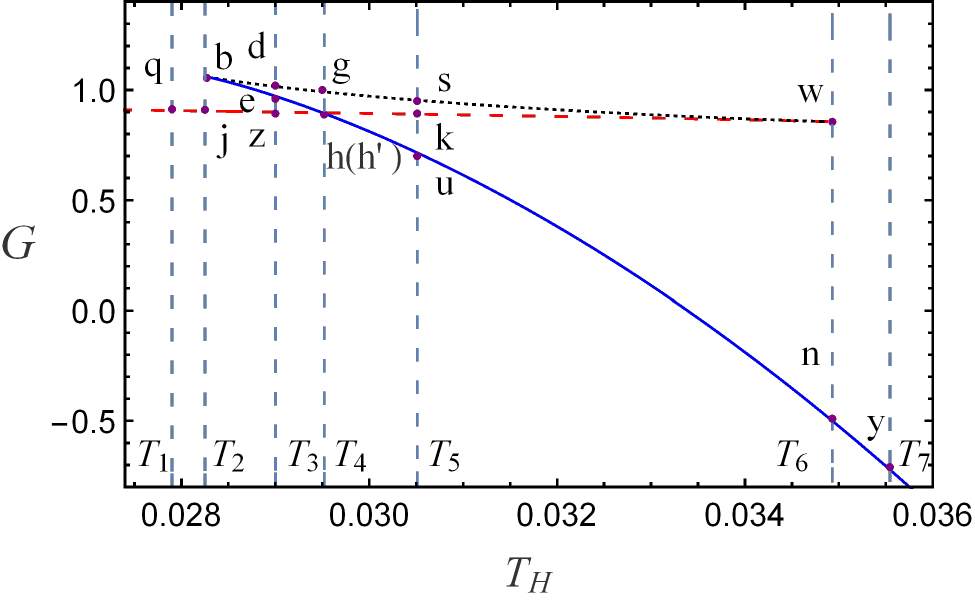}\label{Fig:off-shella}}
  \subfigure[]{\includegraphics[width=6.5cm]{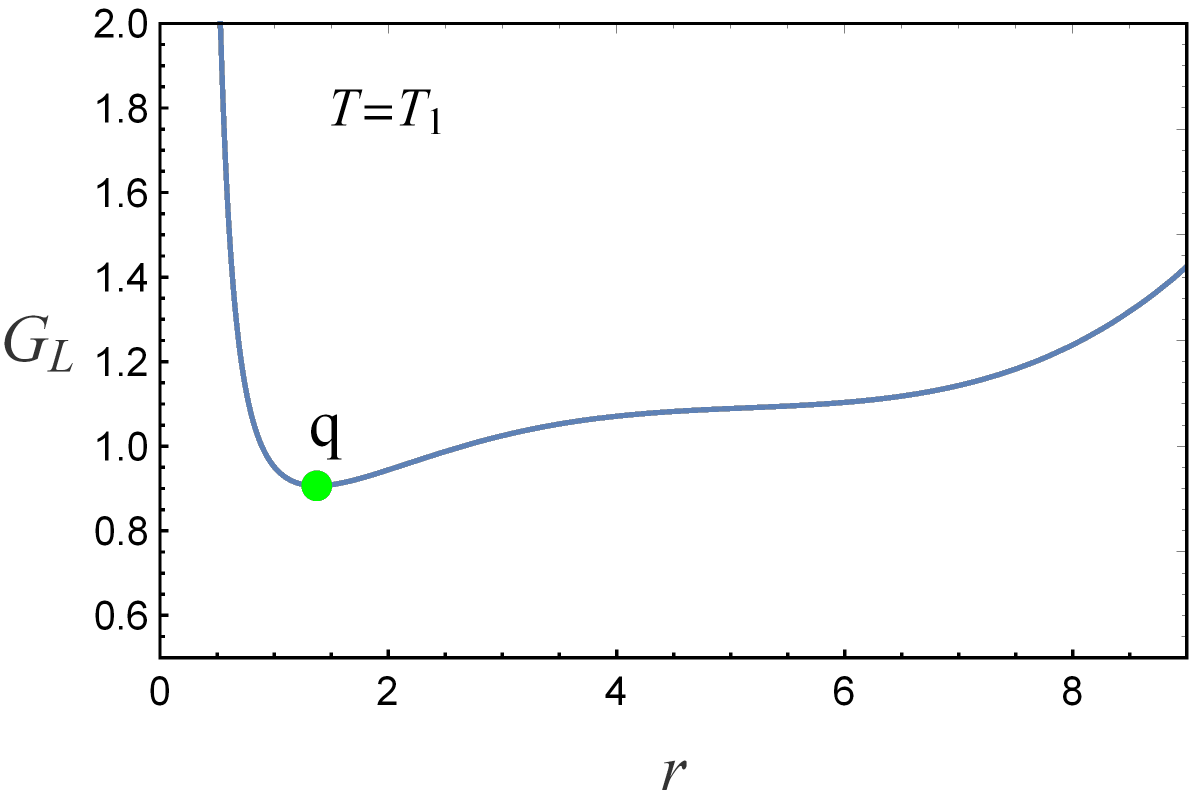}\label{Fig:off-shellb}}
  \subfigure[]{\includegraphics[width=6.5cm]{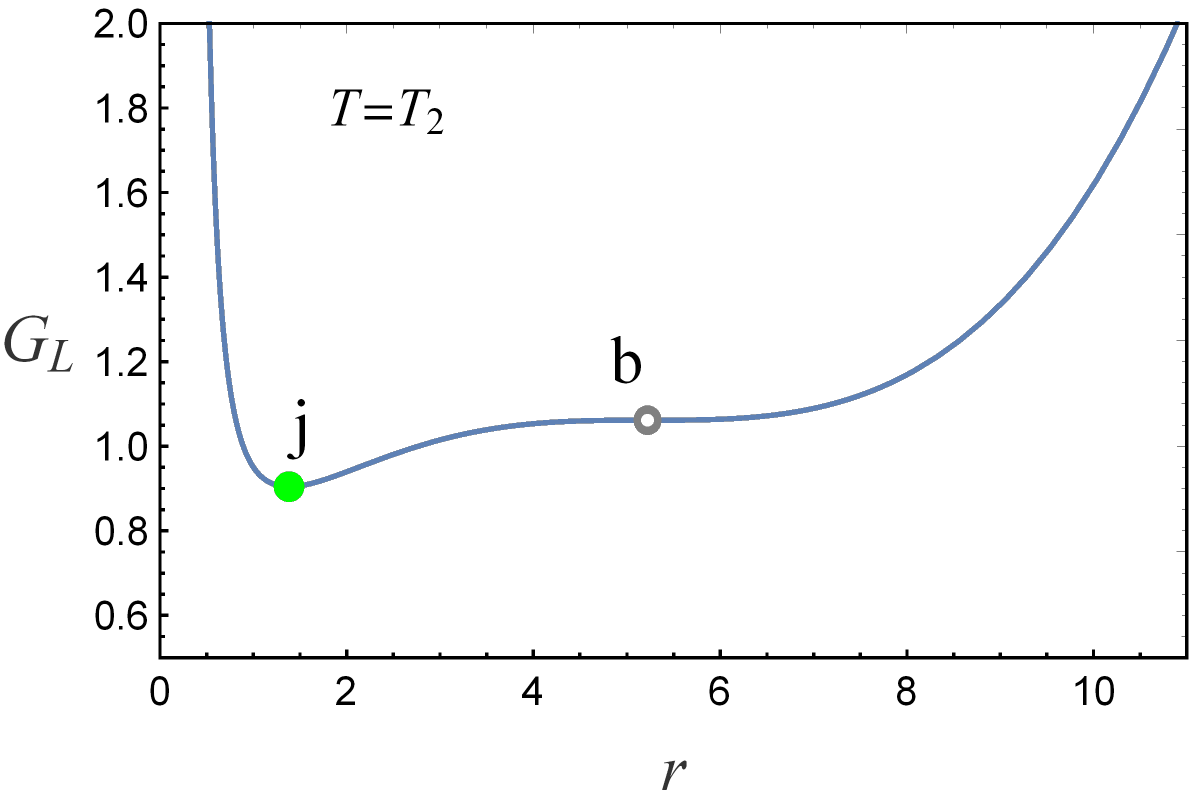}\label{Fig:off-shellc}} \hspace{0.3cm}
  \subfigure[]{\includegraphics[width=6.5cm]{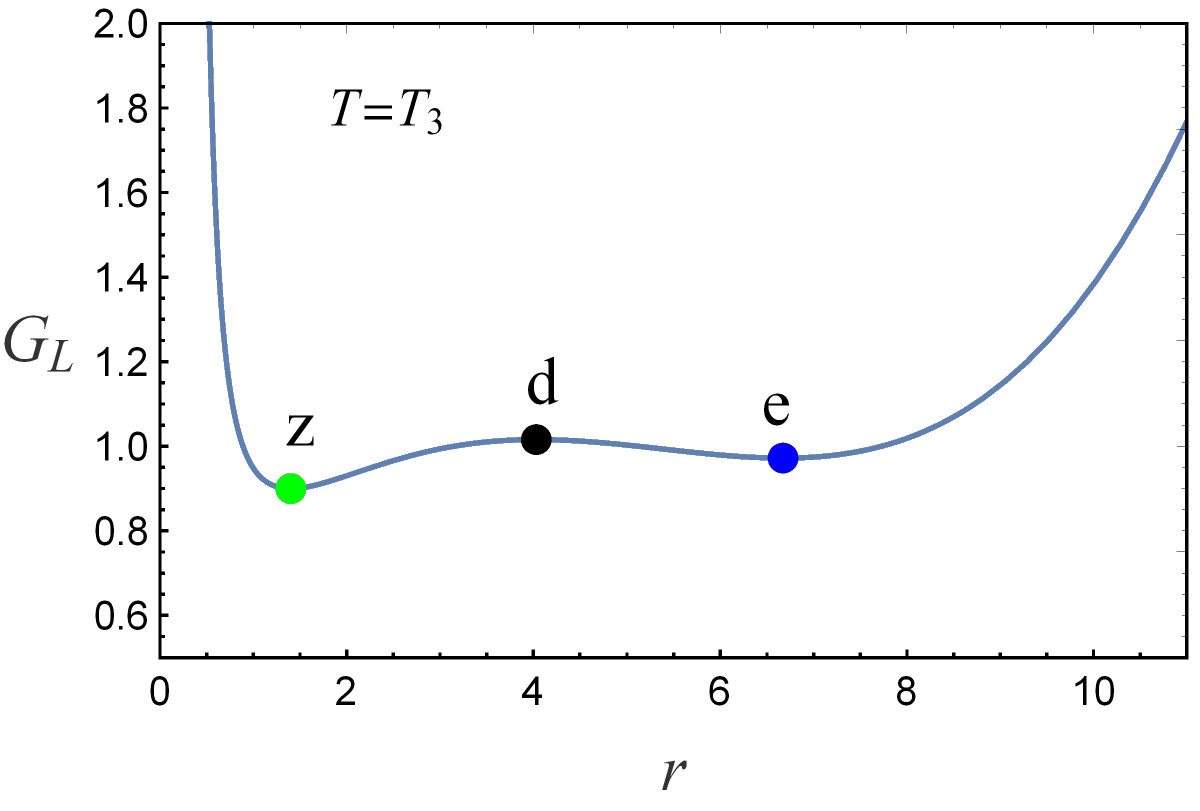}\label{Fig:off-shelld}}
  \subfigure[]{\includegraphics[width=6.5cm]{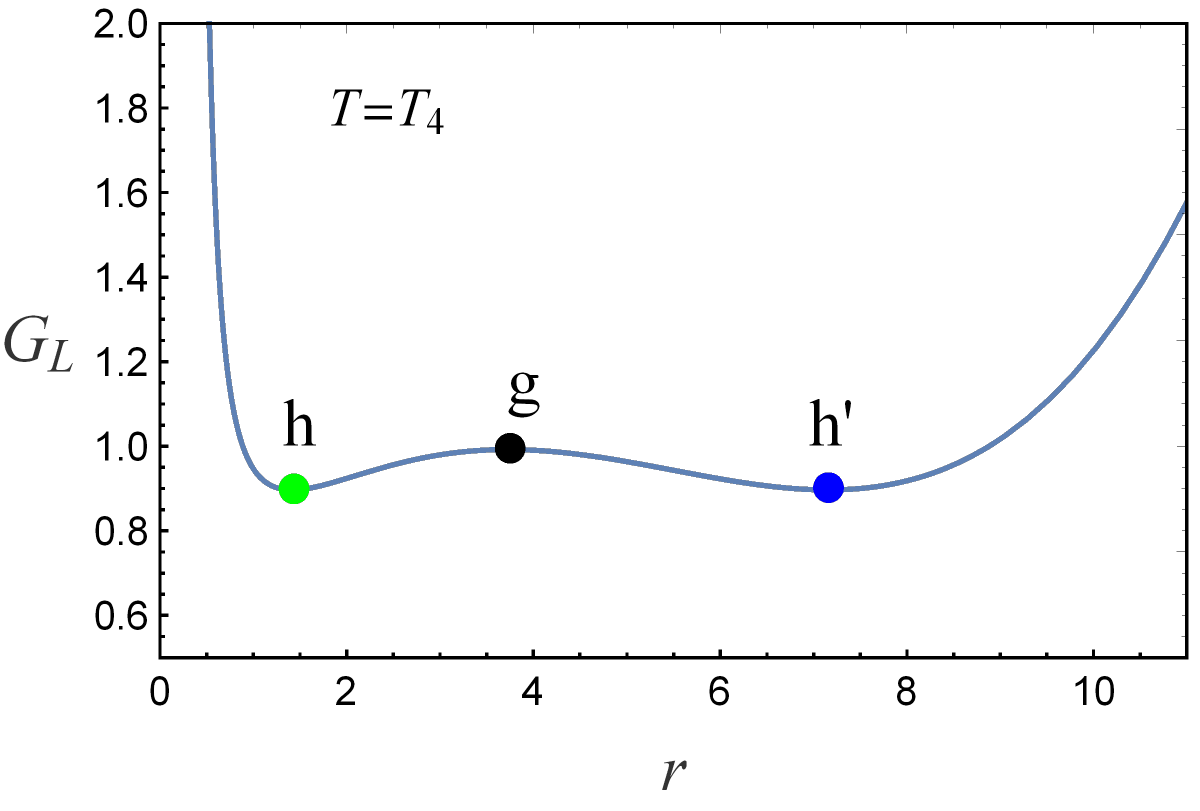}\label{Fig:off-shelle}} \hspace{0.3cm}
  \subfigure[]{\includegraphics[width=6.5cm]{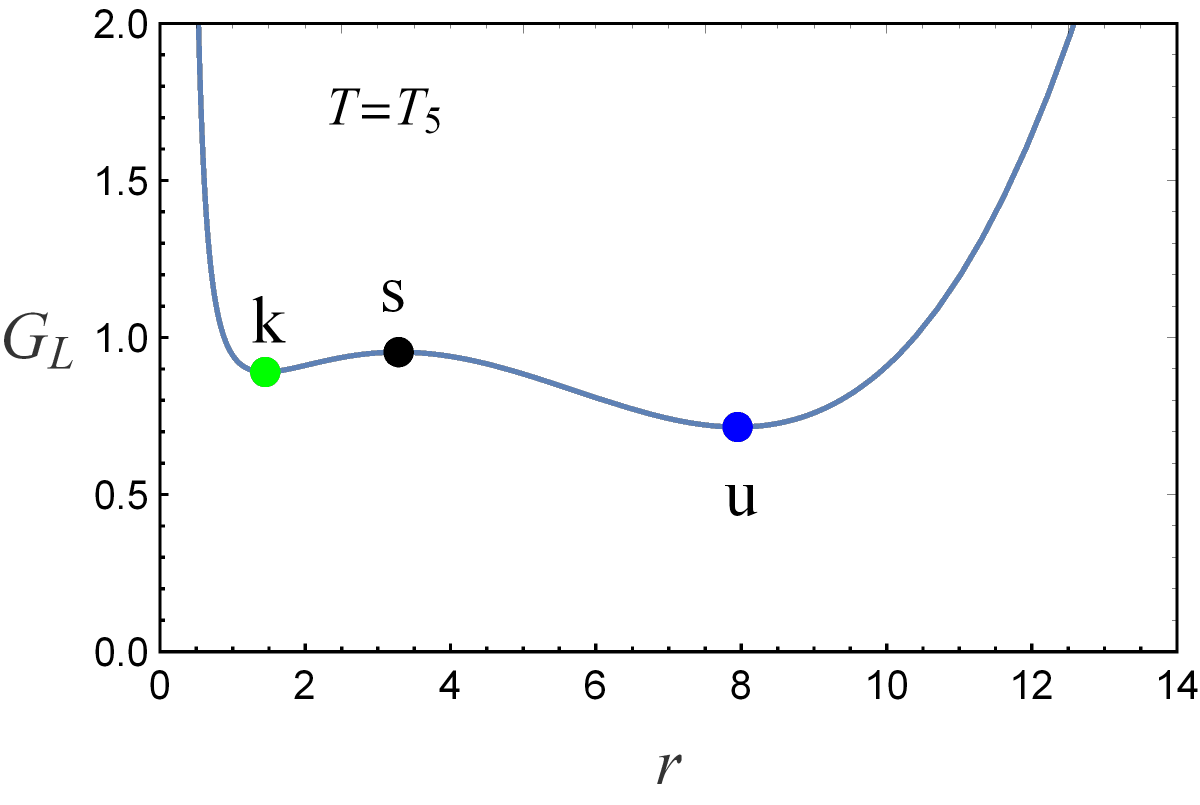}\label{Fig:off-shellf}}
  \subfigure[]{\includegraphics[width=6.5cm]{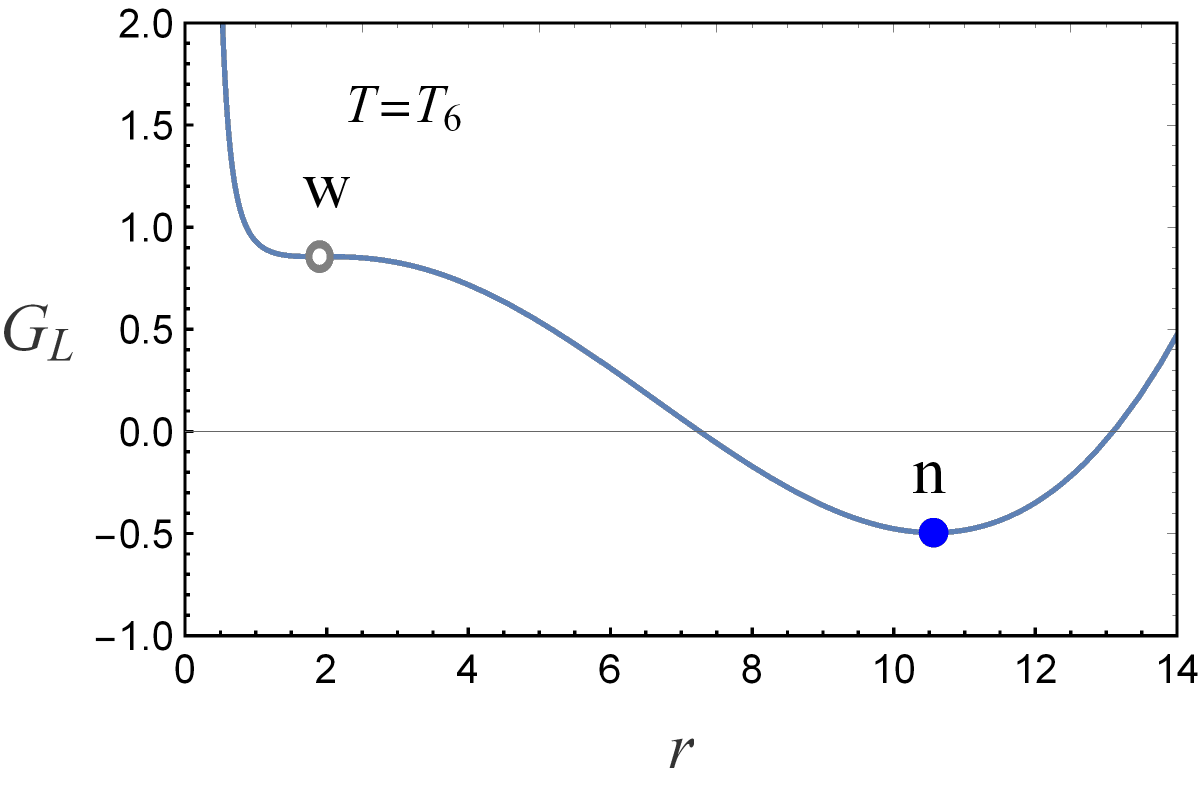}\label{Fig:off-shellg}} \hspace{0.3cm}
  \subfigure[]{\includegraphics[width=6.5cm]{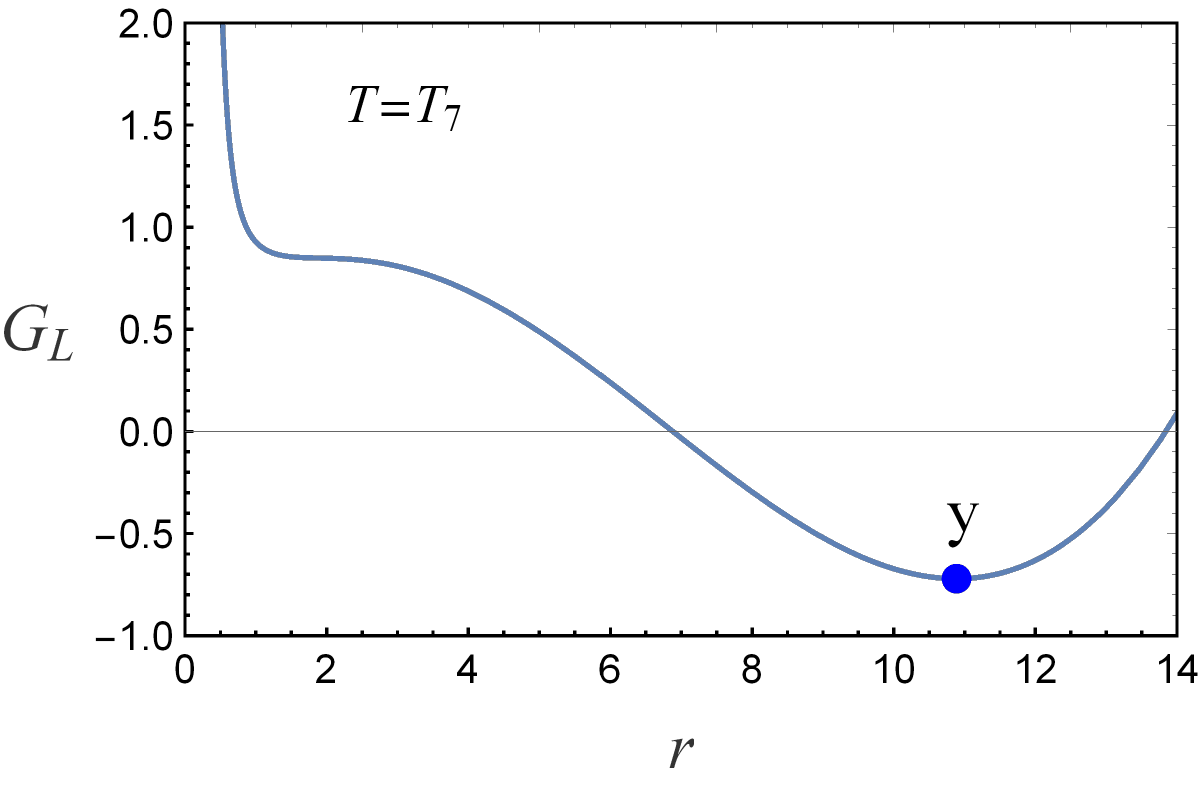}\label{Fig:off-shellh}}
  \caption{The behaviors of the Gibbs free energy with $P = 0.4P_c$ and $a=-1.35$. (a) The Gibbs free energy $G$ vs the black hole temperature $T_{H}$. The solid (blue), dashed (red), and dotted (black) lines correspond the large black hole, small black hole, and intermediate black hole branches, respectively. (b) The Gibbs free energy $G_{L}$ vs the black hole event horizon $r$ with the temperature of the ensemble $T=T_1$. (c) $G_{L}$ vs $r$ with $T=T_2$. (d) $G_{L}$ vs $r$ with $T=T_3$. (e) $G_{L}$ vs $r$ with $T=T_4$. (f) $G_{L}$ vs $r$ with $T=T_5$. (g) $G_{L}$ vs $r$ with $T=T_6$. (h) $G_{L}$ vs $r$ with $T=T_7$.}
  \label{afutyw}
\end{figure}
In Fig. \ref{Fig:off-shella}, we show the Gibbs free energy $G$ as a function of the black hole temperature $T_{H}$. The large, small, and intermediate black hole branches are depicted by the solid (blue), dashed (red), and dotted (black) lines. The system prefers the state of the lowest Gibbs free energy, therefore some states on the small and large black hole branches will be global stable, such as states $q$, $j$, and $u$. Some locally thermodynamic stable black holes, see the states $e$ and $k$ are globally metastable. With the increase of the temperature, the black hole system goes through states $q-j-z-h(h')-u-n-y$.

For the seven different temperatures $T_1\sim T_7$, we display $G_L$ as a function of $r$ in Fig. \ref{afutyw} $(b)\sim (h)$, respectively. When $T=T_1$, the Gibbs free energy $G_L$ has only one extremal point and has the lowest value, which corresponds to a stable black hole state $q$ belonging to the small black hole branch. From Fig. \ref{Fig:off-shellc}, we see that once the temperature $T\leq T_2$, $G_L$ has only one minimum, which corresponds small black hole state. When $T=T_3$, three extremal points emerge, as shown in Fig. \ref{Fig:off-shelld}. Two local minimum points correspond to the locally stable small and large black hole states, as shown the states $z$ and $e$. The local maximum point corresponds to the locally unstable intermediate black hole state $d$. The system will prefer the state $z$ which has the lowest Gibbs free energy. Different from the case $T=T_3$, increasing the temperature such that $T=T_4$, although three extremal points are still given, the two local minimum points have the same Gibbs free energy. The states $h$ and $h'$ are the small and large black hole phases, respectively. This case shows the small-large black hole phase transition, which is a coexistence phase of small and large black holes. Contrary to $T$=$T_3$, for the case of $T=T_5$, the system will prefer the large black state $u$ which has the lowest Gibbs free energy. The state $k$ is a globally metastable small black hole. When $T\geq T_6$, for the $T=T_7$, only one minimum point exists in the Gibbs free energy, which corresponds to a stable large black hole state.

\begin{figure}[H]
  \centering
  \includegraphics[width=8cm]{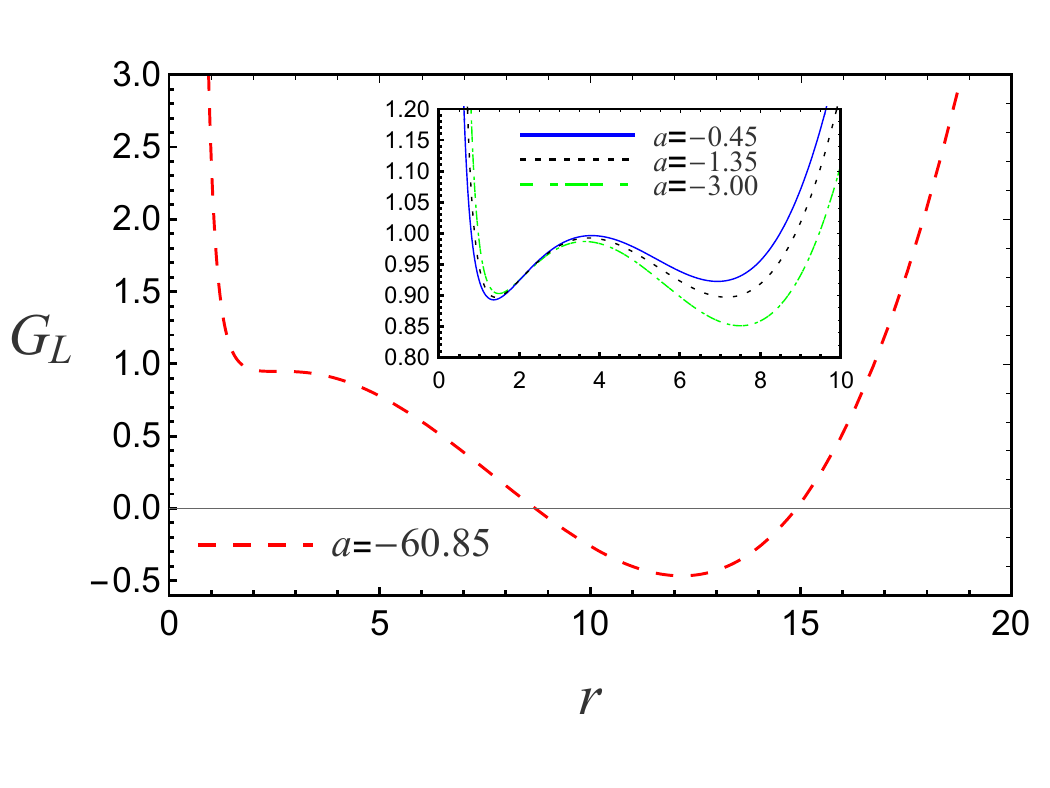}
  \caption{The behaviors of the Gibbs free energy for $P=0.4P_c$ and $T=0.0295$ with different Euler-Heisenberg parameters.}
  \label{GrTaFu}
\end{figure}
For different Euler-Heisenberg parameters, we display $G_L$ as a function of $r$ with $P=0.4P_c$ and $T=0.0295$ in Fig. \ref{GrTaFu}. When $a_{c_{1}}<a< 0$ $(a_{c_{1}}\approx-60.85)$, there are three extreme points. For $a=-0.45$, two local minimum points correspond to the locally stable small and large black hole states. The local maximum point corresponds to the locally unstable intermediate black hole state. The system will prefer the small black hole state which has the lowest Gibbs free energy. Decreasing the Euler-Heisenberg parameter such that $a=-1.35$, the two local minimum points have the same Gibbs free energy. This case describes the small-large black hole phase transition, which is a coexistence phase of small and large black holes. For $a=-3.00$, the system will prefer the large black hole state which has the lowest Gibbs free energy. When $a\leq a_{c_{1}}$ , only one minimum point exists in the Gibbs free energy, which corresponds to a stable large black hole state.

\begin{figure}[H]
  \centering
   \includegraphics[width=6.5cm]{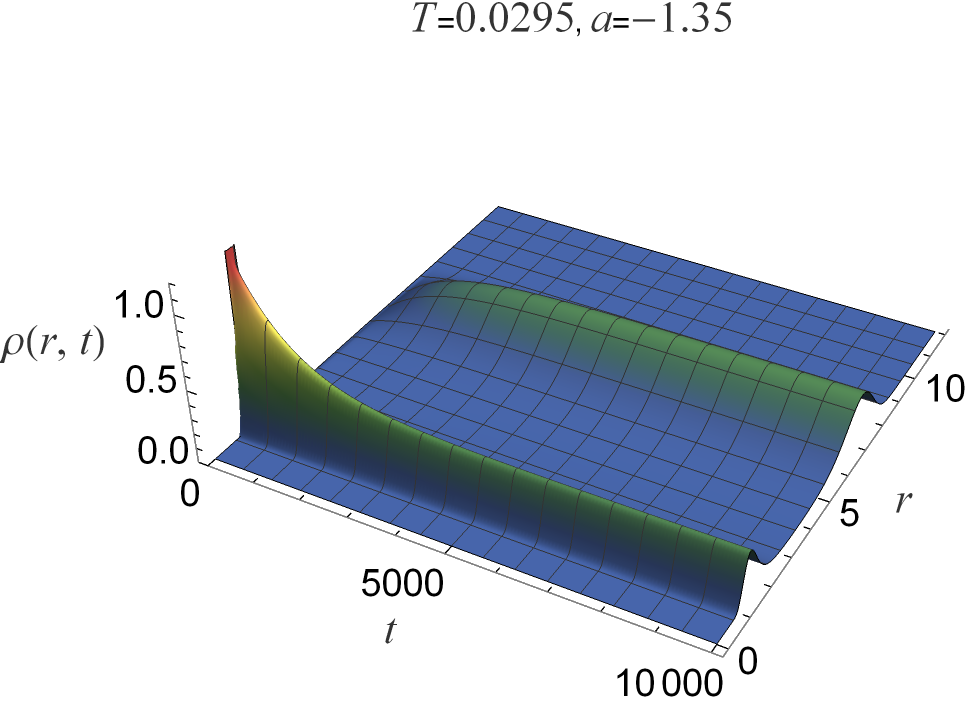}
  \includegraphics[width=6.5cm]{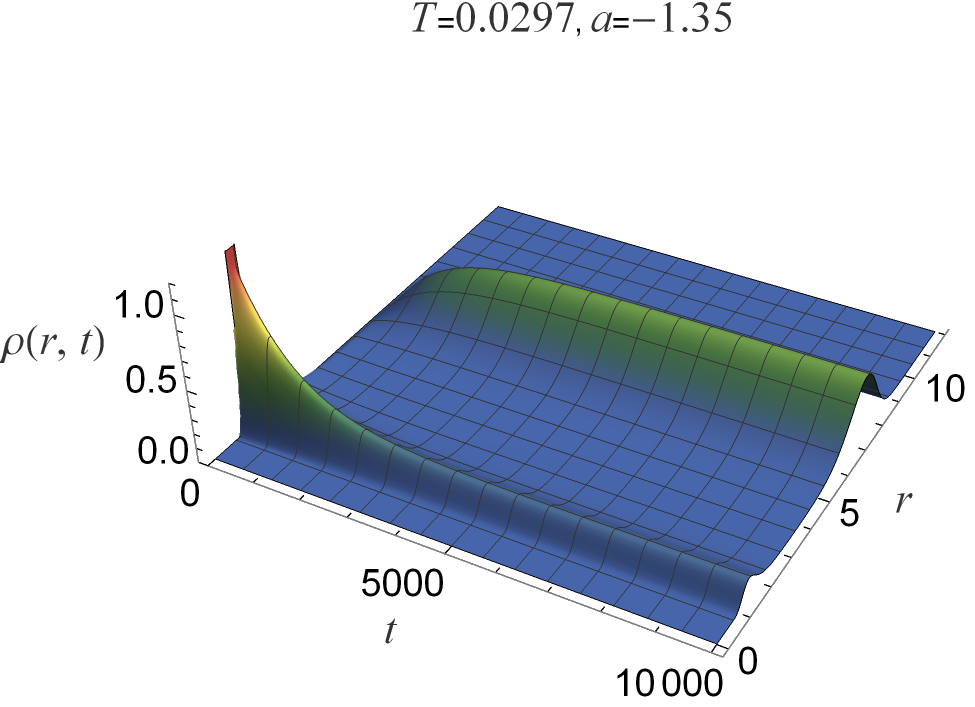}\\\vspace{0.3cm}
  \includegraphics[width=6.5cm]{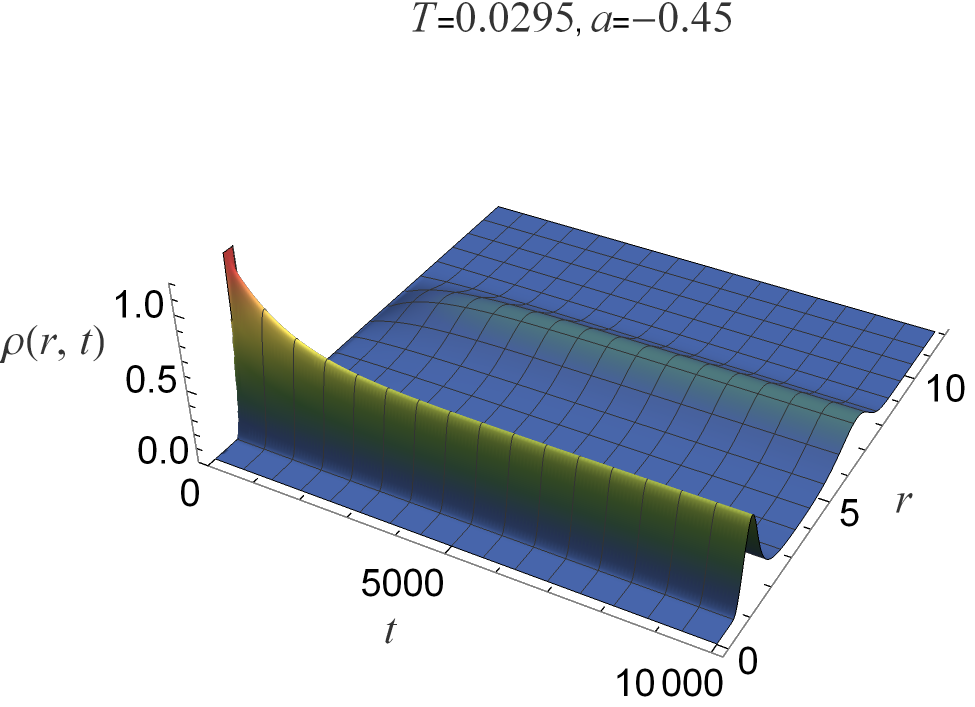}
  \includegraphics[width=6.5cm]{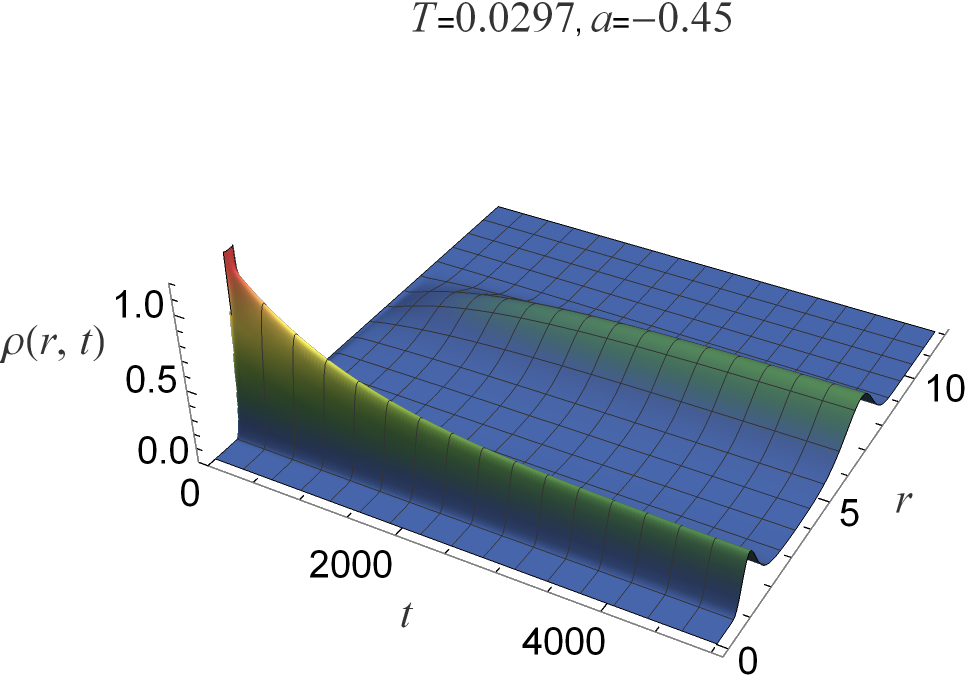}
  \caption{The distributions of probability $\rho(r, t)$ as a function of $r$ and $t$ for different $T$ and $a$. In the left and right columns, $T=0.0295$ and $0.0297$. In the top and bottom rows, $a=-1.35$ and $-0.45$. The initial wave packet is located at the small black hole.}
  \label{aFuS3D}
\end{figure}
\begin{figure}[H]
  \centering
  \includegraphics[width=6.5cm]{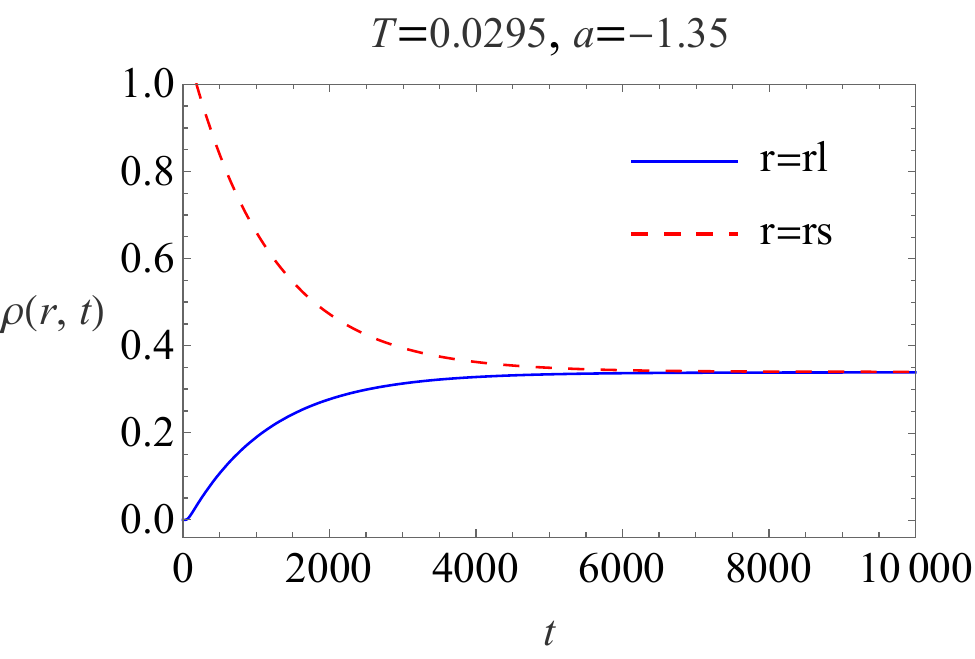}
  \includegraphics[width=6.5cm]{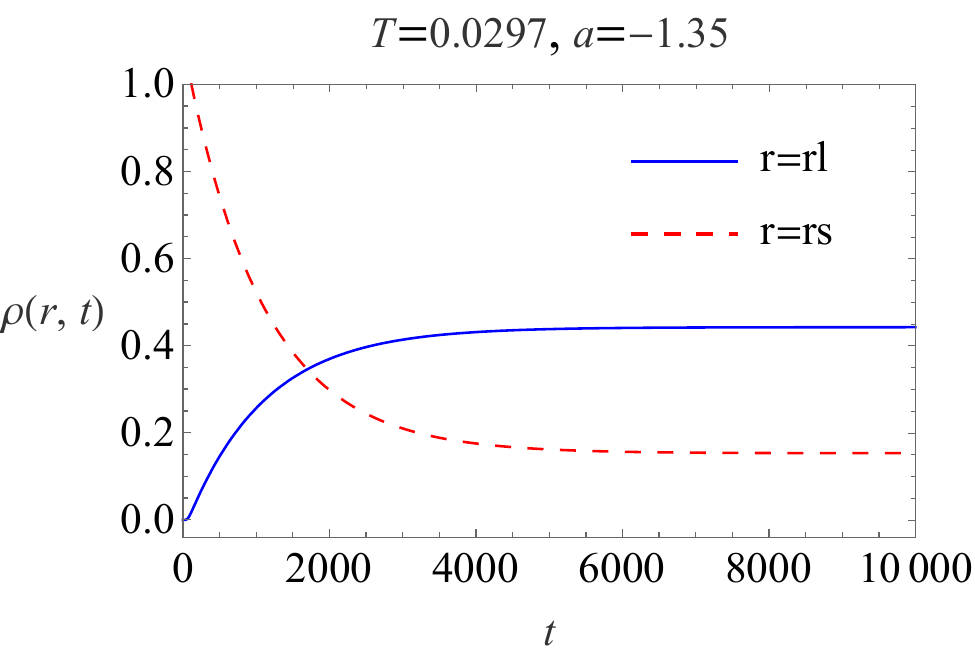}\\
  \includegraphics[width=6.5cm]{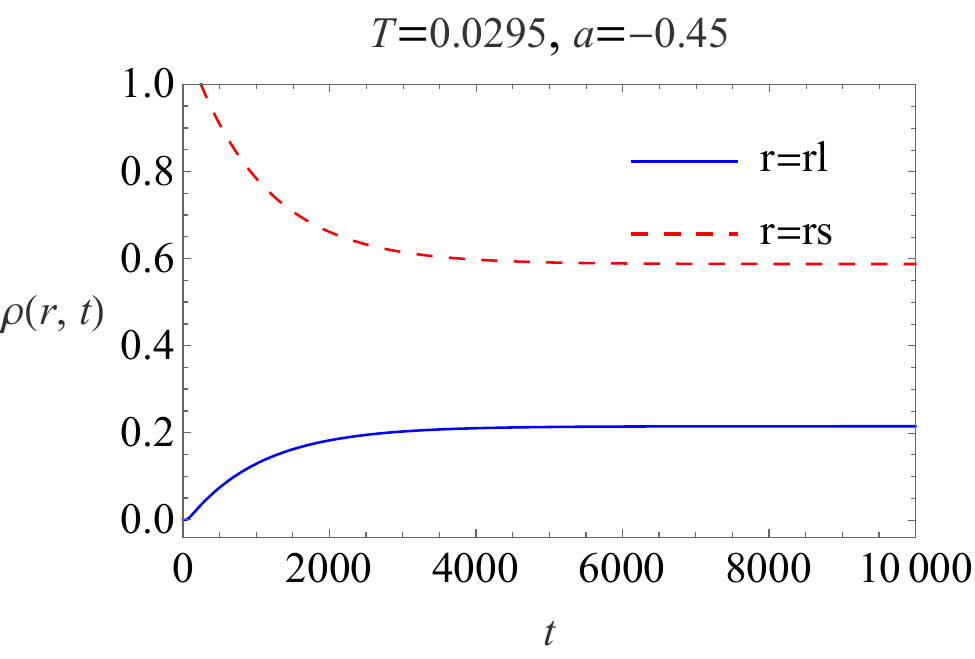}
  \includegraphics[width=6.5cm]{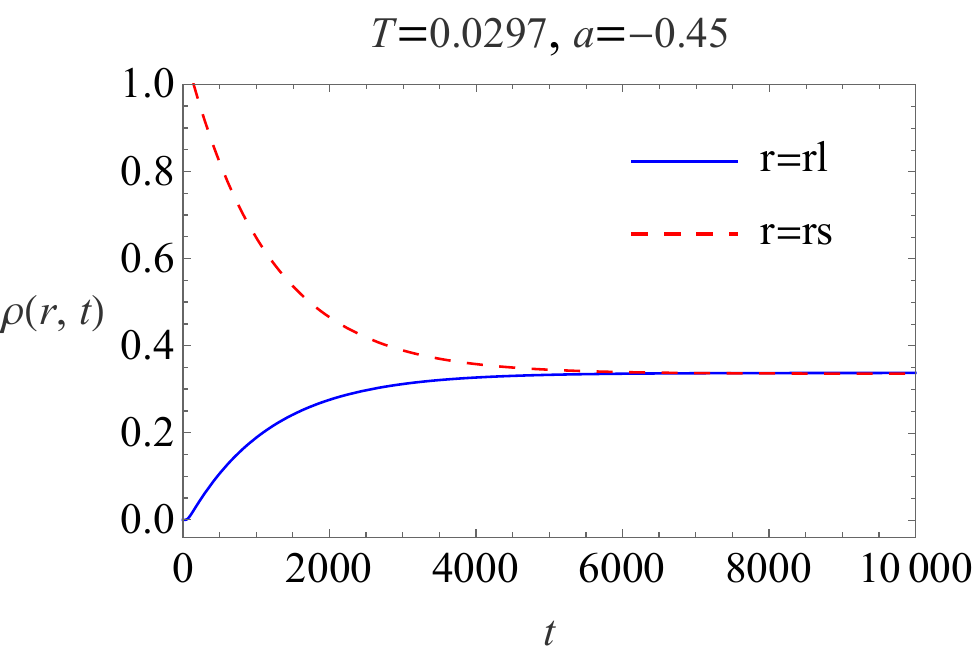}
  \caption{Behaviors of the probability $\rho(r, t)$ as a function of $t$ for different $T$ and $a$. In the left and right columns, $T=0.0295$ and $0.0297$. In the top and bottom rows, $a=-1.35$ and $-0.45$. The initial wave packet is located at the small black hole.}
  \label{aFuS2D}
\end{figure}
\begin{figure}[H]
  \centering
   \includegraphics[width=6.5cm]{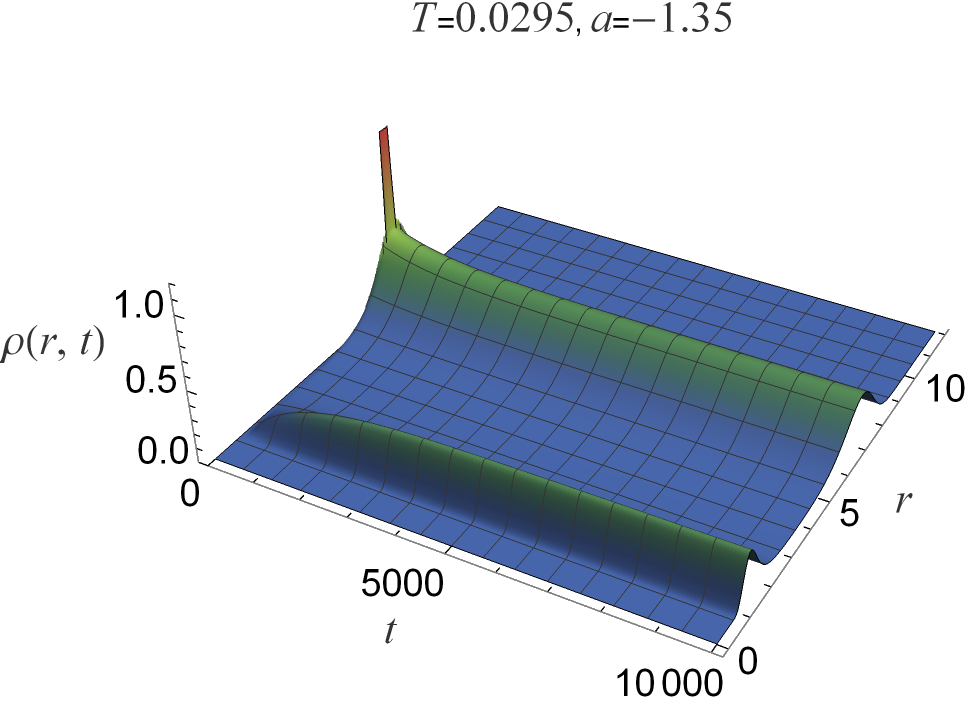}
  \includegraphics[width=6.5cm]{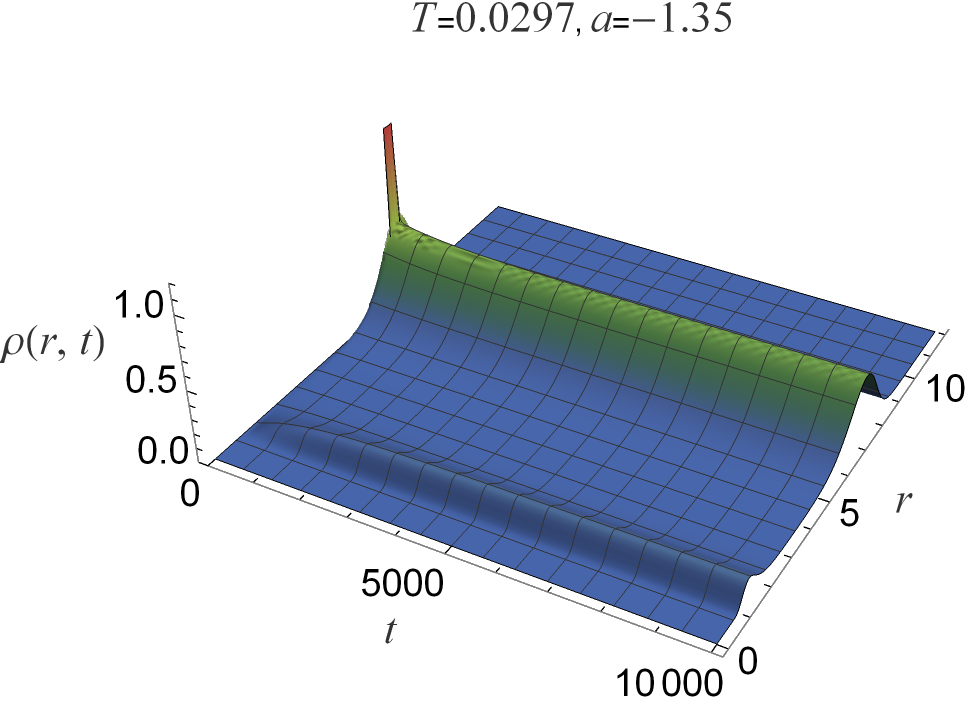}\\\vspace{0.3cm}
  \includegraphics[width=6.5cm]{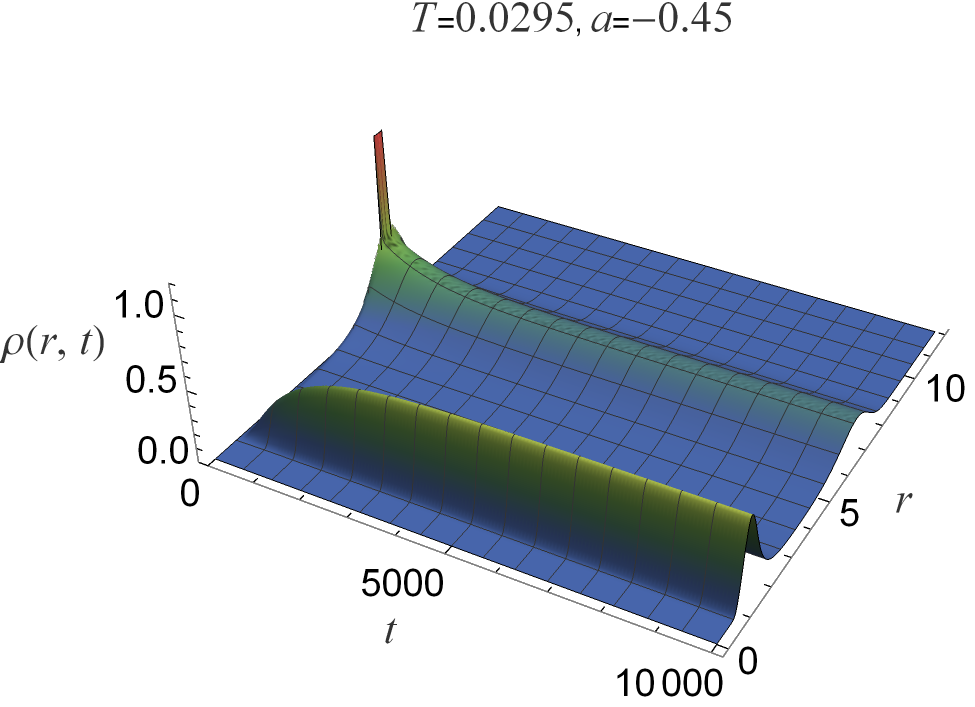}
  \includegraphics[width=6.5cm]{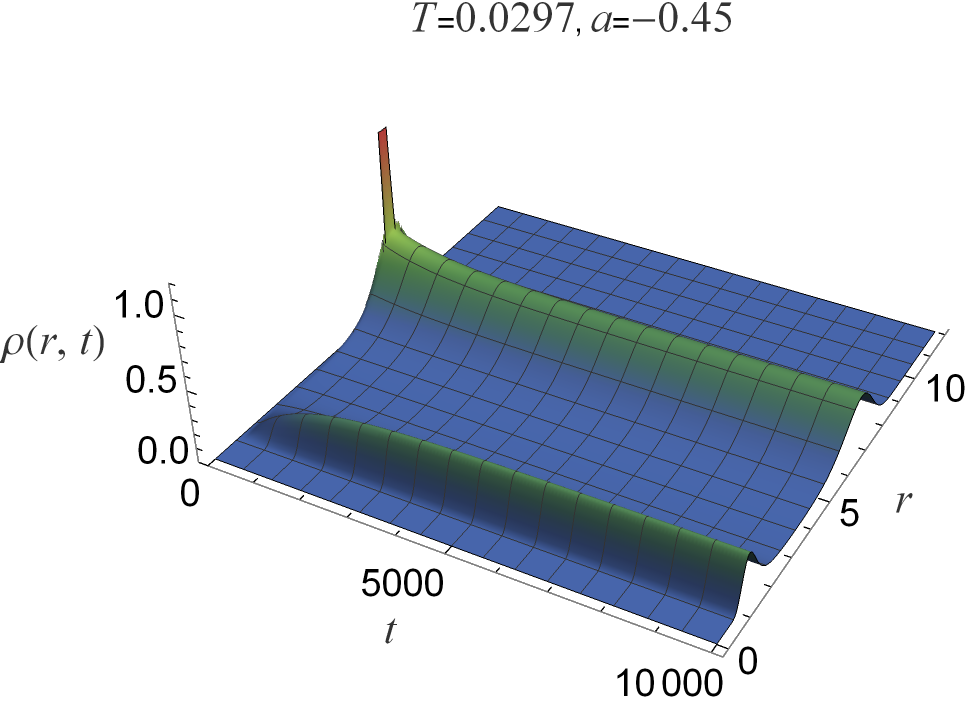}
  \caption{The distributions of probability $\rho(r, t)$ as a function of $r$ and $t$ for different $T$ and $a$. In the left and right columns, $T=0.0295$ and $0.0297$. In the top and bottom rows, $a=-1.35$ and $-0.45$. The initial wave packet is located at the large black hole.}
  \label{aFuL3D}

\end{figure}
\begin{figure}[H]
  \centering
  \includegraphics[width=6.5cm]{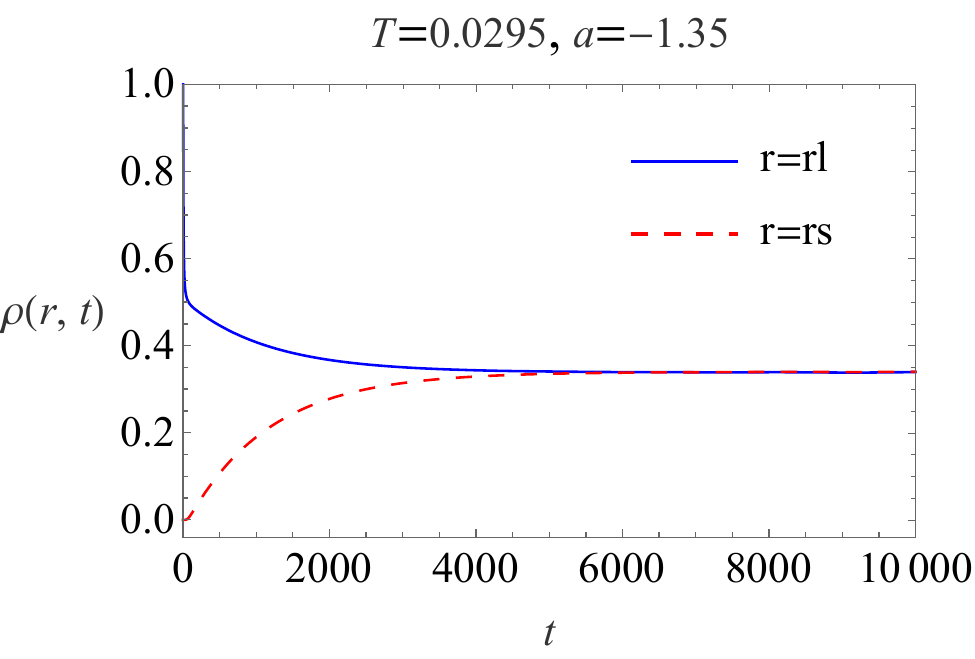}
  \includegraphics[width=6.5cm]{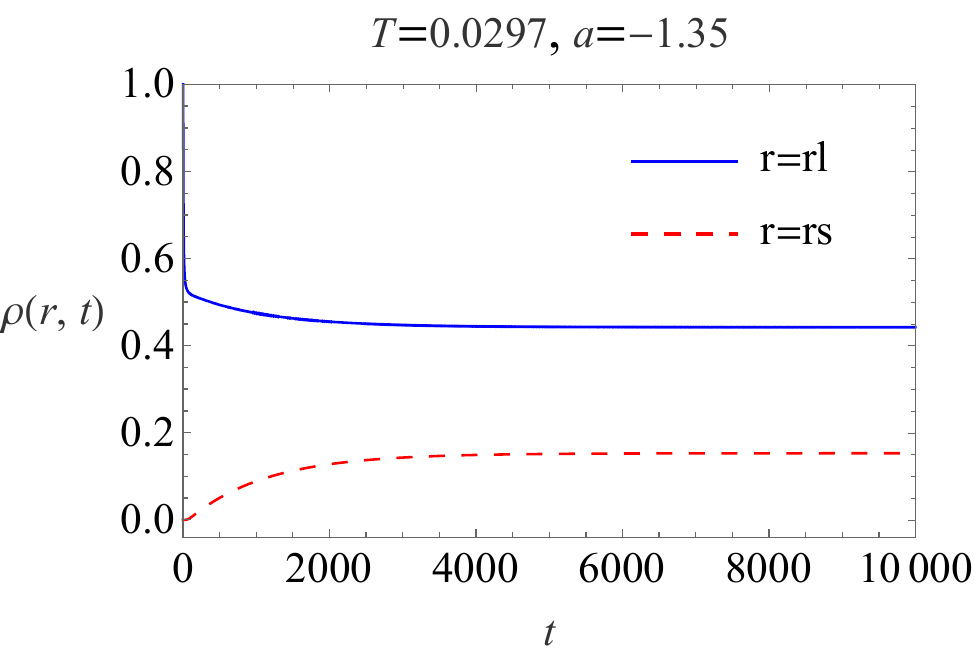}\\
  \includegraphics[width=6.5cm]{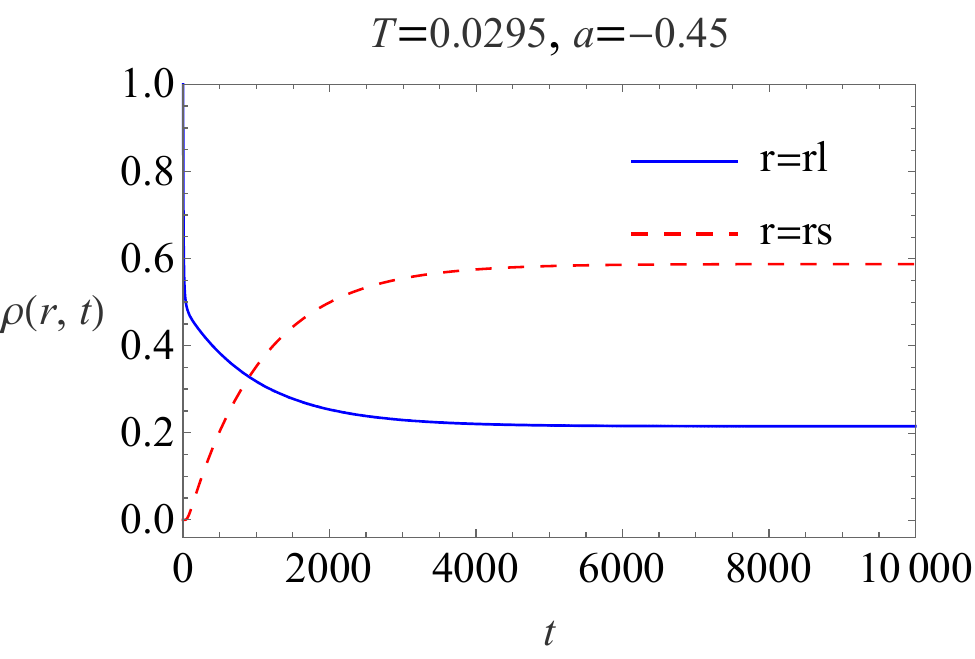}
  \includegraphics[width=6.5cm]{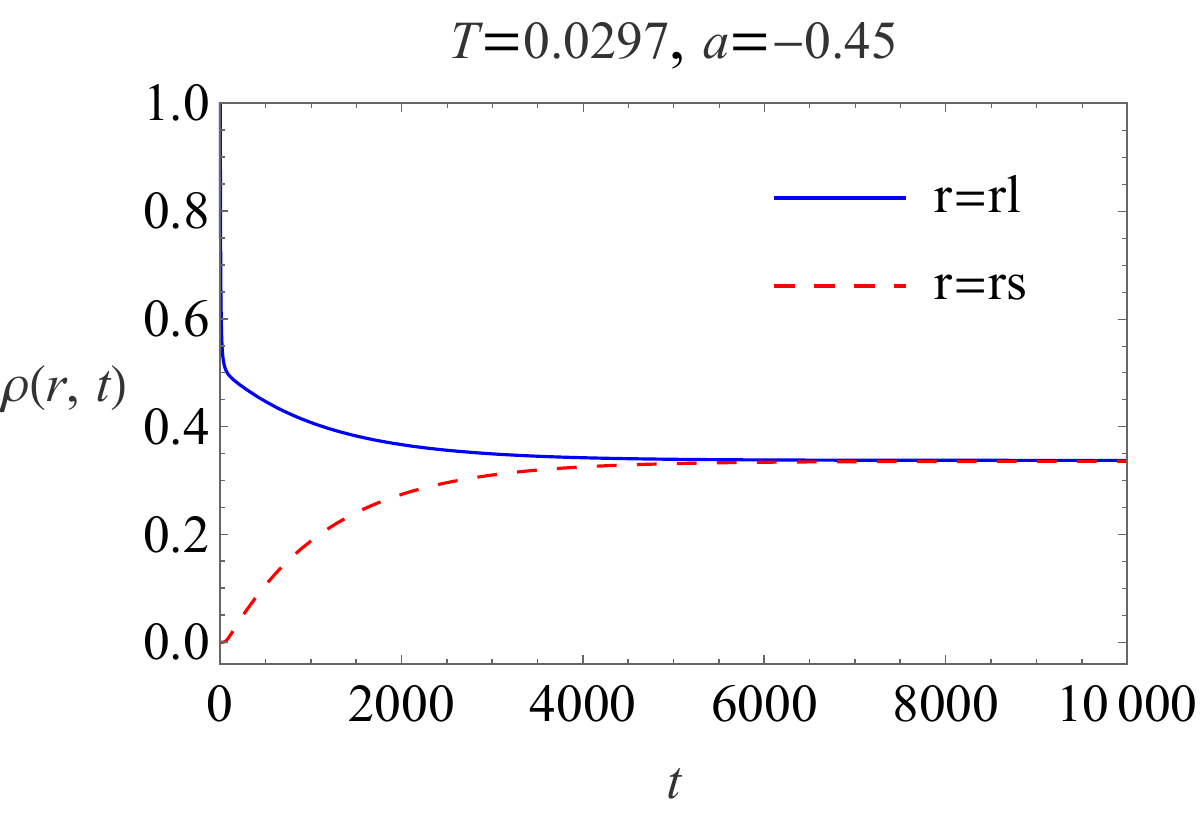}
  \caption{Behaviors of the probability $\rho(r, t)$ as a function of $t$ for different $T$ and $a$. In the left and right columns, $T=0.0295$ and $0.0297$. In the top and bottom rows, $a=-1.35$ and $-0.45$. The initial wave packet is located at the large black hole.}
  \label{aFuL2D}
\end{figure}

For the case of $a<0$, we can ensure the positivity of the black hole mass with $r>0$. Therefore we set the reflecting boundary condition at $r=0.01$ and $r=12$ in order to avoid the numerical instability. We discuss the time evolution of the probability distribution $\rho(r, t)$. When the initial wave packet is located at the small black hole. In Fig. \ref{aFuS3D} and Fig. \ref{aFuS2D}, we exhibit the time dependent behaviors of the probability distribution of the black holes at different $T$ and $a$. We observe that the probability $\rho(r, t)$ of small black hole decreases and finally reaches a stable value with the evolution of time, and a larger stable value can be acquired for a lower temperature $T$ or larger Euler-Heisenberg parameter $a$. Furthermore, we see that it decreases faster in a short time for a higher $T$ or smaller Euler-Heisenberg parameter $a$. The probability $\rho(r, t)$ of large black hole increases and finally attains a stable value with the evolution of time, and a larger stable value can be acquired for higher temperature $T$ or smaller Euler-Heisenberg parameter $a$. The coexistent small and large black hole states can be acquired for $T=0.0295$ and $a=-1.35$, or $T=0.0297$ and $a=-0.45$.
Now, the initial wave packet is located at the large black hole. In Fig. \ref{aFuL3D} and Fig. \ref{aFuL2D}, we plot the probability distribution of the black holes at different $T$ and $a$. We observe that the probability $\rho(r, t)$ of large black hole decreases and finally reaches a stable value with the evolution of time, and a larger stable value can be acquired for a higher temperature $T$ or smaller Euler-Heisenberg parameter $a$. The probability $\rho(r, t)$ of small black hole increases and finally attains a stable value with the evolution of time, and a larger stable value can be acquired for lower temperature $T$ or larger Euler-Heisenberg parameter $a$. We observe that the effect of the Euler-Heisenberg parameter and temperature are similar to the case of the initial wave packet located at the small black hole. The coexistent small and large black hole states can be acquired for the same conditions.

\begin{figure}[H]
  \centering
  \includegraphics[width=6.5cm]{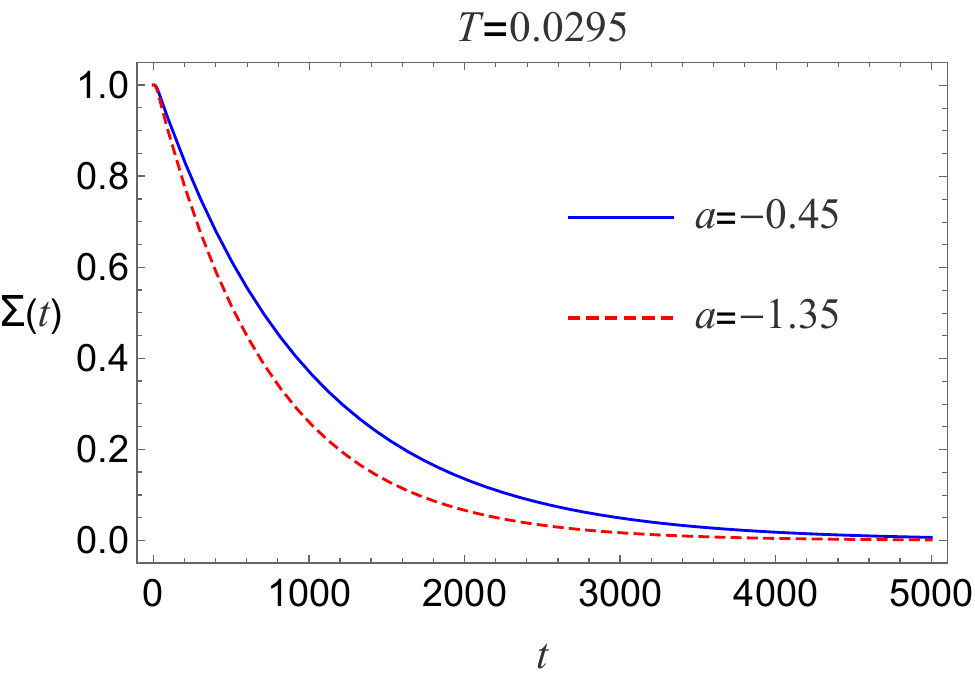}
  \includegraphics[width=6.5cm]{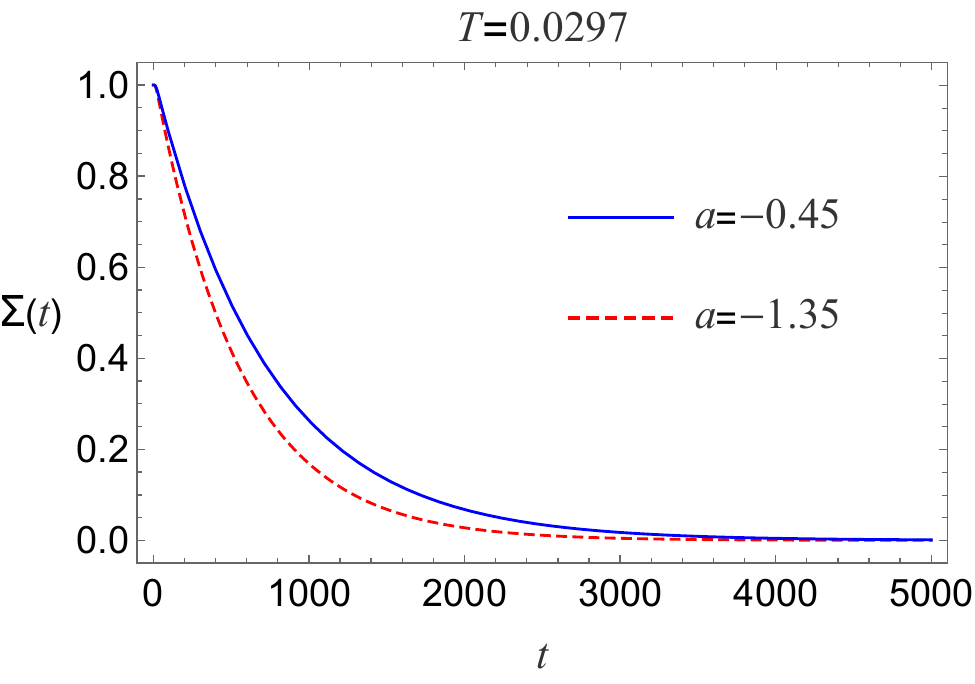}
  \caption{The probability $\Sigma(t)$ for different $T$ and $a$. The solid (blue) and dashed (red) lines represent $a=-0.45$ and $-1.35$. We take $T=0.0295$ (the left panel) and $T=0.0297$ (the right panel). The initial wave packet is located at the small black hole.}
  \label{afus2d}
\end{figure}

The initial wave packet is located at the small black hole. In Fig.~\ref{afus2d}, we present the probability $\Sigma(t)$ for different $a$ with $T=0.0295$ and $T=0.0297$. We see that the probability $\Sigma(t)$ decreases faster for higher temperature $T$. Moreover, the smaller the Euler-Heisenberg parameter $a$ is, the faster the probability decreases.
\begin{figure}[H]
  \centering
  \includegraphics[width=6.5cm]{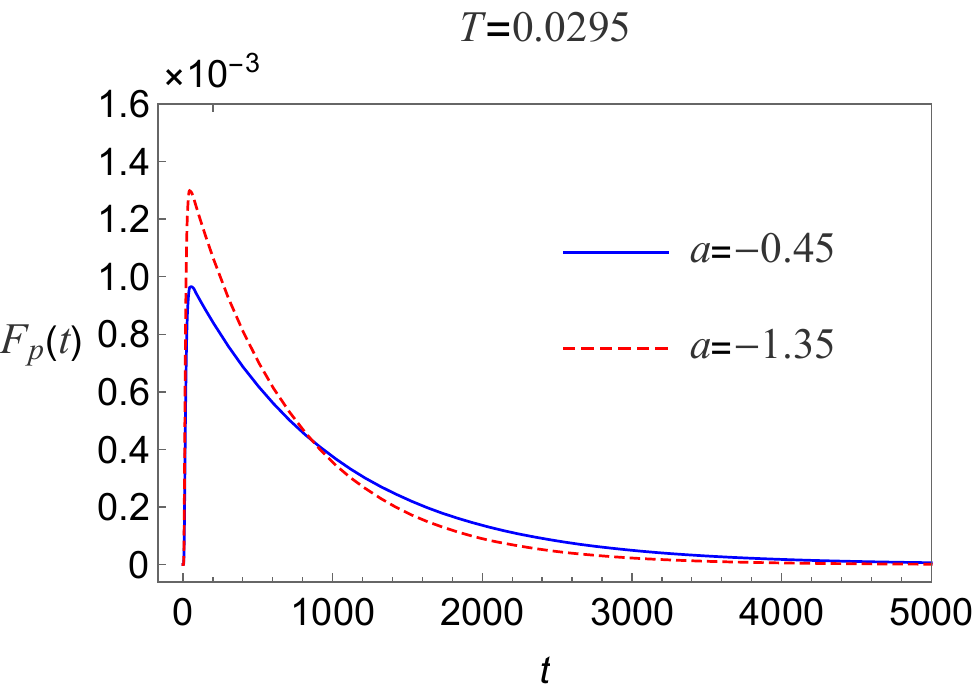}
  \includegraphics[width=6.5cm]{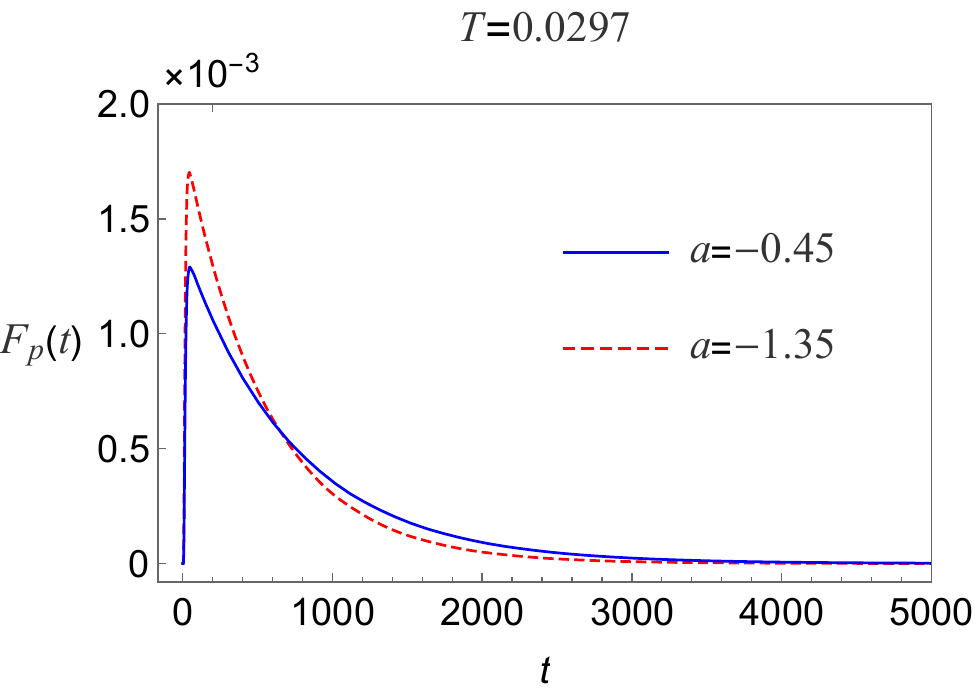}
  \caption{The distributions of the first passage time $F_{p}(t)$ for different $T$ and $a$. The solid (blue) and dashed (red) lines represent $a=-0.45$ and $-1.35$. We take $T=0.0295$ (the left panel) and $T=0.0297$ (the right panel). The initial wave packet is located at the small black hole.}
  \label{afus2df}
\end{figure}
In Fig. \ref{afus2df}, we display the distribution of the first passage time for different $T$ and $a$. There exists a single peak for each curve at a short time. This implies that a considerable fraction of the first passage events takes place at a short time before $F_p(t)$ approaches its exponential decay form. When the temperature increases or Euler-Heisenberg parameter decreases, the peak increases and is shifted to the left.
\begin{figure}[H]
  \centering
  \includegraphics[width=6.5cm]{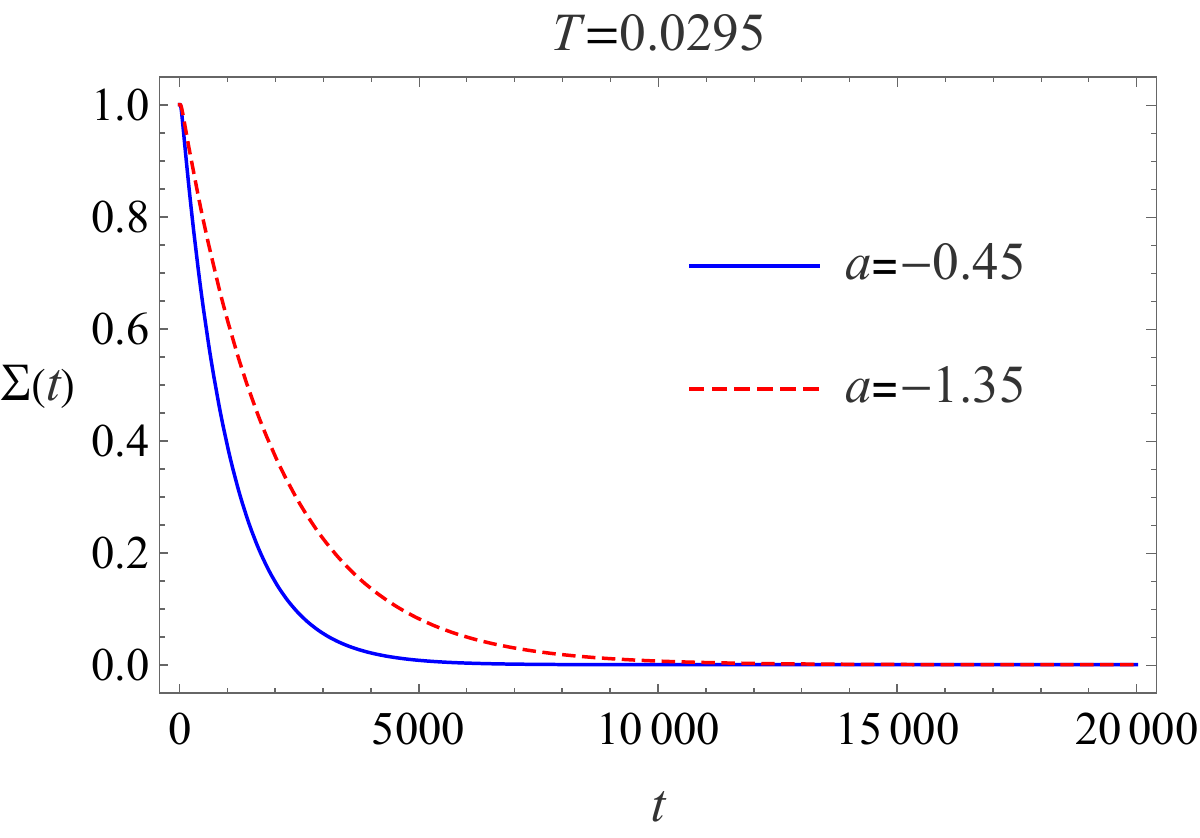}
  \includegraphics[width=6.5cm]{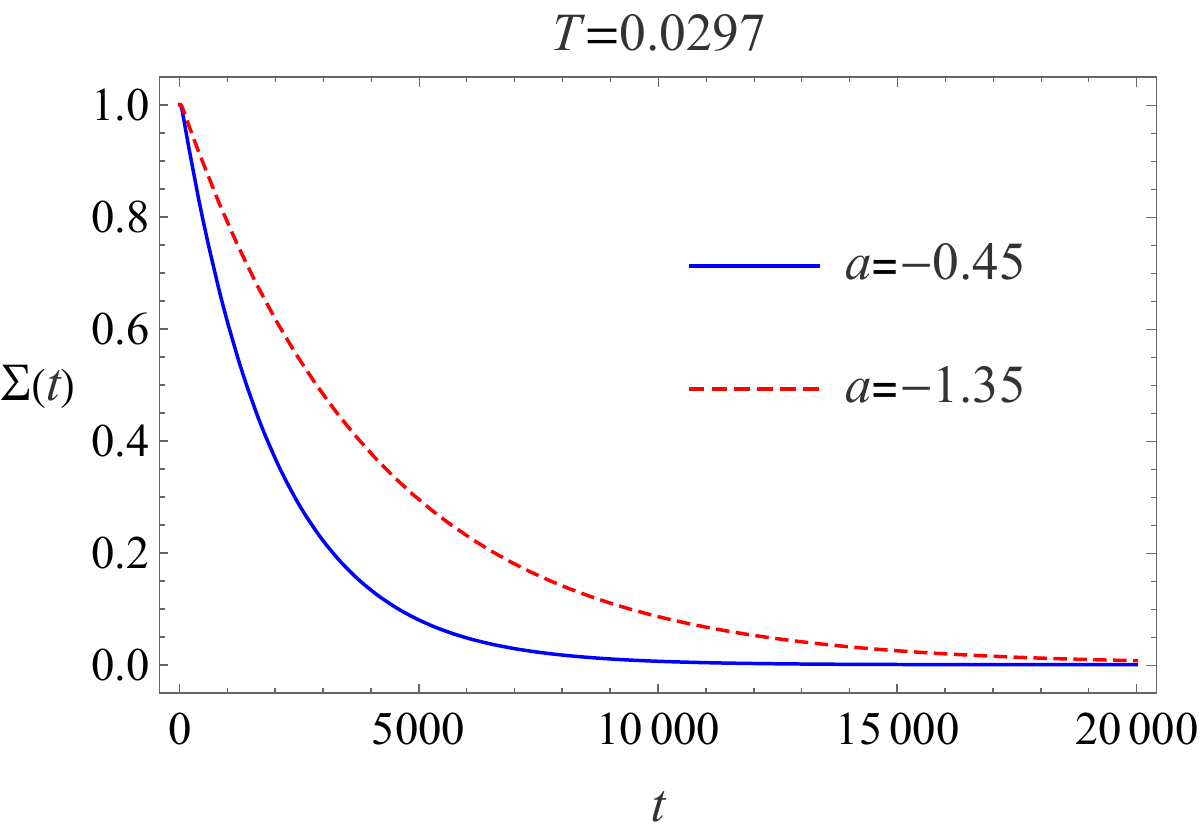}
  \caption{The probability $\Sigma(t)$ for different $T$ and $a$. The solid (blue) and dashed (red) lines represent $a=-0.45$ and $-1.35$. We take $T=0.0295$ (the left panel) and $T=0.0297$ (the right panel). The initial wave packet is located at the large black hole.}
  \label{aful2d}
\end{figure}
Now, the initial wave packet is located at the large black hole. From Fig.~\ref{aful2d}, we observe that the probability $\Sigma(t)$ decreases faster for lower temperature $T$. In addition, the larger the Euler-Heisenberg parameter $a$ is, the faster the probability decreases. This is quite different from the initial small black hole state.
\begin{figure}[H]
  \centering
 \includegraphics[width=6.5cm]{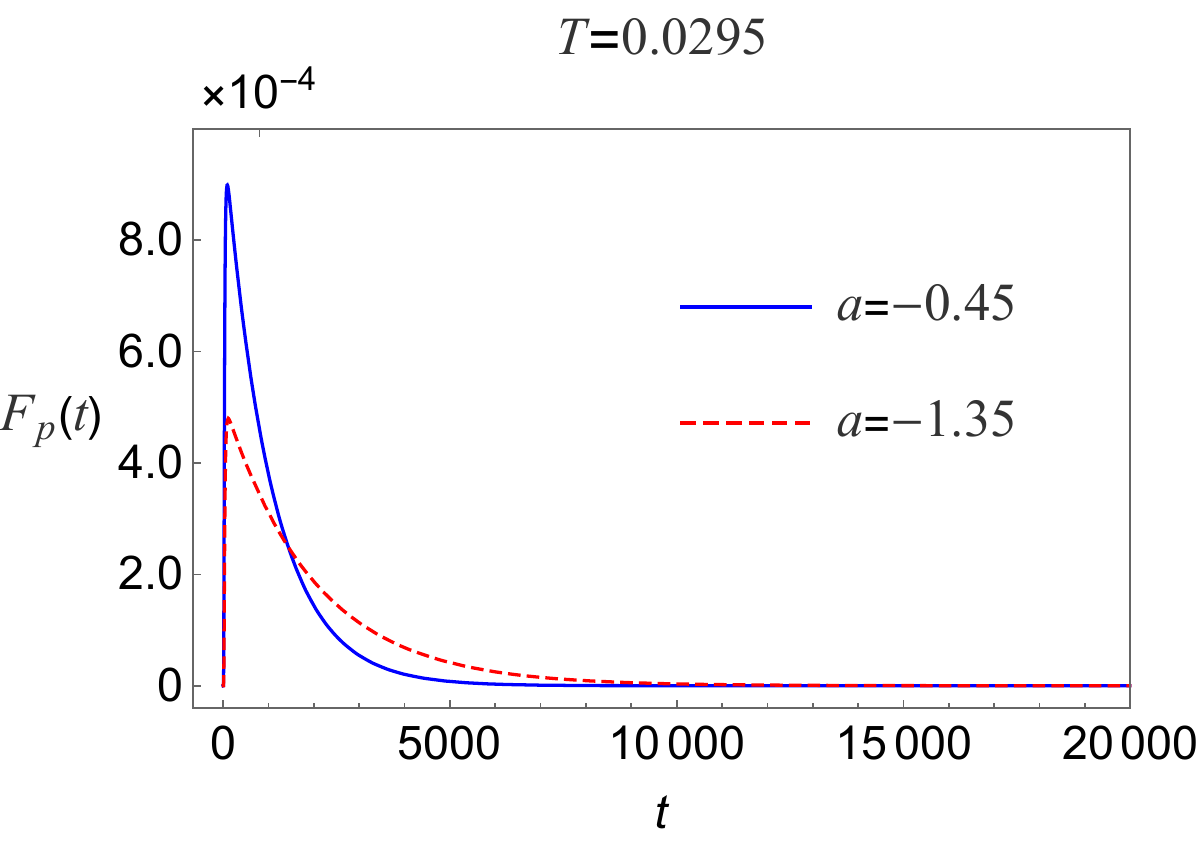}
  \includegraphics[width=6.5cm]{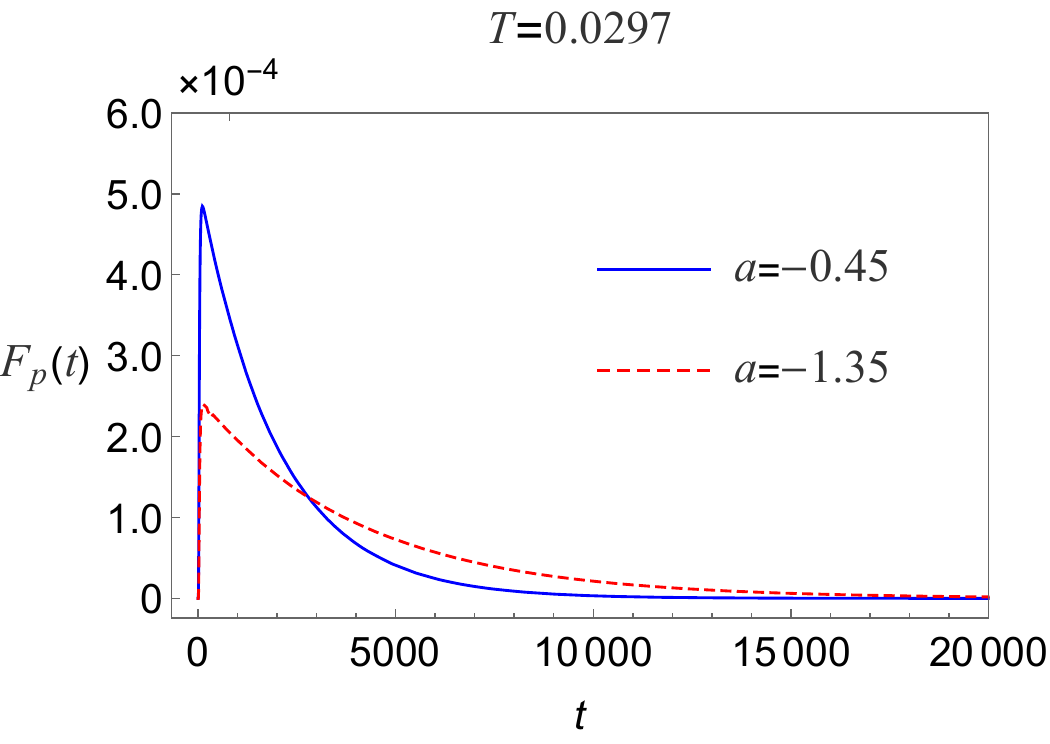}
  \caption{The distributions of the first passage time $F_{p}(t)$ for different $T$ and $a$. The solid (blue) and dashed (red) lines represent the Euler-Heisenberg parameters $a=-0.45$ and $-1.35$, and we take $T=0.0295$ (the left panel) and $T=0.0297$ (the right panel). The initial wave packet is located at the large black hole.}
  \label{aful2df}
\end{figure}
In Fig. \ref{aful2df}, we show the distribution of the first passage time for different $T$ and $a$. There also exists a single peak for each curve at short time. The temperature decreases or Euler-Heisenberg parameter increases, the peak increases. This is also quite different from the initial small black hole state.

\section{conclusions}

We studied the small-large black hole phase transition for the Euler-Heisenberg-AdS black hole on the free energy landscape. For $0< a\leq \frac{32}{7} Q^2 $, we observed that the small-large black hole phase transition can be acquired with a small $a$. The local minimal points correspond to the local stable black holes, and the system prefers the state with lower Gibbs free energy. Furthermore, we studied the probability distribution of the system states by solving the Fokker-Planck equation. For a small $a$, a higher (lower) $T$ corresponds to a larger probability for a large (small) black hole. However, the probability of small (large) black hole will decrease and goes to zero for a large $a$. The coexistent small and large black hole states can be acquired for some conditions. We analyzed the first passage process for the small-large black hole phase transition. We observed that there is a peak for the distribution of the first passage time, which implies that a considerable fraction of the first passage events occurs at a short time. When the temperature increases (decreases) or the Euler-Heisenberg parameter decreases (increases) for the initial wave packet located at the small (large) black hole, the peak of the first passage time increases.

For $a<0$, we observed that the small (large) black hole can switch to the large (small) black hole due to the change of the temperature $T$ or Euler-Heisenberg parameter $a$. A higher (lower) $T$ corresponds to a larger probability for a large (small) black hole. A smaller (larger) $a$ corresponds to a larger probability for a large (small) black hole. The coexistent small and large black hole states can be acquired for some conditions. We also analyzed the first passage process for the small-large black hole phase transition. A higher peak can be acquired for higher (lower) $T$ or smaller (larger) $a$ with initial small (large) black hole state.

\begin{acknowledgments}

This work was supported by the Nature Science Foundation of Shaanxi Province, China under Grant No. 2023-JC-YB-016 and the National Natural Science Foundation of China under Grant No. 11705144.

\end{acknowledgments}

\end{document}